%% file: ACSchroeder.tex
\documentclass[letterpaper]{aa}
\usepackage{graphics,lscape,epsfig}

\newcommand{\HI}{\protect\normalsize H\thinspace\protect\footnotesize
I\protect\normalsize} 
\newcommand{\HII}{\protect\normalsize H\thinspace\protect\footnotesize
II\protect\normalsize}

\newcommand{\B}{$B$}
\newcommand{\II}{$I$}
\newcommand{\J}{$J$}
\newcommand{\K}{$K$}

\newcommand{\IJK}{{$I J K$}}
\newcommand{\JHK}{{$J H K$}}

\newcommand{\bk}{\mbox{($B-K$)}}
\newcommand{\ij}{\mbox{$I-J$}}
\newcommand{\jk}{\mbox{$J-K$}}
\newcommand{\ik}{\mbox{$I-K$}}
\newcommand{\ije}{\mbox{$(I-J)^0$}}
\newcommand{\jke}{\mbox{$(J-K)^0$}}
\newcommand{\ike}{\mbox{$(I-K)^0$}}

\newcommand{\kms}{\,km\,s$^{-1}$}

\newcommand{\etal}{et~al.\ }
\newcommand{\cf}{{cf.}\ }
\newcommand{\eg}{{e.g.},\ }         
\newcommand{\ie}{{i.e.},\ }         
\newcommand{\PKS}{PKS\,1343\,--\,601}

\def\deg{{^\circ}}

\begin{document}

%
\title{The highly obscured region around \PKS\ -- I. 
Galactic interstellar extinctions using DENIS galaxy colours}

\author{Anja C. Schr\"oder \inst{1} 
\and    Gary A. Mamon \inst{2,3}
\and    Ren\'ee C. Kraan-Korteweg \inst{4,5}
\and    Patrick A. Woudt \inst{5}
}
 
\offprints{A. Schr\"oder}
 
\institute{Department 
of Physics \& Astronomy,
              University of Leicester,
              University Road,
              Leicester LE1 7RH,
              United Kingdom
\and 
              Institut d'Astrophysique de Paris (UMR 7095: CNRS \&
	      Universit\'e Pierre \& Marie Curie)
              98 bis Blvd Arago, 
              F--75014 Paris, 
              France
\and 
              GEPI (UMR 8111: CNRS \& Universit\'e Denis Diderot), 
              Observatoire de Paris, 
              F--92195 Meudon, 
              France
\and 
              Depto.\ de Astronom\'\i{a}, 
              Univ.\ de Guanajuato, 
              Ap.\ P. 144,
              Guanajuato, GTO 36000
              Mexico
\and 
              Department of Astronomy, 
              University of Cape Town,
              Private Bag X3,
              Rondebosch 7701,
              South Africa
}
 
\date{Received date; accepted date}
 
\abstract{The highly obscured radio-bright galaxy PKS\,1343\,--\,601 at
Galactic coordinates of $(l,b) = (309\fdg7, +1\fdg8$) has been suspected to
mark the centre of a hitherto unknown cluster in the wider Great Attractor
region. As such it presents an ideal region for a search of galaxies in the
near-infrared (NIR) and an in-depth study of their colours as a function of
extinction. A visual search of a $\sim\!30$ square-degree area centered on
this radio galaxy on images of the NIR DENIS survey (\IJK ) revealed 83
galaxies (including two AGNs) and 39 possible candidates. Of these, 49 are
also listed in the 2MASS Extended Source Catalog 2MASX. Taking the
IRAS/DIRBE extinction values (Schlegel \etal 1998) at face value, the
absorption in the optical ($A_B$) ranges from $\sim\!2^{\rm m}$ to over
$100^{\rm m}$ across the Galactic Plane. Comparing the detections with
other systematic surveys, we conclude that this search is highly complete
up to the detection limits of the DENIS survey and certainly surpasses any
automatic galaxy finding algorithm applied to crowded areas.

The NIR galaxy colours from the $7\arcsec$ aperture were used as a probe to
measure total Galactic extinction. A comparison with the IRAS/DIRBE
Galactic reddening maps suggests that the IRAS/DIRBE values result in a
slight overestimate of the true extinction at such low Galactic latitudes;
the inferred extinction from the galaxy colours corresponds to about 87\%
of the IRAS/DIRBE extinctions. Although this determination still shows
quite some scatter, it proves the usefulness of NIR surveys for calibrating
the IRAS/DIRBE maps in the extinction range of about $2^{\rm{m}} \la A_B \la
12^{\rm{m}}$.

\keywords{Galaxies: clusters: individual -- Galaxies: fundamental
parameters -- Galaxies: photometry -- extinction} 
}

\maketitle
 
 
\section{Introduction}  \label{intro}

Various extragalactic large-scale structures are hidden behind the dust and
stars of the Milky Way, the so-called Zone of Avoidance (ZoA), resulting in
a poor understanding of the dynamics of the nearby Universe; for a detailed
overview see Kraan-Korteweg \& Lahav (2000), Kraan-Korteweg (2005), and the
conference proceedings ``Nearby Large-Scale Structures and the Zone of
Avoidance'' (Fairall \& Woudt 2005). The Great Attractor (GA), an extended
mass overdensity in the nearby Universe, lies for instance close to the
crossing of the Supergalactic plane and the Galactic plane. Its presence
was inferred by the systematic large-scale flow of elliptical galaxies
(Lynden-Bell \etal 1988). Applying the potential reconstruction method of
the mass density field POTENT (Dekel 1994), Kolatt \etal (1995) found its
centre at $(l,b,v) = (320\deg, 0\deg, 4000$\kms ).

Close to the potential well of the GA lies the cluster ACO 3627 $(l,b,v) =
(325\deg, -7\deg, 4848$\kms ). A deep optical galaxy search (Woudt \&
Kraan-Korteweg 2001) revealed this cluster to be as massive and rich a
cluster as the Coma cluster (Kraan-Korteweg \etal 1996, Woudt \etal
2005). It therefore most likely marks the centre of the potential well of
the GA.  However, the GA is an extended region of high galaxy density
(about $40\deg \times 40\deg$ on the sky, see Kolatt \etal 1995), and other
clusters (rich and poor) may well contribute substantially to this mass
overdensity. Identifying them is a challenge as the central part of the
wider GA area lies behind the thickest dust layer of the Milky Way.

About $10\deg$ from the Norma cluster, at $(l,b) = (309\fdg7, +1\fdg8$),
lies the galaxy PKS\,1343\,--\,601 with a recession velocity of 3872\kms\
(West \& Tarenghi 1989). Near-infrared (NIR) observations revealed
PKS\,1343\,--\,601 to be a giant elliptical galaxy, which often reside at
the centre of galaxy clusters. It is also one of the brightest radio
sources in the sky (McAdam 1991): its flux density is only surpassed by
Cygnus A, Centaurus A, Virgo A, and Fornax A. Two of these four radio
sources are situated at the centre of a rich cluster, one in a smaller
cluster, and one in a group of galaxies (Jones \etal 2001). This evidence
motivated Kraan-Korteweg \& Woudt (1999) to investigate by different means
whether PKS\,1343\,--\,601 points to another cluster in the Great Attractor
region. Such a cluster would have a considerable impact on the local
velocity field calculations.

Results are still controversial. A preliminary analysis of the systematic
deep \HI\ search for galaxies with the Parkes Multibeam receiver found a
concentration of galaxies in redshift space around this radio galaxy
(Kraan-Korteweg \etal 2005b). A deep NIR search (\JHK ) of half a degree
radius and a deep \II -band survey of 2 degrees around PKS\,1343\,--\,601
(Nagayama \etal 2004; Kraan-Korteweg \etal 2005a, respectively) have
revealed a distribution of galaxies consistent with a (medium-sized)
cluster around PKS\,1343\,--\,601. X-ray observations with ASCA have only
revealed diffuse emission from PKS\,1343\,--\,601 itself (Tashiro \etal
1998, see also the discussion in Ebeling \etal 2002), which would rule out
a rich cluster.

This paper presents the results of a search for galaxies based on the NIR
DENIS survey (Epchtein \etal 1997) in a much larger but shallower area than
the above ones. The advantages of using the NIR to search for galaxies in
the ZoA are manifold: (i) the NIR is less affected by the foreground
extinction than the optical (the extinction in the \K -band is about 10\%
of the extinction in the \B -band); (ii) the NIR is sensitive to early-type
galaxies, tracers of massive groups and clusters (contrarily to
far-infrared and blind \HI\ surveys); (iii)
the NIR shows little confusion with Galactic objects such as young stellar
objects and cool cirrus sources.

In pilot studies, we have assessed the performance of the DENIS survey at
low Galactic latitudes (Schr\"oder \etal 1997; Kraan-Korteweg \etal 1998;
Schr\"oder \etal 1999; Mamon et al. 2001). We tested the potential of the
DENIS survey to detect galaxies where optical and far-infrared surveys
fail, \ie at high foreground extinctions and in crowded regions; we
established that the NIR colours of galaxies lead to values for the
foreground extinction; and we cross-identified highly obscured galaxies
detected in a blind \HI\ search at $|b| < 5\deg$. Overall, both systematic
NIR surveys DENIS (\IJK ; Paturel \etal 2003, Vauglin \etal 1999) and 2MASS
(\JHK ; Skrutskie \etal 2006, Jarrett \etal 2000a) have proven their
effectiveness in penetrating the ZOA (Jarrett \etal 2000b, Rousseau \etal
2000, Schr\"oder \etal 2000) -- as long as the star density does not exceed
a certain value (Kraan-Korteweg \& Jarrett 2005).

In the following, we will introduce the DENIS survey and the implication of
extinction on galaxy counts in general (Sections~\ref{denis}
and~\ref{counts}, respectively). We will then describe the search area and
the quality of the DENIS data (Sect.~\ref{strips}), and the methods of
galaxy and parameter extraction (Sect.~\ref{param}). In Sect.~\ref{cat} the
catalogue data are described, Sect.~\ref{lit} gives a detailed comparison
with the data of other searches and catalogues in this area, and in
Sect.~\ref{ext} we investigate the extinction in this area using the
derived NIR colours. Conclusions are presented in the final
Sect.~\ref{conclusion}. Throughout the paper, we assume a Hubble constant
of $H_0=70 \rm km\,s^{-1}\,Mpc^{-1}$.

A second paper will provide a detailed discussion of the local environment
of PKS\,1343\,--\,601 using the local galaxy density, the velocity
distribution, and the X-ray luminosity to assess its mass and contribution
to the GA overdensity (Schr\"oder \& Mamon 2006, hereafter Paper II).

\section{The DENIS survey } \label{denis}

The DENIS survey (DEep Near-Infrared Survey of the southern sky) is a
European joint program that simultaneously imaged the sky in the
Gunn-$i$\,($0.82\,\mu$m, hereafter \II ), \J\ ($1.25\,\mu$m) and $K_s$
($2.15\,\mu$m, hereafter \K ) passbands with a resolution of $1\arcsec$ in
\II\ and $3\arcsec$ in \J\ and \K\ (Epchtein 1997, 1998).  The observations
were carried out between 1995 and 2001 with the dedicated 1\,m ESO
telescope at La Silla (Chile). About 92\% of the southern sky ($+2\deg \le
{\rm DEC} \le -88\deg$) has been covered.

DENIS images have a field of view of $12\arcmin \times 12\arcmin$. Exposure
times per field and band are $9\,$s. The observing mode consisted of
step-and-stare scans of 180 images in declination, resulting in
\emph{strips} of $12\arcmin \times 30\deg$. The overlap region between
images are $1\arcmin$ on each side.  Any given $12'\times30^\circ$
\emph{slot} in the sky was usually observed once and, depending on the
quality of the images and/or weather conditions of the previous
observations, repeated.

The reduction process of the DENIS images consisted in bias corrections and
flat-fielding. The latter was done using an iterative fitting of the pixel
response over the night relative to the mean over the image centre
(Borsenberger 1997). The images were then smoothed with kernels the size of
which is a function of the wavelength of the spectral waveband.  Objects
were extracted and measured with {\tt SExtractor} (Bertin \& Arnouts 1996).
For the purpose of the analysis of the ZoA the star/galaxy separation was,
however, performed visually.

To ensure homogeneous quality over the whole survey, the DENIS processing
centre in Paris (PDAC) recently re-processed all strips with the latest
software of the pipeline. The limiting magnitudes for point sources (at a
sensitivity of about $3\sigma$) are $18\fm5$, $16\fm5$, $13\fm5$ for the
\II -, \J -, and \K -bands, respectively, while the completeness limits for
galaxy extraction at high Galactic latitudes are roughly $16\fm5$,
$14\fm8$, and $12\fm0$\footnote{The DENIS \K-band limiting magnitude is
bright because the \K\ background of the DENIS camera is high and dominated
by thermal emission of the instrument.}  (Mamon 1998, 2000).

\section{Extinction effects on galaxy counts}	\label{counts}

\begin{figure}[tb]
\vspace{-2.4cm}
\resizebox{\hsize}{!}{\includegraphics{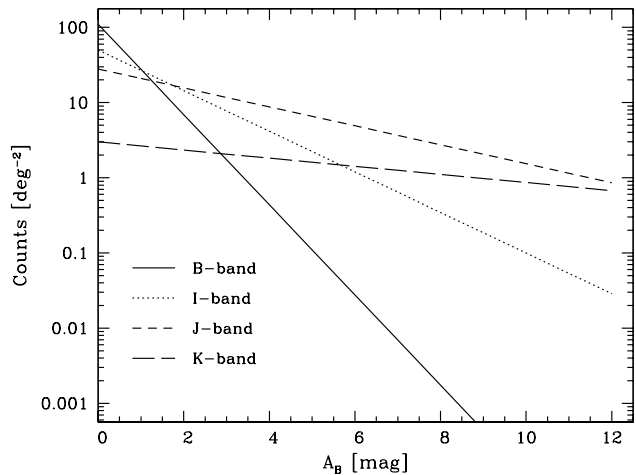}} 
\caption[]{Predicted galaxy counts in \B , \II , \J , and \K\ as a function
of absorption in \B , for highly complete and reliable DENIS galaxy samples
and a $B_J \leq 19^{\rm m}$ optical sample.
  } 
\label{galctsplot}
\end{figure}

In the ZoA, number counts of galaxies decrease due to the increasing
foreground extinction. This effect depends, however, on wavelength.  Using
the formula given in Cardelli \etal (1989), the extinction in the DENIS NIR
passbands are
\begin{equation}
A_I=0.45 A_B \ , \ \ A_J=0.21 A_B \ , \ \ A_K=0.09 A_B \ ,
\label{selext}
\end{equation}
directly implying that the decrease in number counts as a function of
extinction will be considerably slower in the NIR than in the optical.
Figure~\ref{galctsplot} shows the predicted surface number density of
galaxies as a function of Galactic foreground extinction, using the DENIS
\IJK\ galaxy counts for their respective completeness limits as given in
Mamon (1998), and for comparison the $b_J$ galaxy counts of Gardner \etal
(1996) in unobscured regions at the detection limit of $B_{\rm
lim}\!=\!19\fm0$ of the deep \B -band search in this area (Woudt \&
Kraan-Korteweg 2001).

Figure~\ref{galctsplot} indicates that -- given the above number counts and
completeness limits -- the NIR becomes more efficient at $A_B \ga 2^{\rm
m}$ than the optical in revealing galaxies in the ZoA.  The \J -band seems
the most efficient passband at intermediate extinctions ($2^{\rm m} < A_B <
12^{\rm m}$), whereas \K\ becomes superior to \J\ at $A_B \simeq 12^{\rm
m}$. As Kraan-Korteweg (2000) and Woudt \& Kraan-Korteweg (2001) have
shown, their diameter-limited ($D \ge 0\farcm2$) optical ZOA surveys start
to become incomplete at $A_B \ga 3^{\rm m}$. Here, the \J - and \K -bands
will allow a much deeper penetration of the ZoA as long as the star density
does not swamp the fields.

These are very rough predictions and do not take into account any
dependence on morphological type, surface brightness (NIR surveys are for
instance not very sensitive to late and/or low surface brightness spirals),
orientation and crowding, which may lower the number of actually detectable
galaxies (\eg Mamon 1994, Kraan-Korteweg \& Jarrett 2005). One of the
reasons we pursued this NIR in-depth study of galaxies behind the Milky Way
was to investigate this further.

\section{Characteristics of the DENIS data in the cluster area } \label{strips}

The centre of our search area was put at the position of the giant
elliptical PKS\,1343\,--\,601 (RA\,$=13^{\rm h}46^{\rm m}57\fs5$,
Dec\,$=\,-60\deg 22\arcmin 58\arcsec$, J2000). To assess the likelihood
whether the to be identified galaxies form a cluster around
PKS\,1343\,--\,601, we have adopted a search radius that encompasses the
equivalent of at least one Abell radius at the redshift distance of
PKS\,1343\,--\,601. The latter, defined as $r_{\rm Abell}= 1.72/z \arcmin$
(Abell 1958), is $r_{\rm Abell} = 2\fdg25$ for the radial velocity of
$v\simeq 3900$\kms\ (West \& Tarenghi 1989) of the radio galaxy.

We have searched 29 DENIS slots of 37 images each. The total areal coverage
amounts to $\sim\!29.8$ square degrees (\cf Fig.~\ref{clnplot}). Some of
these slots have been observed more than once. In this case the best
quality observation was selected for the visual examination, but
coordinates and photometry was determined from all images that were of
useful quality. Table~\ref{stripstab} gives an overview of the observed
slots and their characteristics.

\input{strips.tab}

{\bf Column 1:} Slot number (according to their designation in the sky).

{\bf Column 2:} Strip number (observations of a given slot with a unique
number according to the observing date). Strip numbers smaller than 3000
were part of the pre-survey (these have no \II -band counterparts and the
exposure time was slightly larger); they were used for verification only
since they are not astrometrically and photometrically calibrated.

{\bf Column 3:} Date of observation (DD/MM/YY).

{\bf Column 4:} Central Right Ascension of the strip (J2000) (recall that
each image has a width of about $12\arcmin$).

{\bf Column 5:} Image numbers for the searched area (only the last three
digits are given for the final image).

{\bf Column 6:} Observed passbands.

{\bf Column 7:} Half-flux radius as calculated by {\tt SExtractor}
multiplied by 2 (an equivalent to the point-spread function, PSF). The
half-flux radii were determined from all stars in the magnitude ranges
$11\fm0 < I < 15\fm0$,
$9\fm0 < J < 13\fm0$,
and $7\fm0 < K < 10\fm5$ 
that do not lie within 50 pixels of the image border. 

The entries in the table are the geometric means of the median half-flux
radii over the given images. When compared with the Full Width Half Maximum
(FWHM) in arcseconds of the PSF calculated by PDAC, the two values compare
well, but the here presented values are slightly larger than the calculated
FWHM (about $0\farcs7$ in \II ). This is mainly due to the pixelization of
the image.

{\bf Column 8:} Seeing quality as estimated from the photometry of stars
(\cf Appendix~A). 1 stands for good seeing (photometry not affected), 2 for
medium seeing (7\arcsec-aperture photometry is affected), and 3 for bad
seeing (all magnitudes affected). The numbers in brackets are an estimate
of the seeing where the astrometry was insufficient for a statistical
comparison of the photometry in the overlap regions.

{\bf Column 9:} Weather conditions. 1 indicates no clouds, 2 unknown
conditions, 3 possible clouds, and 4 clouds.

{\bf Column 10:} Astrometric quality of the strips. The DENIS standard high
accuracy of an {\it rms} of $0\farcs2$ is indicated with a 1,
lower accuracy with 2. For strips where astrometric calibration failed we
determined the coordinates from the DSS\footnote{The STScI Digitized Sky
Survey}-red images (accuracy of about $2-3\arcsec$).

{\bf Column 11:} Order of quality of strips with more than one observation
(1 indicates the best strip, higher numbers designate lower priorities).

All priority 1 strips, except strip 6052, had photometric weather
conditions.  Lower quality images will have a systematic effect on the
search results. It is unlikely to result in a loss of large galaxies, but
in an increase of uncertain galaxies due to the blurring of faint stars
into patches similar in appearance to small and faint galaxies. The
photometry of galaxies obtained under adverse seeing conditions will
obviously also have larger uncertainties.

\section{Galaxy Extraction and Parameters  } 	\label{param}

Extended objects (`nebulae') in general are either galaxies or objects of
Galactic origin (typically emission or reflection nebulae).  In crowded
areas we also find many stars that are blended into elongated objects, or
groups of (unresolved) faint stars that form a diffuse nebula-like
patch. This is usually the main reason why automatic searches for galaxies,
like {\tt SExtractor}, fail at low latitudes. The sharp gradient in surface
brightness with radius can be used to distinguish blended stars from
galaxies. But if the stars are very faint and in addition the seeing
condition is not very good, this criterion becomes less distinct even to
the eye.

We have visually examined all the images. This was done with the DENIS
visualisation package {\tt Denis3d} by E. Copet. It is optimally suited for
such a search for extended objects: apart from the full image in a given
band, it simultaneously displays the \II , \J , and \K\ zoomed-in images
under the cursor.  Using the zoom-windows each whole image was
systematically scanned. The simultaneous inspection of an object in the
3~passbands facilitates the galaxy/star discrimination considerably as the
relative appearance of highly-obscured galaxies in the three bands varies
compared to stars (\cf Fig.~\ref{galctsplot}). To ensure homogeneity of the
search the same cut-values were applied to all images (min\,$=-30$ ADU and
max\,$=70$ ADU, conforming to values of typically $-4.6$ to 10.8\,$\sigma$
in \II , $-6.9$ to 16.0\,$\sigma$ in \J , and $-4.4$ to 10.2\,$\sigma$ in
\K ).

We compared our results closely to the deep \B -band catalogue by Woudt \&
Kraan-Korteweg (2001, hereafter WKK; $B_{lim} = 19\fm0$), \ie our initial
list of candidates was cross-correlated with the \B -band detections. We
found that we had missed a few low surface brightness galaxies. These
generally are very faint in the NIR, but we did recover them in
hindsight. They were subsequently added to our list and classified as {\it
BG}. Only one \B -band galaxy was not visible in the NIR. 

We also identified various extended Galactic objects; they are discussed in
more detail in Sect.~\ref{galactic}.  When classification as a galaxy was
not clear-cut, we labelled the object as an uncertain galaxy.

The automatic extraction package {\tt SExtractor} (Bertin \& Arnouts 1996)
was used to derive \IJK\ Kron-photometry for the visually detected
galaxies.  {\tt SExtractor} computes total magnitudes (`best') as well as
photometry in apertures of 3\arcsec, 5\arcsec, 7\arcsec, 10\arcsec,
20\arcsec, and 40\arcsec.

Colours were determined from the 7\arcsec-aperture magnitudes. This
aperture was chosen to minimise contamination by superimposed stars on the
one hand and the variation in seeing conditions on the other. Since the
radial colour gradient is small in the NIR (\eg Moriondo \etal 2001,
Rembold \etal 2002), the effect of having different fractions of a galaxy
within our fixed aperture will be small and will introduce less uncertainty
in the colour than the increased star subtraction required for larger
apertures.

By going back to the image, a careful analysis was made to test whether
{\tt SExtractor} has deblended all objects in the vicinity of the galaxy
and whether all the parameters agree with each other. As a result we give
quality parameters for the photometry in each band, depending on whether
the 7\arcsec-aperture and/or the total magnitude are estimated to be
uncertain or unreliable.

Although {\tt SExtractor} computes semi-major and semi-minor axes of an
object, we have decided -- for uniformity reasons -- to derive the
diameters by other means since not all our galaxies were extracted by {\tt
SExtractor} and some were contaminated by not-deblended stars. We employed
the contour facility in {\tt ds9} (Joye \& Mandel 2003) to measure the
semi-major axis $A_{\tt ds9}$ out to isophotes of $21\fm75$, $20\fm5$, and
$18\fm0$ in the \II -, \mbox{\J -,} and \K -bands, respectively. The
isophotes were selected to be quarter integers close to values where the
DENIS images in the search area have similar noise characteristics.

\begin{figure}[tb]
\vspace{-1.6cm}
\resizebox{\hsize}{!}{\includegraphics{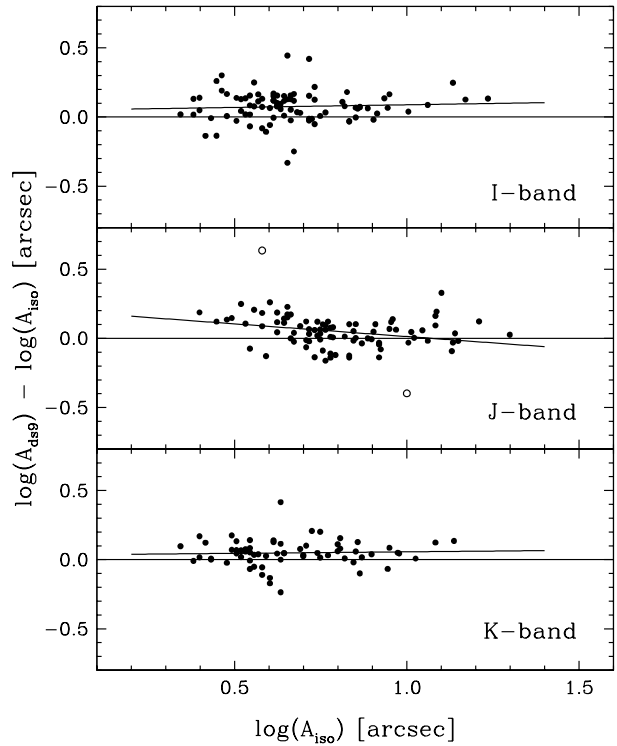}} 
\caption[]{The difference between the two estimates of the isophotal
semi-major axis as determined with {\tt ds9}, $A_{ds9}$, and by {\tt
SExtractor}, $A_{iso}$, is plotted versus the logarithm of $A_{iso}$ for
each passband. The least squares fit is plotted. {\em Open circles} (middle
panel) indicate outliers removed in a second fit.  }
\label{diamplot}
\end{figure}

To compare our diameters with the parameters calculated by {\tt SExtractor}
we used the isophotal area parameter {\tt ISOAREA}:
\[ A_{\rm iso} = \sqrt{{\tt ISOAREA}/\pi/({\tt B}/{\tt A})}, \] 
which is typically 2{\tt A} (with an {\it rms} of $1\farcs8$).
Figure~\ref{diamplot} shows the residuals ($\log{A_{\tt ds9}} - \log{A_{\tt
iso}}$) for each band. They agree well for \II\ and \K, while the \J -band
shows a small but significant deviation:
\begin{eqnarray}
I\!:\, \Delta(\log{A}) = \,\,\,(0.04\pm0.07)\log{A_{\rm iso}} + 0.05\pm0.05
\nonumber \\
J\!:\, \Delta(\log{A}) = (-0.18\pm0.07)\log{A_{\rm iso}} + 0.20\pm0.06
\nonumber \\
K\!: \Delta(\log{A}) = \,\,\,(0.02\pm0.07)\log{A_{\rm iso}} + 0.03\pm0.04.\nonumber 
\end{eqnarray}
Two objects in the \J -band fit show more than $3\sigma$ deviations from
the fit (open circles). In both cases the object is partly blended with
near-by stars which affects the automatic diameter extraction by {\tt
SExtractor}. If we exclude these outliers the fit will improve slightly to:
\[
J\!:\, \log{A_{\tt ds9}} = (-0.12\pm0.06)\log{A_{\rm iso}} + 0.14\pm0.05. 
\]

These comparisons indicate that the isophotes of the semi-major axes
derived with {\tt ds9} agree well with the {\tt SExtractor} limits in the
\II - and \K -bands, whereas in the \J -band {\tt SExtractor} goes slightly
fainter than $\mu_J = 20\fm5$.

We have extracted magnitudes and diameters from all images where the
respective object is visible (at a reasonable distance from the edge) and
averaged those where the quality was acceptable.

We estimated the morphological types by visually inspecting all DENIS
images as well as the DSS-red images.
Fairly accurate classifications are possible at low extinctions (a
comparison with the morphological types determined by WKK show good
agreement). With increasing extinctions the outer spheroid becomes more and
more truncated and a distinction between types more uncertain. At the
highest extinction levels only the bulges of galaxies remain visible,
making it impossible to distinguish between ellipticals and bulges of a
spiral galaxies.

As a guide to the interpretation of the (absorbed) object parameters, the
Galactic foreground extinctions have been determined from the IRAS/DIRBE
maps by Schlegel \etal (1998). The colour excess $E(B-V)$ has been
converted to $A_B$ using $ A_B = R_B\,E(B-V)$, where $R_B = R_V\, A_B/A_V$,
$A_B/A_V = 1.337$ (Cardelli \etal 1989), and $R_V= 3.1$.

Note that the IRAS/DIRBE extinction maps are not properly calibrated at
latitudes $|b|<5\deg$ and therefore only provide an estimate. In
Sect.~\ref{ext}, a first attempt was made at calibrating the IRAS/DIRBE
maps based on the reddened colours of the galaxies detected in this survey.
Moreover, Schlegel \etal (1998) note in their paper that far-infrared point
sources have been removed where source lists exists to correct for
overestimates. This was not done for our search region and we have taken
care to check for possible point sources which could overestimate the local
extinctions. No such source has been found in the here regarded area.

Cameron (1990) has shown that to correct highly absorbed isophotal
magnitudes of galaxies it is necessary to apply both a correction to the
magnitude as well as a correction for the fact that the angular diameter of
an obscured object appears smaller and therefore the isophotal magnitude
appears fainter. Cameron has determined the correction for diameters in the
optical \B -band up to extinctions of $A_B=6^{\rm m}$. The effect of
extinction on the {\tt SExtractor} Kron magnitudes is more difficult to
estimate: Kron `total' magnitudes are computed within an elliptical
aperture the size of which is equal to some constant times the 1st order
moment computed within the detection (or analysis) threshold. The effect of
extinction is to push the entire galaxy intensity profile down by some
factor, which means that the detection isophote is smaller, hence the 1st
order moment is smaller, hence the Kron aperture is smaller and the derived
magnitude is fainter.

To analyse this effect in detail was beyond the scope of this paper, and we
have not attempted a diameter correction for the magnitudes. Throughout the
paper extinction-corrected magnitudes mean a correction for magnitudes
only, unless noted otherwise. On the other hand, the colours given refer to
a fixed aperture where a diameter correction is unnecessary for all
galaxies larger than the aperture.

\section{The catalogue } \label{cat}

Table~\ref{galtab} lists all the galaxies and galaxy candidates in the
searched area. The columns are as follows.

{\bf Column 1:} Identity consisting of the lettering DZOA (for DENIS-ZOA)
followed by the DENIS slot number and a consecutive number for each initial
candidate.

{\bf Column 2:} Total number of sightings of same galaxy (\eg on overlap
regions or repeat observations). A plus denotes an additional observation
from the pre-survey (these observations were used for verification only).

{\bf Column 3:} Right Ascension and Declination (J2000). A colon after the
coordinates indicates a lower positional accuracy (\cf
Table~\ref{stripstab}). They were determined from the DSS-red images.

{\bf Column 4:} Galactic longitude $l$ and latitude $b$ in degrees.

{\bf Column 5:} \B -band extinction $A_B$ as derived from the reddening
values $E(B-V)$ of the IRAS/DIRBE maps (Schlegel \etal 1998) using $ A_B =
4.14\, E(B-V)$. Note that the extinction values are not calibrated for
$|b|<5\deg$ and may be unreliable. For the subsequent analysis, the
extinctions in the NIR bands were calculated using eqs~[\ref{selext}].

{\bf Column 6:} Classification of the candidate: {\it DG} stands for DENIS
galaxy; {\it UG} stands for an uncertain galaxy; {\it BG} indicates a
galaxy that was identified after consulting the WKK catalogue.  {\it NG}
stands for a \B -band galaxy that was found to be non-galaxian with DENIS.

{\bf Column 7:} Visibility of the galaxy in the \B -, \II -, \J -, and \K
-bands respectively, where 1 stands for a positive identification and 0 for
a non-detection.

\input{gal_denis.tex}

{\bf Column 8:} Morphological type of the galaxy as estimated from the
appearance in all three NIR passbands as well as the DSS-red image.  `E':
elliptical galaxy; `S': spiral galaxy (with no possible differentiation
between early- and late-type spiral); `SE': early-type spiral; `SM'
medium-type spiral; `SL': late-type spiral (including irregulars). A
question mark is given if a distinction between an elliptical galaxy and
the bulge of a spiral galaxy was not possible.  A dash is used where the
object is not a galaxy (class {\it NG}).

{\bf Columns 9\,--\,11:} Total magnitudes and errors in \II , \J , and \K ,
as derived using {\tt SExtractor} (with the {\tt MAG\_AUTO} parameter).
Note that a galaxy may be visible in one of the NIR bands but have no
magnitudes depending on the ability of {\tt SExtractor} to detect or
deblend the object from its neighbours.

{\bf Columns 12\,--\,14:} The colours and errors obtained from the 7\arcsec
-aperture magnitudes as derived using {\tt SExtractor} and corrected for
reddening using the DIRBE/IRAS extinction values: \ije , \ike , and \jke .

{\bf Column 15:} Quality of the photometry for the \II -band (first digit),
\J -band (second digit), and \K -band (last digit): 0 means good
photometry, 2 uncertainty in the 7\arcsec-aperture magnitude, 3 uncertainty
in the total magnitude, and 4 uncertainty in both. 5 means the
7\arcsec-aperture magnitude is unreliable, 6 stands for unreliability in
the total magnitude, and 7 means both are unreliable. A 9 indicates that
the strip was non-photometric (strip 6052) and all photometry is unreliable
(the estimated extinction due to the clouds in this case is $\sim0\fm25 -
0\fm3$). Magnitudes with quality parameters 5~--~9 have been excluded in
the further analysis.

{\bf Columns 16\,--\,18:} The major diameters of the galaxies in \II , \J ,
and \K\ in arseconds, derived at an isophote of $21\fm75$, $20\fm5$, and
$18\fm0$, respectively. These isophotes have a comparable background noise
in all three bands across the entire search area.

\begin{figure}[tb]
\vspace{-2.cm}
\resizebox{\hsize}{!}{\includegraphics{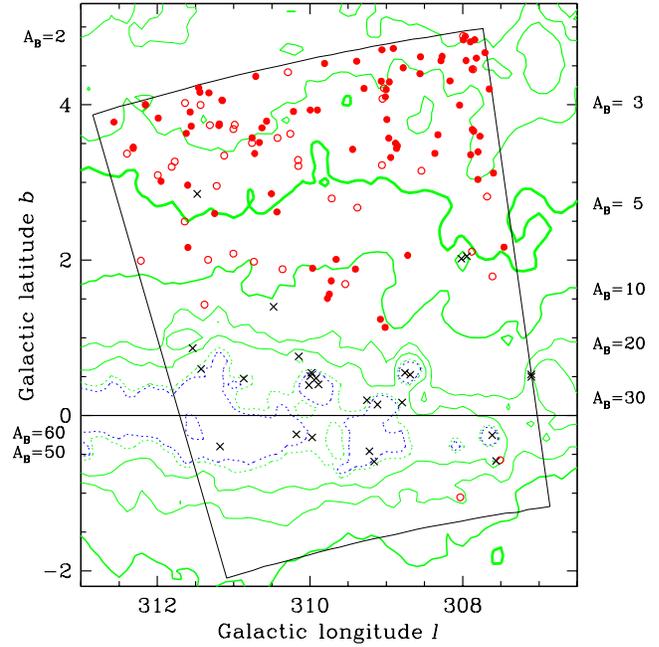}} 
\caption[]{Distribution of galaxies \emph{(circles)} and Galactic objects
\emph{(crosses)} in the searched area (\emph{tilted rectangle}).
\emph{Filled circles} are galaxies visible with DENIS, and \emph{open
circles} stand for uncertain galaxies. Extinction contours according to the
IRAS/DIRBE maps are displayed as labelled. The galaxy PKS\,1343\,--\,601
($l=309\fdg7, b=+1\fdg8$, $A_B=12\fm3$) is located close to the centre.  }
\label{clnplot}
\end{figure}

We discovered 83 galaxies (plus 38 uncertain candidates) on the 1073
searched images. 79 (33) of them are visible in the \II -band, 82 (35) in
\J , and 67 (25) in \K .  Figure~\ref{clnplot} shows the distribution of
the detected galaxies: filled circles stand for definite galaxies, open
circles show uncertain candidates. Crosses denote Galactic objects. The
contours depict the extinctions at $A_B=2^{\rm m}$, $3^{\rm m}$, 5$^{\rm
m}$, $10^{\rm m}$, $20^{\rm m}$, $30^{\rm m}$, $50^{\rm m}$, and $60^{\rm
m}$, as indicated.

\subsection{Notes on individual objects } \label{notes}

\begin{figure*}[p]
\resizebox{16cm}{!}{\includegraphics{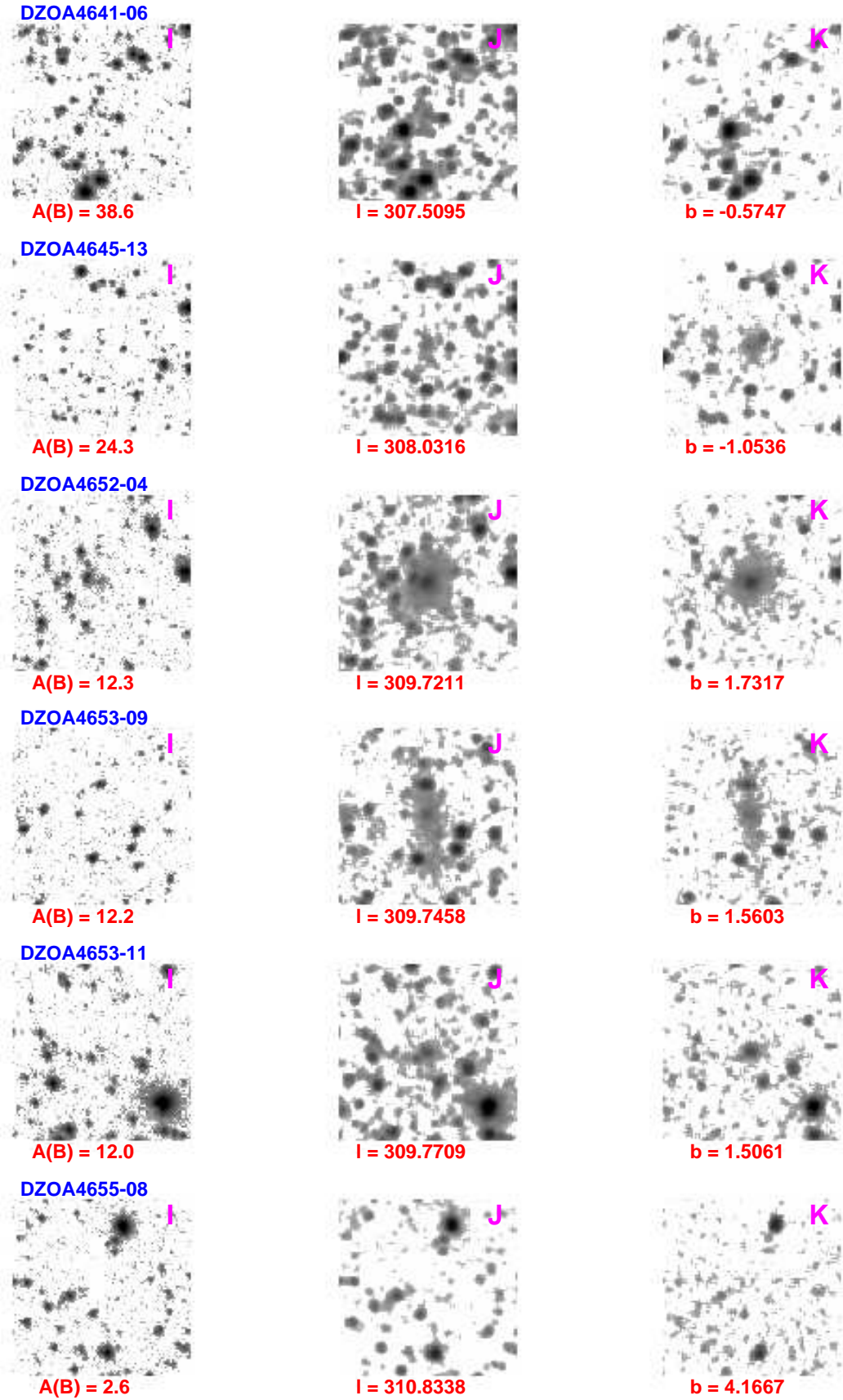}} 
\caption[]{\II -, \J -, and \K -band images (left, middle, and right hand
column, respectively) of the galaxies discussed in Sect.~\ref{notes}; the
names are given at the top, \B -band extinctions and Galactic coordinates
at the bottom of the respective row of images of the galaxy. The minimum
and maximum cut values for the \II -band are 1.0 and 300, respectively, for
the \J -band they are 0.05 and 70, respectively, and for the \K -band they
are 0.1 and 70, respectively.  }
\label{notegalplot}
\end{figure*}

In this section we describe some particularly interesting galaxies. 
Thumbnail images of these galaxies are presented in Fig.~\ref{notegalplot}
with the \II -, \J -, and \K -band image from left to right. The name of
the galaxy is printed at the top and the \B -band extinction and Galactic
coordinates at the bottom. The \J - and \K -band images have been smoothed,
and the cut values have been calculated separately for each image in
dependence of the background.

\begin{description}
\itemsep 5pt
\item[{\bf DZOA4641-06 and DZOA4645-13}:] Both these objects, classified as
uncertain galaxies, have very high extinctions ($A_B = 39^{\rm m}$ and
$24^{\rm m}$, respectively). They are only visible in the \K -band and
could be Galactic nebulae (\eg \HII\ regions). However, Galactic objects
generally also have very red stars in the immediate surroundings (\cf
images in Appendix~\ref{nebimages}), which is not the case here.  Moreover,
an intrinsically bright galaxy at the cluster distance would indeed still
be visible in the \K -band: the extinction-corrected
magnitudes are $K^0 = 8\fm8$ and $8\fm3$, respectively. For comparison, the
central elliptical galaxy is only slightly brighter at $K^0 = 8\fm0$, which
would mean that these two galaxy candidates could lie at the distance of
the cluster or closer.

The image of DZOA4645-13 indicates that it may even consist of a pair of
galaxies; note, however, that the photometry refers to a single object.

\item[{\bf DZOA4652-04/PKS\,1343\,--\,601}:] The giant elliptical galaxy in
the centre of the search area has an extinction corrected \K -band
magnitude of $8\fm0$ and is the brightest galaxy (after extinction
correction) in the whole search area. Since this galaxy is not visible on
the ESO/SERC \B -band plates we have derived an extinction-free \B -band
magnitude of $\sim\!11\fm8$ using $(B-K)^0 = 3\fm83$ (Girardi \etal 2003)
for elliptical galaxies.

Tashiro \etal (1998) present ASCA observations of the galaxy and its
surroundings. They have also observed PKS\,1343\,--\,601 with XMM. Both
observations show a point-like emission from the galaxy centre, indicating
it to have an active nucleus. A faint jet to the south-east in the
direction of the radio lobes is clearly visible on the XMM and Chandra
images. There is also faint diffuse emission connected with the
galaxy. This is interpreted by Tashiro \etal as emission due to inverse
Compton scattering in the radio lobes.  It remains uncertain, whether part
of the diffuse X-ray emission could actually be due to hot galaxy cluster
gas. This will be discussed more deeply in Paper II.

To determine an X-ray flux for the galaxy, we used the XMM data from the
three cameras (37\,ks). We fitted a power law modified by absorption to the
spectrum (it shows a negligible Fe line), and find a photon index of
$1.58\pm0.02$, and a hydrogen column density of $N_{\rm H} = (1.65\pm
0.03)\times 10^{22}$\,cm$^{-2}$. We derive an unabsorbed X-ray flux [$1 -
10$\,keV] of $7.66 \times 10^{-12}$\,erg\,cm$^{-2}$\,s$^{-1}$ with a
reduced $\chi^2$ of 1.0. This leads to an integrated bolometric luminosity
of $4.1 \times 10^{42} h_{70}^{-2} \rm erg\,s^{-1}$. The latter is
suggestive of a weak AGN (\cf DZOA4653-11 below). Adopting the above
derived \B -band magnitude we find $\log L_B/L_{\sun} = 11.0$ and $\log
L_X/L_B = 31.6$. Note that this value depends on the adopted $(B-K)$ color 
(a change of $0\fm1$ leads to a change of 0.1 in $\log
L_B/L_{\sun}$).

\item[{\bf DZOA4653-09}:] This is the second largest galaxy within the
cluster centre area. With an extinction corrected \K -band magnitude of
$9\fm1$ it is about $1^{\rm m}$ fainter than PKS\,1343\,--\,601. The
extinction for both galaxies is very similar, but while PKS\,1343\,--\,601
is visible in all passbands, DZOA4645-13 is only visible in the \J - and \K
-bands. This suggests a lower surface brightness than for elliptical
galaxies. DZOA4653-09 is therefore more likely to be an early or medium
type spiral: in the \K -band the bulge is quite distinct and shows an
elongated faint halo.

This galaxy was not detected in the blind \HI\ Parkes Multibeam survey of
the ZoA (Schr\"oder \etal 2005, Henning \etal 2005); it is therefore
unlikely to be an \HI -rich spiral at the cluster distance. On the other
hand, being close to the centre of the suspected cluster, this galaxy may
be \HI -deficient.

\item[{\bf DZOA4653-11}:] This galaxy was discovered in the Galactic Plane
optical identification programme of XMM-Newton serendipitous sources
carried out by the Survey Science Centre (Motch \etal 2003, Watson \etal
2001). It was identified as a highly obscured AGN by Michel \etal (2004)
though a significant part of the extinction is Galactic. It also has an
entry in the serendipitous 1XMM catalogue (XMM-Newton Survey Science
Centre, 2003).

A deep \II -band image obtained for the optical identification programme
shows a bright bulge and very faint halo, which implies an early type
spiral galaxy (see Fig.~1 in Piconcelli \etal 2006).  The optical spectrum
shows a prominent H$\alpha$ line. Together with the [\ion{N}{ii}] lines
($\lambda\lambda \, 6548,6583$) and [\ion{S}{ii}] line ($\lambda 6731$)
this provides a heliocentric velocity of $v=3628 \pm 12$\,\kms, putting
DZOA4653-09 right at the cluster distance (see also the velocity
determinations in Schr\"oder \etal 2005, Masetti \etal 2006, and Piconcelli
\etal 2006).

The XMM-Newton spectrum is highly absorbed due to both intrinsic as well as
Galactic absorption: using as a simple model a power law modified by
absorption and a faint iron line (44\,eV equivalent width), we found a
photon index of $1.20\pm0.02$ and a hydrogen column density of $N_{\rm H} =
(2.02\pm 0.04)\times 10^{22}$\,cm$^{-2}$. The Galactic column density at
this point is only $N_{\rm H} = 1.07\times 10^{22}$\,cm$^{-2}$, hence there
is significant intrinsic absorption (\cf PKS\,1343\,--\,601).

Piconcelli \etal (2006) discuss this galaxy in more detail. They conclude
that DZOA4653-09 is likely to be an intermediate Seyfert. Their more
complex model gives an absorbed X-ray flux [$1 - 10$\,keV] of $3.72 \times
10^{-11}$\,erg\,cm$^{-2}$\,s$^{-1}$ with a reduced $\chi^2$ of 0.94.
According to Elvis \etal (1994) the intrinsic hard band luminosity
($L_{2-10}$) is about 3\% of the bolometric luminosity for quasars, but
Gandhi \& Fabian (2003) find a value closer to 10\% for Seyferts. Using the
latter we derive a bolometric luminosity of $1.5 \times 10^{44} h_{70}^{-2}
\rm erg\,s^{-1}$. We estimate the \B -band magnitude to be $13\fm0$ using
the extinction-corrected \K -band magnitude of $9\fm48$ and the colour
$(B-K)^0 = 3\fm50$ (Girardi \etal 2003) for early-type spiral galaxies
which is corrected for Galactic extinction as well as internal absorption
(we assume the internal absorption to be negligible in the \K -band). We
find $\log L_B/L_{\sun} = 10.5$ and $\log L_X/L_B = 33.7$, a typical value
for AGNs.

\item[{\bf DZOA4655-08}:] This galaxy is the only one in the entire search
area that was found in the \B -band but is {\it not} visible on the DENIS
images. For completeness reasons we have included it in our
catalogue. DZOA4655-08 is a late type spiral and lies at a low extinction
of $A_B=2\fm5$. This is consistent with the predictions in
Fig.~\ref{galctsplot} to find more galaxies in the optical than in the NIR
at low extinctions levels ($A_B < 2\fm2$; note this limit depends quite
strongly on morphological type and surface brightness).

\end{description}

\subsection{Galactic objects } \label{galactic}

\input{neb_denis.tex}

Separating Galactic objects from galaxies at high extinction levels can be
confusing: Galactic nebulae, like galaxies, can be bright at longer
wavelengths, either because of the large extinction or because of their
intrinsically very red colours (\eg young stellar objects).  Most of them
are found at very low latitudes and very high extinctions ($A_B>50^{\rm
m}$) and because of this are unlikely to be galaxies, but some were found
at lower extinctions.  We used similarities in form and colour in
comparison with other Galactic objects and definite galaxies to classify
cases at medium extinctions. In addition, a couple of Planetary Nebulae
were found at low extinctions. These can be usually recognised by their
sharp and sometimes irregular edges.

We have listed all objects believed to be Galactic in Table~\ref{nebtab},
and thumbnail images of them are shown in Appendix~\ref{nebimages}. Columns
1\,--\,7 are as described in Table~\ref{galtab}. The coordinates are either
the estimated centre of the nebulosity or indicate the most prominent
feature connected with it. Note that the extinctions given in Column~5 are
total extinctions along the line-of-sight through our Galaxy, \ie the
actual extinction for a Galactic object can be anywhere between zero and
the total extinction. Where possible, the classification in Column~6
indicates Planetary Nebulae (PN), reflection nebulae (RN), \HII\ regions,
and young stellar objects (YSO) which show one or several very red stars
and some diffuse emission connected with it. The visibility in the \B -band
in column~7 refers to the SSS\footnote{The SuperCOSMOS Sky Survey (Hambly
\etal 2001)}-blue images. Column~8 gives the name in the literature as
found with NED\footnote{the NASA/IPAC Extragalactic Database}: 2MASX stands
for the 2MASS\footnote{see
http://irsa.ipac.caltech.edu/Missions/2mass.html} all-sky extended source
catalogue (Two Micron All Sky Survey team, PMN for Parkes-MIT-NRAO Radio
Survey (Griffith \etal 1994), AM for Arp \& Madore (1987), and KK for
Karachentseva \& Karachentsev (2000).

\section{Comparison with other catalogues } \label{lit}

We have searched NED for known galaxies in our search area. Apart from
2MASS and WKK counterparts only few others had been observed before, most
of them in the radio. They are all listed in Table~\ref{littab}. For quick
reference we repeat the DENIS-ID (Col.~1), coordinates (Col.~2), extinction
(Col.~3), class (Col.~4), and visibility (Col.~5) from the columns~1, 3, 5,
6, and~7 in Table~\ref{galtab}, respectively. Column~6 gives the 2MASS-ID
from the extended source catalogue (Two Micron All Sky Survey team, 2003);
Col.~7 gives the WKK name and the last column gives other IDs from other
catalogues. These are IRAS for the IRAS Point Source Catalog (IRAS PSC;
Joint IRAS Science Working Group 1988), PMN for Parkes-MIT-NRAO Radio
Survey (Griffith \etal 1994), HIZOA for Juraszek \etal (2000), HIZSS for
the \HI\ Parkes ZoA Shallow Survey (Henning et al. 2000), NW04 for Nagayama
\etal (2004), PKS for the Parkes Catalog of radio sources (Wright \&
Otrupcek, 1990) 4U for the Fourth Uhuru Catalog of X-ray Sources (Forman
\etal 1978), and 1XMM for the First XMM-Newton Serendipitous Source
Catalogue (XMM-Newton Survey Science Centre, 2003).

\input{gal_lit.tex}

\subsection{$B$-band} \label{wkk}

\begin{figure}[tb]
\vspace{-2.cm}
\resizebox{\hsize}{!}{\includegraphics{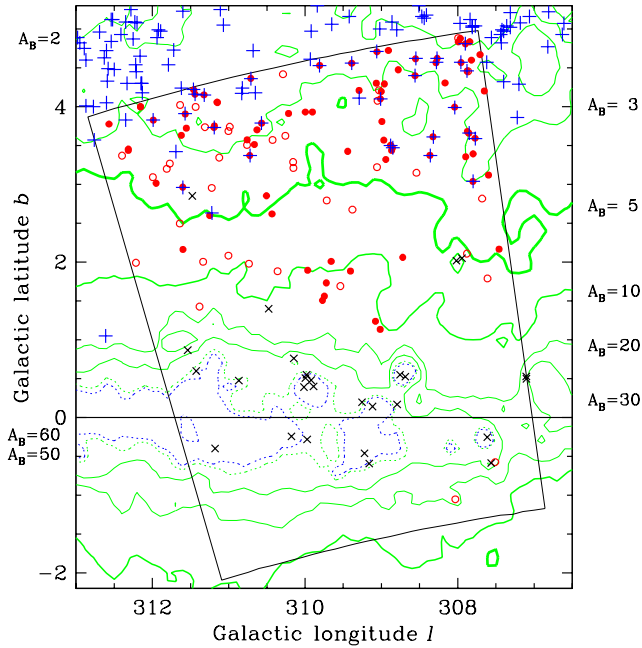}} 
\caption[]{Same as Figure~\ref{clnplot}, but with \B -band galaxies
(\emph{plus signs}) from Woudt \& Kraan-Korteweg (2001) overlaid.  }
\label{wkkplot}
\end{figure}

The WKK catalogue lists 35 galaxies within our search area. One galaxy
(WKK2589, DZOA4655-08) is not visible on the DENIS images. It is a small,
very low surface brightness galaxy, probably Sm or Irr. Five \B -band
galaxies, all classified as uncertain galaxies by WKK, were identified as
(blended) stars with the higher spatial resolution of the DENIS \II -band
images (they are classified as {\it NG} in
Table~\ref{galtab}). Figure~\ref{wkkplot} shows the WKK galaxies as plus
signs overlaid over the plot shown in Fig.~\ref{clnplot}.

As expected from Fig.~\ref{galctsplot} most \B -band galaxies are found in
the low extinction regions. The highest extinction for a \B -band galaxy in
the searched area is $A_B=5\fm2$ (DZOA4641-07, WKK2301). The completeness
limit for $B^0=15\fm5$ and $D^0=60\arcsec$ of the WKK-catalogue is
$A_B=3^{\rm m}$ (Woudt \& Kraan-Korteweg 2001).

The \B -band magnitudes are isophotal and comparable to the $B_{25}$
magnitudes. To derive (Galactic) extinction-corrected colours $(B-I)^0$,
$(B-J)^0$, and $(B-K)^0$ for our galaxies we have therefore used the total
\II -, \J -, and \K -band Kron-magnitudes.  Note that these are more
uncertain than colours from a fixed aperture due to problems with star
subtraction. In addition, at high extinctions it is necessary to correct
the diameters as well as the magnitudes (Cameron 1990). However, the
diameter correction for NIR data has not been investigated to date (\cf
Sect.~\ref{param}); hence, we have only applied a magnitude
correction. Since the diameter correction affects mainly the disk we have
not attempted to derive colours for the SL-types.

\begin{table}[ht]
\normalsize
\caption{Comparisons of colours }
\label{bcoltab}
\footnotesize
\tabcolsep 1mm
\begin{tabular}{l@{\extracolsep{2mm}}c@{\extracolsep{2mm}}c@{\extracolsep{2mm}}c@{\extracolsep{5mm}}
c@{\extracolsep{2mm}}c@{\extracolsep{2mm}}c@{\extracolsep{5mm}}c@{\extracolsep{2mm}}c@{\extracolsep{2mm}}c}
\noalign{\medskip}
\hline
\noalign{\medskip}
Type & \multicolumn{3}{c}{$(B-I)^{corr}$} &
\multicolumn{3}{c}{$(B-J)^{corr}$}  & \multicolumn{3}{c}{$(B-K)^{corr}$} \\
  & mean & $\sigma$ & n &  mean & $\sigma$ & n &  mean & $\sigma$ & n \\
\noalign{\medskip}
\hline
\noalign{\medskip}
\multicolumn{10}{l}{\it This work (with Galactic extinction correction only):} \\ 
E  & 2.31 & 0.01 &\phantom{0}2 & 3.60 & 0.20 &\phantom{0}2 & 4.40 & 0.03 &\phantom{0}2 \\
SE & 1.70 & 0.74 &\phantom{0}7 & 2.63 & 0.93 &\phantom{0}7 & 3.53 & 1.21 &\phantom{0}6 \\
SM & 1.70 & 0.71 &\phantom{0}9 & 2.82 & 0.84 &\phantom{0}8 & 3.93 & 0.80 &\phantom{0}6 \\
\noalign{\medskip}
\multicolumn{10}{l}{\it Girardi et al. 2003 (with full corrections):} \\ 
E+S0  &&&&&&& 3.83 & 0.03 & 145 \\
S0+SE &&&&&&& 3.79 & 0.03 & 157 \\
SM    &&&&&&& 2.98 & 0.03 & 507 \\
\noalign{\medskip}
\multicolumn{10}{l}{\it Moriondo et al. 2001 (with full corrections):} \\ 
E  & 2.40 &      &\phantom{0}1 & 2.86 & 0.47 &\phantom{0}3 & 3.80 & 0.46 &\phantom{0}3 \\
SE & 1.81 & 0.28 &\phantom{0}4 & 2.53 & 0.20 &\phantom{0}8 & 3.50 & 0.36 &\phantom{0}8 \\
SM & 1.41 & 0.38 &          15 & 2.23 & 0.32 &          17 & 3.07 & 0.34 &          17 \\
\noalign{\medskip}
\multicolumn{10}{l}{\it Moriondo et al. 2001 (with Gal.\ extinction
correction only):} \\ 
E  & 2.40 &      &\phantom{0}1 & 2.87 & 0.48 &\phantom{0}3 & 3.81 & 0.47 &\phantom{0}3 \\
SE & 2.06 & 0.33 &\phantom{0}4 & 2.81 & 0.29 &\phantom{0}8 & 3.81 & 0.34 &\phantom{0}8 \\
SM & 1.69 & 0.30 &          15 & 2.59 & 0.32 &          17 & 3.49 & 0.35 &          17 \\
\noalign{\smallskip}
\hline
\noalign{\smallskip}
\end{tabular} 		
\normalsize		
\end{table}		

The mean corrected colours, the standard deviation and the number of
galaxies are given in Table~\ref{bcoltab} along with respective values
taken from Girardi \etal (2003; field and group galaxies) and from Moriondo
\etal (2001; galaxies in the Pisces-Perseus cluster), where the latter two
are corrected both for Galactic extinction as well as for internal
absorption. Since Moriondo \etal give only NIR photometry we have extracted
the two correction terms as well as the corrected \B - and \II -band data
from the LEDA\footnote{the Lyon-Meudon Extragalactic DAtabase, see {\tt
http://leda.univ-lyon1.fr}.}  database. Both samples agree very well for
\bk , despite the large difference in the number of galaxies.

We have not attempted to correct our data for internal absorption since the
inclinations as well as the morphological classifications are uncertain at
these high foreground extinction levels. We have therefore turned this
correction off for the Moriondo \etal sample (see last entry in
Table~\ref{bcoltab}) to better compare it with our sample. The colours
agree reasonably well within the errors, except for the sample of
elliptical galaxies where only a couple of galaxies contribute to the mean
value: the standard deviation for the two ellipticals in our sample is very
small and they are clearly not representative.

The colours of our SM-sample are nominally redder than the colours of our
SE-sample. On the other hand, the standard deviations of the two subsamples
are at least twice as large as the ones for the Moriondo \etal samples.
Obviously, this effect is partly due to the large uncertainties in
morphological types of our sample.  In addition, while for large samples
the statistical mean is little affected by the (unknown) individual
inclination-dependent corrections, small samples may be highly biased, and
the statistical means as well as the scatter may vary widely solely due to
a different distribution in inclinations.

Consequently, we can only say that the colours of these two (small)
subsamples appear to be similar. For example, if we include the colours of
one galaxy found south of our search area (WKK2503 at $l=308\fdg41$ and
$b=-3\fdg38$) to the SM-sample, we find $(B-I)^0=1.43$ instead of 1.70,
$(B-J)^0=2.53$ instead of 2.82, and $(B-K)^0=3.59$ instead of 3.93, which
agrees much better with the Moriondo \etal colours.

Furthermore, Monnier Regaigne \etal (2003) quote $(B-K) = 4.0$ for a
predominantly early-type sample and colours as blue as $(B-K) = 2.7$ for
infrared low surface brightness galaxies (while they corrected the \B -band
data for Galactic extinction and internal absorption, the corrections for
the \K -band data were assumed to be negligible). These values denote the
large range of \bk\ colours that can be found for galaxies in general,
which implies in turn that small subsamples are sensitive to any
uncertainties in morphological types.

\subsection{2MASS $J$- and $K$-band comparison} \label{2mass}

\begin{figure}[tb]
\vspace{-2.cm}
\resizebox{\hsize}{!}{\includegraphics{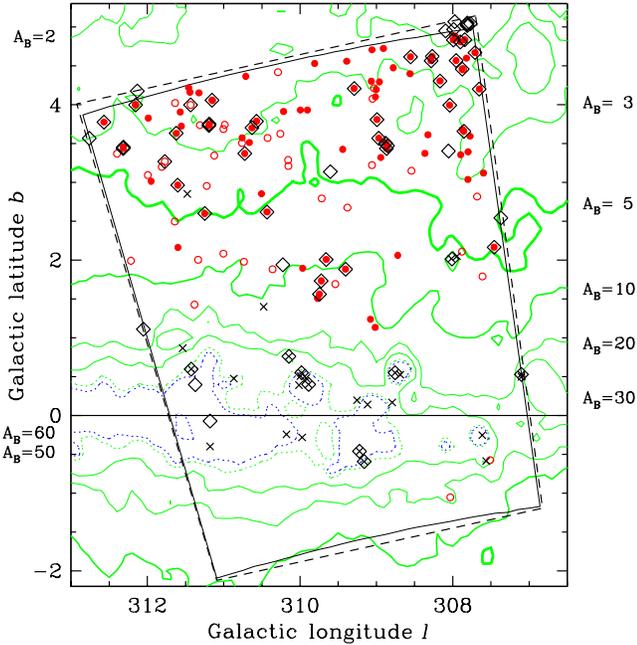}} 
\caption[]{Same as Figure~\ref{clnplot}, but with 2MASS galaxies
(\emph{diamonds}) overlaid. The 2MASS extraction region is marked with a
\emph{dashed} rectangle.  }
\label{2massplot}
\end{figure}

We have extracted all objects from the 2MASS all-sky extended source
catalogue (Two Micron All Sky Survey team, 2003) within the region
$13.45<$\,RA\,$<14.11$ and $-63.8<$\,Dec\,$<-57.4$. There are 65 objects,
11 of which are just outside the searched DENIS
area. Figure~\ref{2massplot} shows the distribution of the extended 2MASS
objects (diamonds) together with their extraction region with a dashed
line.

Forty nine objects are common in both data sets, and five 2MASS objects
were not found with DENIS.  Of the 49 objects in common, 9 are classified
by us as Galactic objects, and two as an uncertain galaxy. Of the five
2MASS objects that were not found in the DENIS search, visual inspection of
DENIS images indicates that two are at very high extinctions: one is very
probably a Galactic object, while the other seems to be a small star very
close to a bright star (this can also be seen on the 2MASS \J -band
image). The third object is probably a faint double star with a very small
spatial separation (it is slightly elongated but does not appear to be
diffuse on the 2MASS images), while the last two are very small but bright
galaxies that were not recognised as such in the DENIS search.

Of the 45 DENIS galaxies that were not in the 2MASS extended source
catalogue most will be detected in the 2MASS point source catalogue, except
for a few late type galaxies that are too faint in \J\ and \K\ and could
only be found through the DENIS \II -band.

This gives a rough estimate of the reliability of 80\% for the automatic
2MASS extraction for galaxies in this highly confused and obscured region,
while the completeness of the 2MASS galaxy extraction in this area is
slightly less than 50\%.

\begin{figure}[tb]
\resizebox{\hsize}{!}{\includegraphics{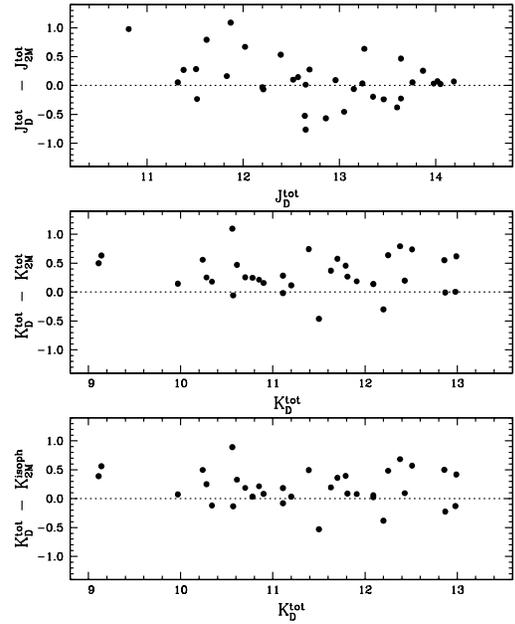}} 
\caption[]{The difference of total DENIS and 2MASS magnitudes is plotted
versus DENIS total magnitudes in $J$ (\emph{upper panel}) and in $K$
(\emph{middle panel}), and the difference between DENIS total and 2MASS
isophotal $K$-band magnitudes is shown in the \emph{bottom panel}.  }
\label{d2mplot}
\end{figure}

We compare the \J - and \K -band magnitudes of the 2MASS galaxies with our
magnitudes in Fig.~\ref{d2mplot}. The upper panel shows the differences in
the \J -band with a mean of $0\fm10\pm0\fm07$ and a standard deviation of
$0\fm42$. In the \K -band (middle panel) the mean of the differences is
$0\fm31\pm0\fm05$ with a standard deviation of $0\fm32$, which indicates
that there is a substantial offset between the two catalogues in the \K
-band. This offset can be explained by the large difference in magnitude
limits of the two surveys: the $10\sigma$ sensitivity limit for extended
sources with 2MASS are $13\fm1$ and $13\fm9$ for \K\ and \J , respectively
(Jarrett \etal 2000a), while the much deeper $3\sigma$ extraction limit for
extended sources with DENIS are only $12\fm0$ and $14\fm8$, respectively
(Mamon 1998, 2000). Note, though, that these limits refer to high Galactic
latitudes only.  If we use the 2MASS isophotal \K -band magnitudes (at
$20^{\rm m}$ per square arcsecond) instead, the mean of the differences is
only $0\fm19\pm0\fm05$ with a standard deviation of $0\fm31$ (bottom
panel).

We find a similar difference between the 2MASS major diameter at the \K
-band $20^{\rm m}$-per-square-arcsecond isophote and the DENIS major
diameter that we have derived with {\tt ds9} at the \K -band $18^{\rm
m}$-per-square-arcsecond isophote in such a way that the 2MASS diameters
are systematically larger, as expected given that their isophote is 2
magnitudes fainter. Hence, the 2MASS total magnitudes are indeed expected
to be brighter than those we derive.  In particular, we find
\[ \log{D_{\rm D}}\!-\!\log{D_{\rm 2M}} = (-0.73\pm0.03)\, \log{D_{\rm D}} + (0.12\pm0.03) \]
where the diameters are given in arcseconds. The {\it rms} difference is
$0.04$.

Using the 2MASS isophotal $J$-band the change in offset is negligible (the
mean is now $0\fm04\pm0\fm07$ with $\sigma = 0\fm34$; not shown) as
expected by the smaller difference in limiting magnitudes. This means, that
the DENIS \J -band photometry is very close to the isophotal magnitude at
$20^{\rm m}$ per square arcsecond.

\subsection{$K$-band } \label{taka}

\begin{figure}[tb]
\vspace{-2.cm}
\resizebox{\hsize}{!}{\includegraphics{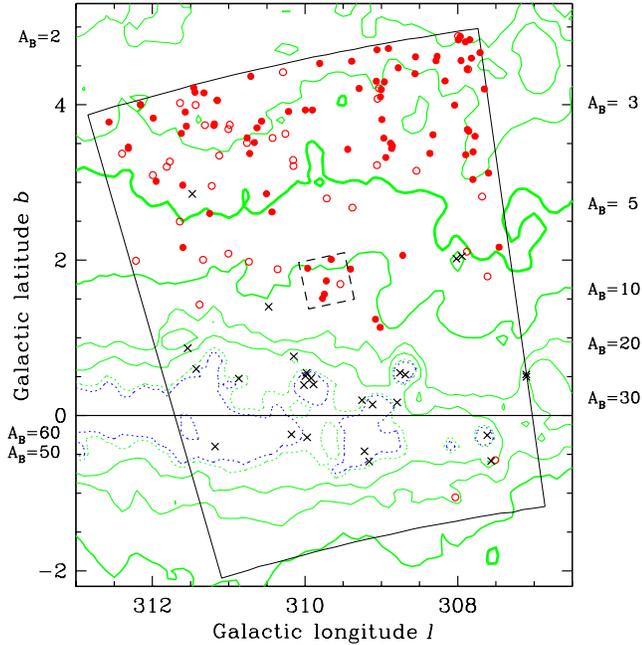}} 
\caption[]{Same as Figure~\ref{clnplot}, but with the area searched by
Nagayama et al.\ overlaid as a small (\emph{dashed}) rectangle.  }
\label{takaplot}
\end{figure}

Nagayama \etal (2004) have obtained deep \K -band images of an area of
$36\arcmin \times 36\arcmin$ around PKS\,1343\,--\,601.
Figure~\ref{takaplot} shows this area as a dashed rectangle overlaid on
Fig.~\ref{clnplot}. The plot does not show the detections by Nagayama \etal
since the spatial density of these detections is very high due to the much
deeper imaging. All six DENIS galaxies in this region (including one
uncertain candidate) have been detected by them as well.

Comparing the \K-band photometry of these 6 objects we find an offset of
$0\fm30$ with a standard deviation of $0\fm67$; excluding one outlier at
$\Delta({\rm mag}) \simeq 1\fm4$ the offset is reduced to $0\fm09$ with a
standard deviation of $0\fm47$. The large offset is mainly due to the fact
that Nagayama \etal derive isophotal magnitudes at a surface brightness of
$20^{\rm m}$ per square arcsec (\cf previous section). In fact, their \K
-band photometry agrees well with the 2MASS $K_{20}$ magnitudes.

\subsection{21\,cm wavelength } \label{hi}

A complementary way to find galaxies in the Galactic plane is through blind
\HI\ surveys, which have two major advantages: (i) the \HI\ emission of the
interstellar gas in galaxies is not affected by Galactic extinction; and
(ii) the \HI\ observation usually produces an accurate radial velocity. On
the other hand, \HI\ surveys miss elliptical and many lenticular galaxies,
and at low Galactic latitude \HI -faint galaxies may be missed due to the
increased noise caused by strong Galactic continuum sources.

The blind \HI\ Parkes ZoA survey, conducted with the multibeam receiver on
the Parkes telescope (Staveley-Smith \etal 2000), surveyed the entire
southern ZoA ($212\deg \le \ell \le 36\deg$, $|b| \leq 5\deg$) in the
velocity range $-1200$ to 12\,700\kms\ with an integration time of 25
minutes.  We have extracted detections from two sub-surveys: (i) The
shallow \HI\ ZoA survey (hereafter HIZSS; Henning \etal 2000) comprises 8\%
of the integration time of the full survey and has a sensitivity of
15\,mJy\,beam$^{-1}$ after Hanning smoothing. (ii) Juraszek \etal (2000;
hereafter JS00) have studied the area of the GA ($308\deg \la \ell \la
332\deg$) at 16\% of the integration time of the full survey with a
sensitivity of 20\,mJy\,beam$^{-1}$ before Hanning smoothing.  The
theoretical sensitivity after Hanning smoothing
($\sim\!14$\,mJy\,beam$^{-1}$) is only marginally better than the one of
the shallow survey despite the increase in integration time since the
sensitivity depends on the longitude (continuum sources cause ripples in
the \HI\ spectra), and the GA region lies closer to the Galactic centre
than the average field in the shallow survey.

\begin{figure}[tb]
\vspace{-2.cm}
\resizebox{\hsize}{!}{\includegraphics{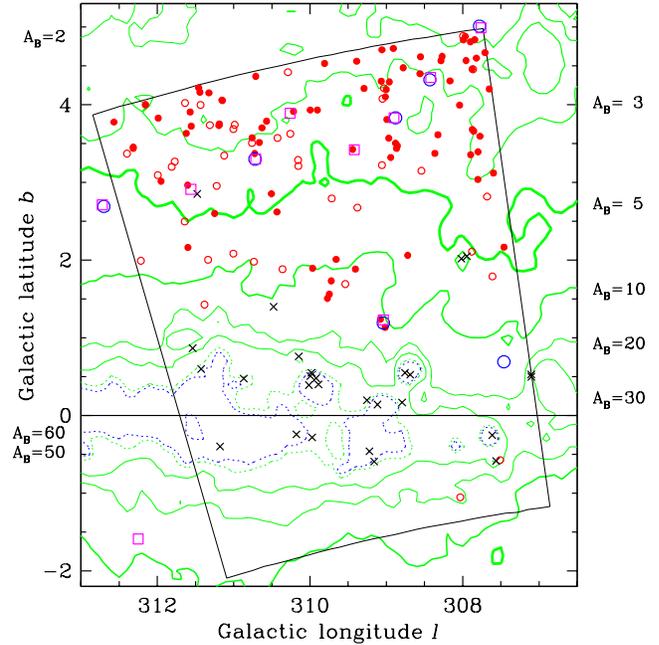}} 
\caption[]{Same as Figure~\ref{clnplot}, but with galaxies detected with
HIZSS (\emph{large open circles}) and JS00 (\emph{large open squares})
overlaid.  }
\label{hiplot}
\end{figure}

There are 12 detections in total in the DENIS search area. They are shown
in Fig.~\ref{hiplot}, where large open circles stand for the HIZSS survey
and large open squares for JS00. Note that the position uncertainty is
about 4\,arcmin. Four galaxies were detected in common by the two surveys,
and one HIZSS galaxy at $l=307.46$, $b=0.69$ was not detected by JS00 but
appears in the full deep survey. Of the eight \HI\ galaxies, four were
detected with the DENIS survey, while the other four are neither visible in
the blue nor in the NIR.

A discussion of the velocity distribution is given in Schr\"oder \etal
(2005) and Paper II.

\section{Extinction and NIR colours }  \label{ext}

Colours of nearby galaxies, unaffected by k- and evolutionary corrections,
are independent of distance.  Hence, NIR colours can be used as an
independent means to derive extinctions in low latitude areas where the
IRAS/DIRBE maps of Schlegel \etal (1998) are not properly calibrated.  For
our analysis we have applied the extinction correction according to the
IRAS/DIRBE maps to all colours and look for dependencies in the residuals.

\begin{figure}[tb]
\vspace{-.8cm}
\resizebox{\hsize}{!}{\includegraphics{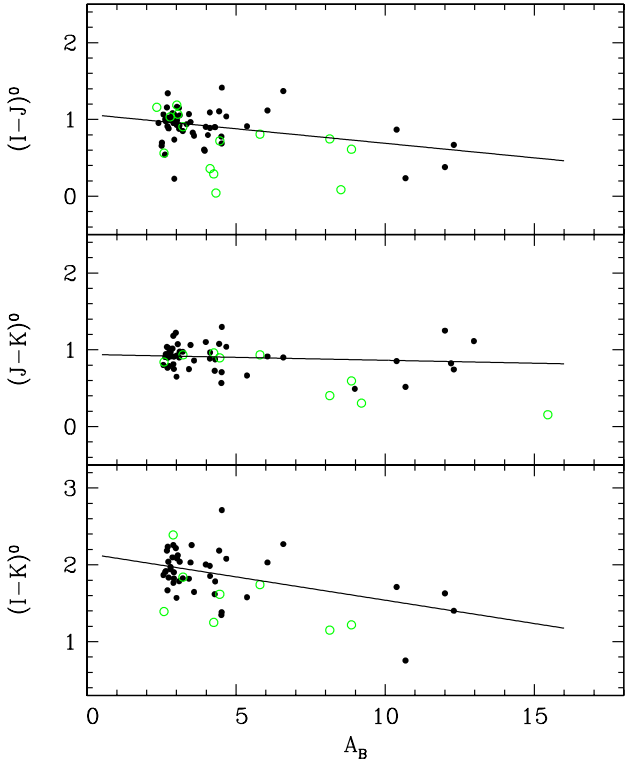}} 
\caption[]{The NIR colours, corrected for extinction according to the
IRAS/DIRBE maps, are plotted versus extinction in the $B$-band.
\emph{Filled circles} are galaxies, while \emph{open circles} are uncertain
galaxies. The \emph{line} represents the least squares fit to the filled
circles only.  }
\label{colextplot}
\end{figure}

Figure~\ref{colextplot} shows the results (filled circles are galaxies,
open circles uncertain candidates): galaxies at higher extinctions are
clearly too blue, \ie \emph{the extinctions of the IRAS/DIRBE maps are
overestimated at low galactic latitudes}. The least squares fits to the
filled circles (\ie excluding the uncertain candidates) give:
\begin{eqnarray}
(I-J)^0 &=& (-0.04\!\pm\!0.01) A_B + (1.07\!\pm\!0.06), \ 
\label{ImJ0} \\
(J-K)^0 &=& (-0.01\!\pm\!0.01) A_B + (0.94\!\pm\!0.05), \ 
\label{JmK0} \\
(I-K)^0 &=& (-0.06\!\pm\!0.02) A_B + (2.15\!\pm\!0.09), \ 
\label{ImK0} 
\end{eqnarray}
where the {\it rms} deviations of the data points from the best fit line
are $\sigma\!=\!0.21$, 0.18, and 0.28, respectively.  A Spearman Rank test
indicates probabilities of
0.11, 0.21, and 04 that the slopes of the fits in \ij , \jk , and \ik ,
0.occur by chance, respectively.
 
For a generic colour $C$, eqs~[\ref{ImJ0}\,--\,\ref{ImK0}] can be written
as
\begin{equation}
C^0 = a A_B + b \ ,
\label{linearC0}
\end{equation}
the reddening equation is simply
\begin{equation}
C = C^0 + \left ({E\over A_B}\right ) A_B \ ,
\label{reddening}
\end{equation}
and the true extinction-corrected colour is
\begin{equation}
\widetilde {C^0} = C - \left ({E\over A_B}\right )\,\widetilde A_B \ .
\label{truecolor}
\end{equation}
Combining eqs.~[\ref{linearC0}], [\ref{reddening}], and [\ref{truecolor}],
one gets
\begin{equation}
\widetilde {C^0} = a\,A_B + b + \left (A_B - \widetilde A_B \right )\,\left
({E\over A_B}\right ) \ .
\label{colorsol}
\end{equation}

If the opacity is overestimated by some constant factor, that is, if the
extinction, $A_B$, is overestimated by some constant additive term, there
would be no slope in Fig.~\ref{colextplot}, i.e., $a=0$ for $\widetilde
{C^0}$ to be independent of $A_B$ in equation~(\ref{colorsol}).
 
We therefore now assume that the true extinction is a constant $f$ times
the IRAS/DIRBE value of Schlegel et al.:
\begin{equation}
\widetilde A_B = f\,A_B \ ,
\label{ABtrue}
\end{equation}
and that the selective extinctions in the DENIS bands given in
eqs~[\ref{selext}] are exact (we use tilde signs for `true' quantities).
Equation~(\ref{colorsol}) becomes
\begin{equation}
\widetilde {C^0} = \left [a + (1-f)\, \left ({E\over A_B}\right )\right
]\,A_B + b \ ,
\label{C0}
\end{equation}
So for the extinction-corrected colour to be independent of $\widetilde
A_B$, and hence $A_B$ (from eq.~[\ref{ABtrue}]), the quantity in brackets
in eq.~[\ref{C0}] must be zero, yielding
\begin{equation}
f = 1 + {a \over E/A_B} \ .
\label{fofa}
\end{equation}
With the selective extinctions of eqs~[\ref{selext}] and the $a$
coefficients from eqs~[\ref{ImJ0}\,--\,\ref{ImK0}], eq.~[\ref{fofa}] yields
$f=0.84\pm0.05$ using $I-J$, $f=0.93\pm0.08$ using $J-K$, and
$f=0.83\pm0.05$ using $I-K$. Combining the first two, independent,
estimates of $f$ (see, \eg appendix A of Sanchis \etal 2004), we conclude
that
\begin{equation}
\widetilde A_B = (0.87\pm 0.04)\,A_B \ ,
\label{Atrue1}
\end{equation}
\ie \emph{in this area of the Galactic plane, the true reddening appears to
be 13\% lower than the IRAS/DIRBE estimate of Schlegel et al}.

\subsection{Caveats}

With a sample like the present one, where the uncertainties are large, it
is important to understand the possible biases and different explanations
for the effect seen. In the following we address the most important
caveats.

\subsubsection{Patchiness}         \label{patch}

One of the problems with attempting to calibrate the extinctions in the
Galactic plane is the relatively poor spatial resolution of $6\arcmin$ of
the IRAS/DIRBE maps, which misses extinction variations on smaller angular
scales. As a consequence it is possible that we miss galaxies in the
high-extinction patches and therefore underestimate the overall true
extinction.

We have tested this by restricting our least squares fit to the points
below $A_B=10^{\rm m}$, assuming that we find all galaxies around
$A_B=10^{\rm m}$ and that the variations in $A_B$ are not very large (this
is confirmed by the fact that no considerable changes in star counts in the
\II -band are apparent at these extinction levels). The least squares fits
to our data of Fig.~\ref{colextplot}, restricted to $A_B < 10^{\rm m}$, now
show considerably larger variations and errors:
\begin{eqnarray}
(I-J)^0 &=& (+0.04\!\pm\!0.03) A_B + (0.83\!\pm\!0.11), 
\label{ImJ0lowA} \\
(J-K)^0 &=& (-0.04\!\pm\!0.02) A_B + (1.07\!\pm\!0.08), 
\label{JmK0lowA} \\
(I-K)^0 &=& (-0.00\!\pm\!0.04) A_B + (1.95\!\pm\!0.16), 
\label{ImK0lowA} 
\end{eqnarray}
where the {\it rms} deviations of the data points from the best fit line
are similar to the fit to the full range in extinction, namely
$\sigma\!=\!0.20$, 0.16, and 0.26, respectively.  The resulting corrections
for the extinction are now (eq.~[\ref{fofa}]): $f=1.15\pm0.13$ using $I-J$,
$f=0.64\pm0.17$ using $J-K$, and $f=0.99\pm0.12$ using $I-K$. Combining the
first two (independent) estimates of $f$, which are, however, not
consistent anymore, we find
\begin{equation}
\widetilde A_B = (0.96\pm 0.10)\,A_B \ .
\end{equation}
Therefore, \emph{the IRAS/DIRBE extinctions of Schlegel et al. are
consistent with the NIR colours of the galaxies of our sample, restricted
to $A_B < 10^{\rm m}$}.

The major drawback of this method is that our restricted sample has only
very few galaxies above $A_B=5^{\rm m}$ and none below $A_B=2\fm2$, hence,
the range in extinctions is too small to provide meaningful fits. With a
larger sample one could also compare the scatter in colour at low and at
high extinctions: the scatter in colour at any given extinction is expected
to be inflated by the uncertainty in the extinction correction; in other
words, the extinctions would have considerably large error bars. A galaxy
in a higher extinction patch would be redder than expected, and if we start
missing galaxies at the highest extinctions it means we are biased towards
the blue. We would therefore expect the scatter in colour to decrease at
the high extinction end because we would miss more and more of the red
galaxies and hence be biased towards the blue as the fits indicate.

We could test this on a different project, the search for NIR counterparts
of galaxies found in a blind \HI\ survey (Schr\"oder \etal 2005). This
sample covers the extinction range $1^{\rm m}<A_B<8^{\rm m}$ with better
statistics, and we do not find such a decrease in scatter. In fact, the
dependence on extinction found in this project confirms our first result
(eq.~[\ref{Atrue1}]), \ie
\begin{equation}
\widetilde A_B = (0.79\pm 0.05)\,A_B \ .
\end{equation}

A comparison of the estimates of $f$ for the colour \ik\ gives more
consistent results (the sample of galaxies here is slightly different and
comprises only of galaxies with measurements in all three passbands):
$f=0.83\pm0.05$ from eq.~[\ref{ImK0}], and $f=0.84\pm0.08$ from Schr\"oder
\etal (2005). The estimate of $f$ from the low extinction sample, \ie
$f=0.99\pm0.12$ (eq.~[\ref{ImK0lowA}]), agrees within the errors.

We are therefore confident that the variation of the extinction on a small
scale (the x-axis error) at these levels is small and does not affect our
fit significantly.

\subsubsection{Morphological segregation}

Rich clusters usually show a morphological segregation: one finds
predominantly early-type (red) galaxies in the centre and late-type (blue)
galaxies in the outskirts of a cluster.

\begin{figure}[tb]
\vspace{-0.8cm}
\resizebox{\hsize}{!}{\includegraphics{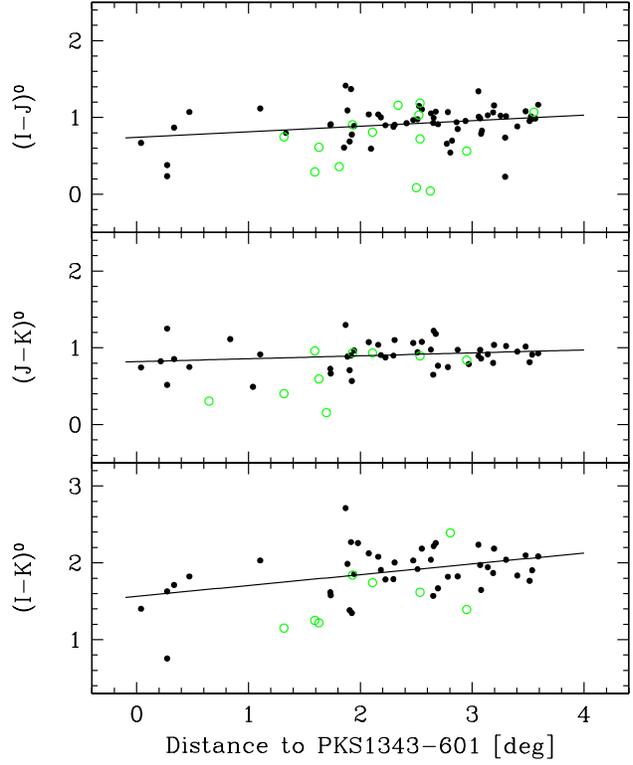}} 
\caption[]{The NIR colours, corrected for extinction according to the
IRAS/DIRBE maps, as a function of distance from PKS\,1343\,--\,601 in
degrees. \emph{Filled circles} are galaxies, while \emph{open circles} are
uncertain galaxies. The \emph{line} represents the least squares fit to the
filled circles.  }
\label{coldistplot}
\end{figure}

To investigate this we have plotted the colours corrected for extinction
according to the IRAS/DIRBE maps as a function of {\it distance} to
PKS\,1343\,--\,601 in degrees (Figure~\ref{coldistplot}; filled circles are
galaxies, open circles uncertain candidates), assuming that
PKS\,1343\,--\,601 is at or very near the centre of the cluster.

\begin{figure}[tb]
\vspace{-0.8cm}
\resizebox{\hsize}{!}{\includegraphics{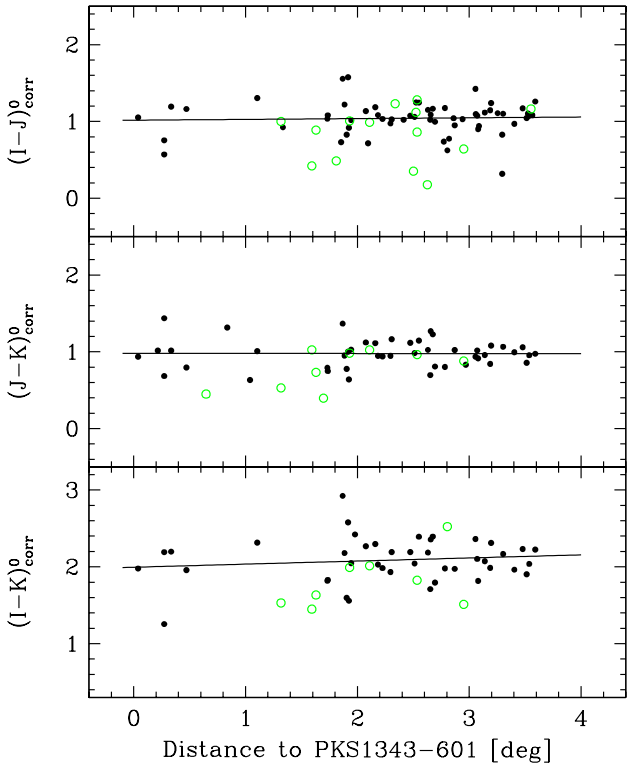}} 
\caption[]{Same as Figure~\ref{coldistplot}, but with a correction factor
of $f=0.87$ applied to the extinctions.  }
\label{coldist2plot}
\end{figure}

Contrary to the expectation we find that the galaxies close to
PKS\,1343\,--\,601 are clearly bluer than those at a larger distance.
However, since the cluster seems to be much smaller than expected (see
Paper II where we derive a virial radius of $\sim\!0\fdg3$), we do not
expect to see any colour change due to morphological segragation much
beyond that. In fact, if we exclude the inner $0\fdg6$ from the fit we do
not find any significant dependence on distance anymore. On the other hand,
we cannot achieve a proper fit using only the six galaxies within the inner
$0\fdg6$ since the scatter is too large. We therefore cannot determine
whether we see a colour change due to morphological segregation.

To explain why the inner six galaxies are bluer than the rest we refer to
Fig.~\ref{clnplot} which shows that these galaxies are all located at the
highest extinction levels where we still find galaxies.  In other words,
Fig.~\ref{coldistplot} shows only the extinction effect seen in
Fig.~\ref{colextplot} in a slightly different form.  To demonstrate this,
we have corrected the colours for the extinction as determined in
eq.~[\ref{Atrue1}], and Fig.~\ref{coldist2plot} with the corrected colours
shows no further dependence on extinction, \ie the slopes are now
consistent with zero. No difference can be seen between the galaxies within
the virial radius and the rest, and we conclude that
any possible effect of morphological segregation does not affect our fit
significantly.

\subsubsection{Survey magnitude limits}

The limiting magnitude of the DENIS survey depends on the passband, see
Sect.~\ref{denis}. As a consequence we will start losing red and faint
galaxies when the observed (extincted) colour approaches the difference
between the limiting magnitudes.  We therefore expect a bias towards blue
galaxies at higher extinctions, exactly as we see in Fig.~\ref{colextplot}.

To test how large is this effect (since only the faint galaxies are
affected) we have performed a series of Monte Carlo simulations of a sample
of galaxies subjected to different extinctions: The galaxy magnitudes were
selected from a Euclidean galaxy count function (\cf Fig.~\ref{galctsplot})
using the magnitude limits for galaxies given in Sect.~\ref{denis}; the
(intrinsic) colour distribution was assumed to be uniform around a mean of
$1\fm05$ and $1\fm00$ for \ij\ and \jk , respectively, with a half-width of
$0\fm25$ and $0\fm05$, respectively. The galaxy positions were taken from a
uniform distribution of galactic latitudes, and extinctions were derived by
a cubic spline fit between observed latitudes and extinctions.  The
resulting extinction corrected colours were then plotted versus extinction
to imitate Fig.~\ref{colextplot} and a slope was fitted.

\begin{figure}[tb]
\vspace{-.8cm}
\resizebox{\hsize}{!}{\includegraphics{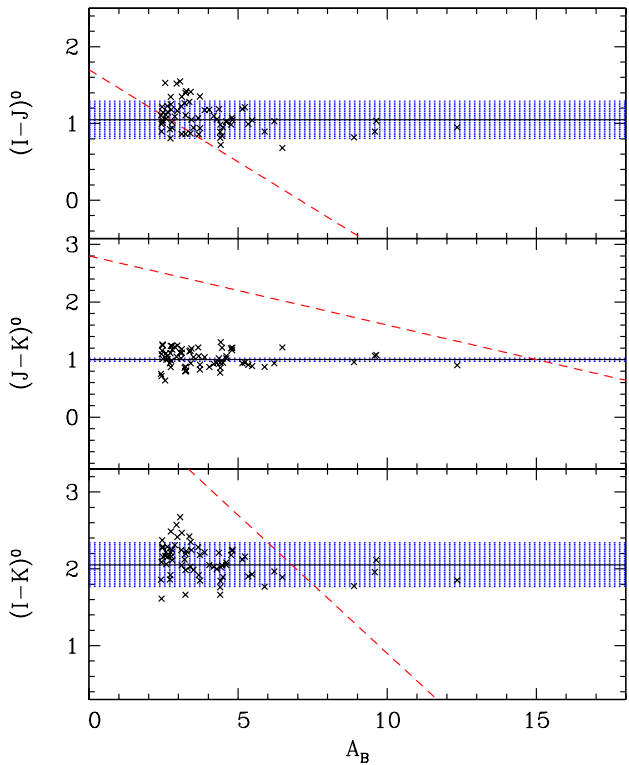}} 
\caption[]{Monte Carlo simulation of galaxies with their
extinction-corrected colours displayed in the same format as
Fig.~\ref{colextplot}. The dashed lines indicate the limiting colour above
which faint galaxies will be missed; the shaded regions depict the
$1\sigma$-distributions of the intrinsic colours.  }
\label{colextsimplot}
\end{figure}

One of the simulations is shown in Fig.~\ref{colextsimplot}, which is given
in the same format as Fig.~\ref{colextplot}: the $1\sigma$ colour
distributions are indicated by the shaded regions, and the sloped dashed
lines show the (observed) colour at the magnitude limits (that is, above
these lines we lose faint galaxies). It is quite obvious that the line in
\ij\ intersects the galaxy colours already at extinctions of $A_B \simeq
2^{\rm m}$. It is therefore not possible to simply avoid this bias by
restricting the fit to the lower extinction ranges.

For 1000 simulations (using 5000 galaxies each as input which resulted in a
similar number of galaxies `detected' as in reality) we find that a slope
of $-0.04$ ($-0.01$, and $-0.06$) as observed in Fig.~\ref{colextplot} is
observed only 0.3\% (2.6\%, and 0.1\%) of the time by chance for \ij\ (\jk
, and \ik ), respectively, assuming that the Schlegel \etal extinctions are
correct. Hence, we can say that the overestimate of extinction by Schlegel
\etal that we find is statistically significant.

We have also investigated how sensitive these results are to a deviation in
the intrinsic NIR colours: Though the mean \jk\ colour of $1\fm0$ is fairly
well established from the 2MASS survey (\eg Jarrett \etal 2003), variations
can be found, depending on the precise selective transmission of the
telescope optics, filter and detector.  We extracted the DENIS catalogue of
LEDA galaxies (Paturel \etal 2005) where the mean colours for galaxies
(type G) with $|b| > 45\deg$ are $1\fm07$ and $1\fm13$ for $\langle
I-J\rangle$ and $\langle J-K\rangle$, respecitively.  For comparison, Mamon
\etal (1998) find $1\fm15$ and $1\fm05$, respecitively, for bright high
latitude DENIS galaxies,
while our galaxy sample, after correction for the effect seen in
Fig.~\ref{colextplot}, gives $1\fm07$ and $0\fm98$, respecitively
(eqs~[\ref{ImJ0}] and~[\ref{JmK0}]).

For slightly redder colours ($1\fm15$ and $1\fm05$, respecitively), the
simulations show that 0.1\%, 3.0\%, and 0.0\% of the slopes are more
negative, and for a slightly bluer colour set ($0\fm95$ and $0\fm90$,
respecitively), they give 0.3\%, 3.0\%, and 0.0\% of the slopes more
negative than observed in Fig.~\ref{colextplot}. Therefore, the {\em
uncertainty} in intrinsic colour does not have a large affect on the
slopes.

In addition, we also tested the magnitude completeness limits as derived
from the catalogue by Paturel et al., namely $16\fm1$, $14\fm7$, $13\fm2$
for the \II -, \J -, and \K -bands, respectively. Using the first colour
set ($1\fm05$ and $1\fm00$, respecitively) we find a slightly higher
percentage of the simulations, namely 1.1\%, 3.5\%, and 0.5\%, resulted in
slopes more negative than the ones given in
eqs~[\ref{ImJ0}\,--\,\ref{ImK0}].

Hence, we conclude that despite uncertainties in the mean intrinsic colours
as well as in the exact magnitude completeness limits, the simulations show
that the overestimate of extinction by Schlegel \etal that we find is
statistically significant given the magnitude selection effects.

\subsection{Discussion}

We have shown that in our search area the extinction estimates of the
DIRBE/IRAS maps by Schlegel \etal (1998) are slightly overestimated, which
cannot be fully explained by (i) patchiness of the extincting material in
the sky, (ii) morphological segregation within the assumed cluster, or
(iii) selection effects at the magnitudes limits of the survey.

While the fractional error in the reddening estimates of the DIRBE/IRAS
maps is 16\% (Schlegel et al.), we find a systematic overestimation of
$1/f=15\%$ ($f = 0.87$), thus fairly large in comparison with the
DIRBE/IRAS uncertainty on a single point.
When comparing reddening values derived from Mg$_2$ indices of a large
sample of elliptical galaxies, Schlegel \etal find that the highest
reddening values of their maps appear to be overestimated.  Inspecting
their Figure~6 we can say that a slope of $f-1 = -0.13$ according to our
findings is compatible with their data. It has to be stressed, though, that
the sparsity of points at $E(B-V) > 0.2$ in their figure makes it difficult
to estimate a slope at all.
Schlegel \etal state clearly (in their Appendix C) that their predicted
reddenings at low Galactic latitudes ($|b|<5\deg$) are not to be trusted
because of (i) the possibility that contaminating sources may exist (our
search area is such a case though no such sources have been found by us),
and (ii) a calibration at higher extinctions was not attempted.

Since then, various attempts have been made to calibrate the DIRBE/IRAS
maps at different latitudes and using different objects and methods. Most
of these find that the DIRBE/IRAS maps are overestimated at low latitudes.

Nagayama \etal (2004), who have also searched for galaxies in this cluster
(see Sect.~\ref{taka}), have determined extinction values using \jk\
colours of foreground giant stars. They find that the extinction in the \K
-band, $A_K$, is systematically lower by $0\fm4$ than the DIRBE/IRAS
values. This corresponds to a factor of $f\simeq 0.67$ for the average
extinction in this area. Their extinction map shows a higher spatial
resolution than the DIRBE/IRAS maps, and one can see that the variation of
their extinction along a line of equal extinction in the DIRBE/IRAS maps is
only moderate ($\sigma \left(A_K\right) \simeq 0\fm3$). This indicates that
the intrinsic patchiness in this area does not lead to large variations in
extinction on a small spatial scale and hence has not caused us to
predominantly miss galaxies in supposedly very high extinction patches (see
Sect.~\ref{patch}).

Dutra \etal (2003b) have derived reddening values from the spectral
continuum in the optical of elliptical galaxies in three different areas of
the sky. In combination with previous work by this group (Dutra \etal
2003a), where they used bulge giants within 10 degrees of the Galactic
centre to derive an $A_K$ extinction map, they find a calibration factor of
$f \sim 0.75$ for reddening values of up to $E(B-V) = 1.6$ ($A_B=6\fm6$).

Choloniewski \& Valentijn (2003b) used the surface brightness of galaxies
in the ESO-LV catalogue (Lauberts \& Valentijn 1989) in the $R$- and \B
-band (Choloniewski \& Valentijn 2003a) and find a calibration factor of
$f = 0.71$.

Other work in the literature include Arce \& Goodman (1999), who find $0.67
< f < 0.77$ for reddening in the Taurus dark cloud complex, and Schr\"oder
\etal 2005 with $f=0.79$, as discussed in Sect.~\ref{patch}. On the other
hand, Fingerhut \etal 2003 find a much \emph{higher} extinction in the
direction of Maffei 1, which translates to $f = 1.3$. However, this is a
single object and seems to be the exception.

We can therefore conclude that our calibration factor of $f=0.87$ derived
from a fairly high extinction area
($2^{\rm{m}}\!\la\!A_B\!\la\!12^{\rm{m}}$) is in reasonable agreement with
the results of other investigations ($0.67<f<0.79$) at various extinction
levels in the sense that the DIRBE/IRAS maps overestimate the extinction at
low Galactic latitudes where they are uncalibrated.

We have shown that there is no obvious evidence that we predominately find
galaxies in low extinction patches in the investigated range of
extinctions, but better statistics are needed to completely solve this
question.  One possible way would be to use galaxies at moderate extinction
levels (where we can be sure to go reasonably deep to find all galaxies in
the NIR) in an area of high spatial density of galaxies, \eg the Norma
cluster in the centre of the GA region, to look for a strong spatial
variation in the reddening estimates by other means. One could also
investigate larger areas by comparing shallow surveys with deeper ones, \eg
using the UKIDSS (Warren, 2006) and VISTA (McPherson, Craig, \& Sutherland
2003) projects, to determine whether the fainter `new' galaxies in the
deeper surveys show all significantly larger reddening values. In fact, the
comparison of our `shallow' search with the deeper search by Nagayama
(2004) already indicates that a strong spatial variation cannot be the only
explanation: the shallow survey finds actually a higher absolute extinction
than the deeper one, contrary to the argument for patchiness.

However, independent of the cause, a calibration factor of about $f=0.87$
applied to the DIRBE/IRAS maps will improve significantly the actual
extinction correction for galaxies at low latitudes.

\section{Conclusion}  \label{conclusion}

We have investigated a $\sim\!30$ square-degree area around the
radio-bright galaxy PKS\,1343\,--\,601 using the NIR survey DENIS.  We have
found 83 galaxies and 39 possible candidates. The searched area covers a
large range of extinctions (from $\sim\!2^{\rm m}$ to over $100^{\rm m}$
where the search area crosses the Galactic plane). The galaxy counts
reflect the dependence on extinction, as qualitatively shown in
Fig.~\ref{galctsplot} (\ie the effect is less pronounced for the longer
wavelengths), as well as the underlying large-scale structures: there is a
clear enhancement of galaxies around the radio galaxy, indicating that
there is a cluster or at least a group of galaxies.

Our detections compare well with those of the 2MASS survey: while we find
more galaxies at intermediate extinctions due to the \II -band, which is
more sensitive to late-type spiral galaxies than the \J - and \K -bands,
2MASS goes deeper in the \K -band. However, our search is more reliable,
since the 2MASS extended source catalogue is derived automatically and
includes without distinction both galaxies and exenteded objects in our
Galaxy.

Using the NIR galaxy colours from the $7\arcsec$ aperture, we find that the
extinction derived from the IRAS/DIRBE maps (Schlegel \etal 1998) is
overestimated by about 15\%, that is, the ratio of true extinction as
derived using NIR colours to IRAS/DIRBE extinction is $f = \widetilde
A_B/A_B = 0.87\pm0.04$. Other findings in the literature confirm this
result.

We investigated several possible explanations: (i) patchiness of obscuring
material which results in finding galaxies only in areas of below-average
extinction, (ii) morphological segregation within the dense system of
galaxies around PKS\,1343--\,601, and (iii) selection effects at the
magnitudes limits of the survey. None of these affect the here presented
galaxy colours on a statistical significant level. We therefore conclude
that the paucity of calibration of the IRAS/DIRBE at low latitudes is the
major cause of the overestimate.

The second paper of this series (Schr\"oder \& Mamon 2006) will discuss the
question whether PKS\,1343\,--\,601 is part of a cluster or a mere group of
galaxies, making use of the galaxy distribution, velocity distribution and
X-ray emission.

\acknowledgements 
The authors are grateful to the DENIS teams in Chile and at PDAC for all
the efforts in observing and reducing the data, in particular
J. Borsenberger, G. Simon, P. Textier for preferential treatment of DENIS
data needed for this project. We thank Ch. Motch for making unpublished
data available to us.

The DENIS project has been partly funded by the SCIENCE and the HCM plans
of the European Commission under grants CT920791 and CT940627. It is
supported by INSU, MEN and CNRS in France, by the State of
Baden-W\"urttemberg in Germany, by DGICYT in Spain, by CNR in Italy, by
FFwFBWF in Austria, by FAPESP in Brazil, by OTKA grants F--4239 and
F--013990 in Hungary, and by the ESO C\&EE grant A--04--046.  This research
has made use of the NASA/IPAC Infrared Science Archive (2MASS) and the
NASA/IPAC Extragalactic Database (NED), which are operated by the Jet
Propulsion Laboratory, California Institute of Technology, under contract
with the National Aeronautics and Space Administration, as well as the
Lyon-Meudon Extragalactic Database (LEDA), supplied by the LEDA team at the
Centre de Recherche Astronomique de Lyon, Obs.\ de Lyon. Furthermore, this
research has also made use of the Digitized Sky Surveys (produced at the
Space Telescope Science Institute under U.S. Government grant NAG W-2166)
and the SuperCOSMOS Sky Surveys (produced by the Wide Field Astronomy Unit
at the University of Edinburgh).  A. Schr\"oder gratefully acknowledges
financial support from the European Marie Curie Fellowship Grant.



\appendix

\section{Photometry} \label{photom}

\begin{figure}[tb]
\resizebox{\hsize}{!}{\includegraphics{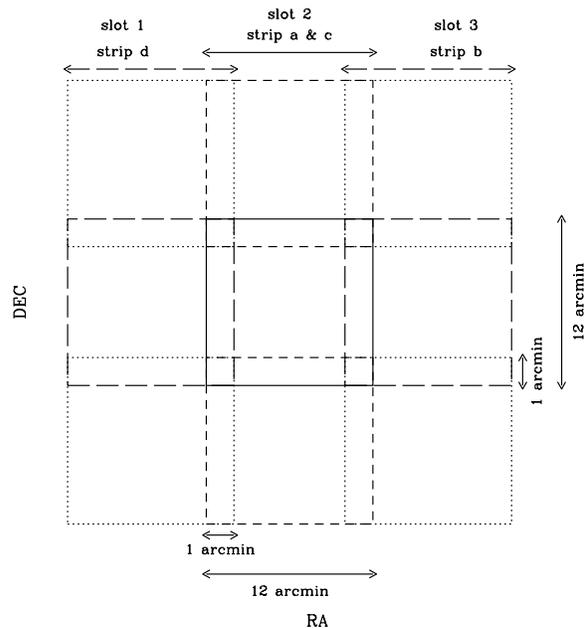}} 
\caption[]{The different overlap regions of an image are shown here in
exaggerated form. Each image of slot 2 has been observed twice (strip a and
c). There are 1-arcminute-wide overlap regions in RA (between adjacent
slots) and in DEC (between images of the same strip).  }
\label{overlapplot}
\end{figure}

We have tested the quality of the photometry by matching all objects
extracted by {\tt SExtractor} in the overlap regions between (a) images
within one slot (overlap in DEC), (b) images of adjacent slots (overlap in
RA), and (c) re-observations of the same slot (overlap of the whole
images).  Figure~\ref{overlapplot} shows a sketch of these overlap regions
in exaggerated form.  The matching was done automatically and was only
possible for observations where high quality astrometry was available (see
Col.~10 in Table~\ref{stripstab}). For each pair of strips we have then
plotted the differences in the total magnitudes as well as in the
7\arcsec-aperture magnitudes for the matched objects versus magnitude,
time, RA, and DEC (the latter two depend on the kind of overlap regions) to
look for systematic dependences.

In the following we will discuss the results in detail.  A summary of the
global effects of photometric and seeing conditions on the photometry of
each strip is given in Table~\ref{stripstab}.

\begin{itemize}

\item We used the magnitude differences of matched objects in the RA
overlap regions, plotted versus time (\ie scan direction), to look for
variation in atmospheric extinction. These can be distinguished from normal
scatter by the fact that they occur in all three passbands with the same
amplitude. Abrupt changes as small as $\Delta$mag$=0\fm1$ were immediately
obvious (\eg strip 9426), while other strips showed a constant significant
offset over the whole range of images (about 12 minutes observing time),
\eg strip 3823.

\item Bad seeing conditions can also show rapid variations of the magnitude
difference versus time, but in this case the 7\arcsec -aperture magnitudes
are much more strongly affected than the total magnitudes, and there are
marked differences between the passbands (\eg strip 9461).

\item In the \II -band a small number of points show an offset of
$\pm\sim0\fm25$ from the mean for the following strips: 7659, 9426, 3625,
7747, 6371, 7462, 7532, 7484, 3976. A simple check on the images showed
that the affected objects were scattered among the `normal' objects in the
overlap region.  There seems to be a dependence on RA in such a way that
the left quarter to half of the image is not affected.  No dependence on
DEC across the image could be found. No apparent reason for such deviations
could be found.

\item Small but negligible variations in the magnitude differences due to
the pattern of the $2 \times 2$ CCDs that make the image could be found
both in RA and DEC.

\item Occasionally an increased scatter for the very brightest objects was
found, mainly due to saturation.

\item Small offsets of the mean magnitude difference in the DEC overlap
regions within a strip indicate a slight gradient across an image.

\item For strips between $\sim$\,6000 and $\sim$\,7800 the \J -band shows a
marked lack of points in the RA overlap region in two places: just to the
right of the middle of the image, and in the right-hand corner. This was
due to the lack of detections by {\tt SExtractor} around black areas in the
CCDs.

\end{itemize}

A total recalibration of all the observations in the survey is planned by
PDAC which will remove most of the intrinsic effects that we have noticed.

\section{Images of Galactic objects } \label{nebimages}

Thumbnail images of the Galactic nebulae, presented and discussed in
Sect.~\ref{galactic}, are displayed in Figure~\ref{galnebplot}. There are
two objects per row, each with the \II -, \J -, and \K -band image shown
from left to right; the name is printed on top and the total \B -band
extinction and Galactic coordinates at the bottom of the images. The \J -
and \K -band images have been smoothed, and the cut values have been
calculated separately for each image in dependence of the background.

\begin{figure*}[p]
\resizebox{\hsize}{!}{\includegraphics{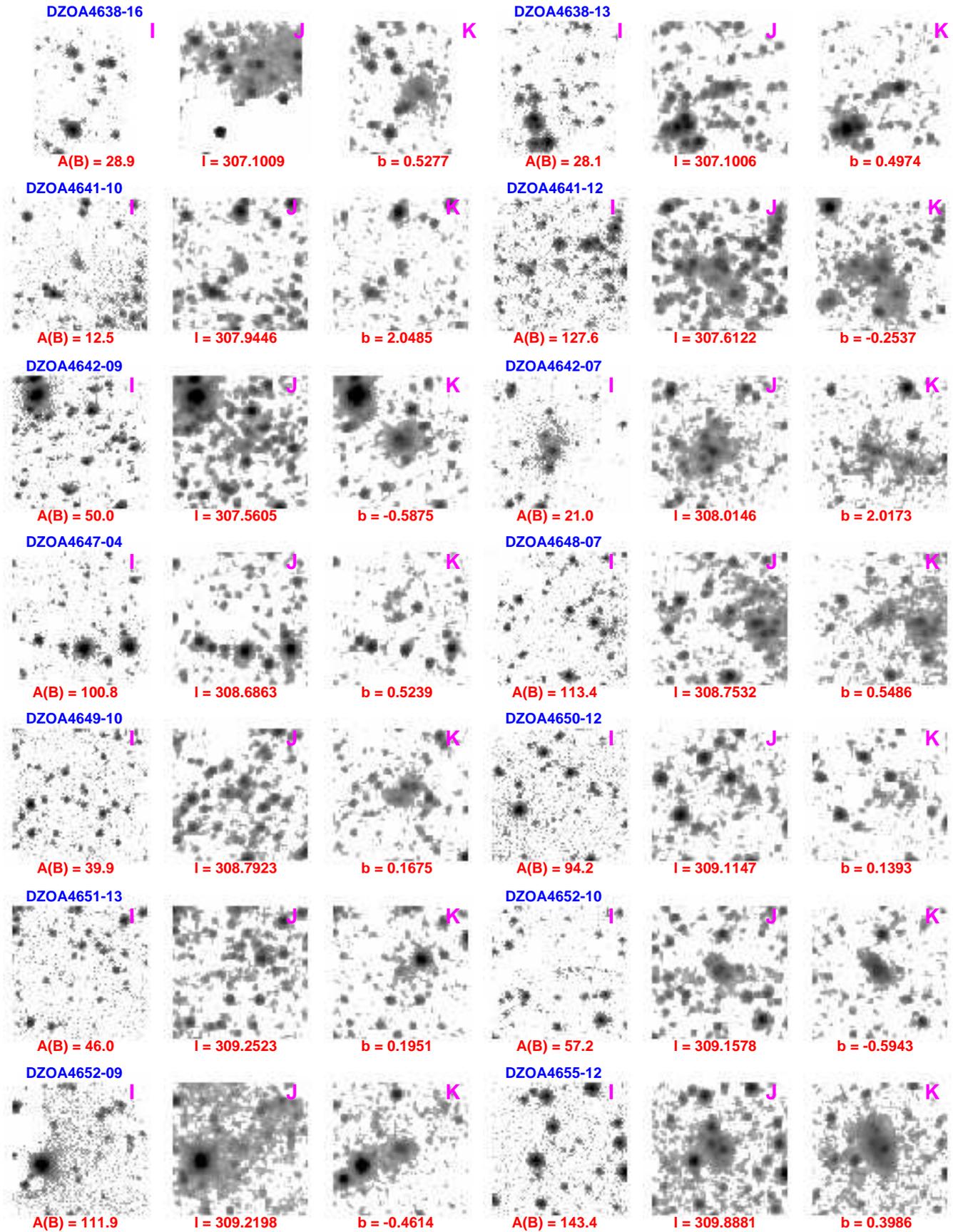}} 
\caption[]{\II -, \J -, and \K -band images (left, middle, and right hand
column, respectively) of the Galactic nebulae discussed in
Sect.~\ref{galactic}; names, total \B -band extinctions, and Galactic
coordinates are given for each set. The minimum and maximum cut values for
the \II -band are 1.0 and 300, respectively, for the \J -band they are 0.05
and 70, respectively, and for the \K -band they are 0.1 and 70,
respectively.  }
\label{galnebplot}
\end{figure*}
\addtocounter{figure}{-1} 
\begin{figure*}[p]
\resizebox{\hsize}{!}{\includegraphics{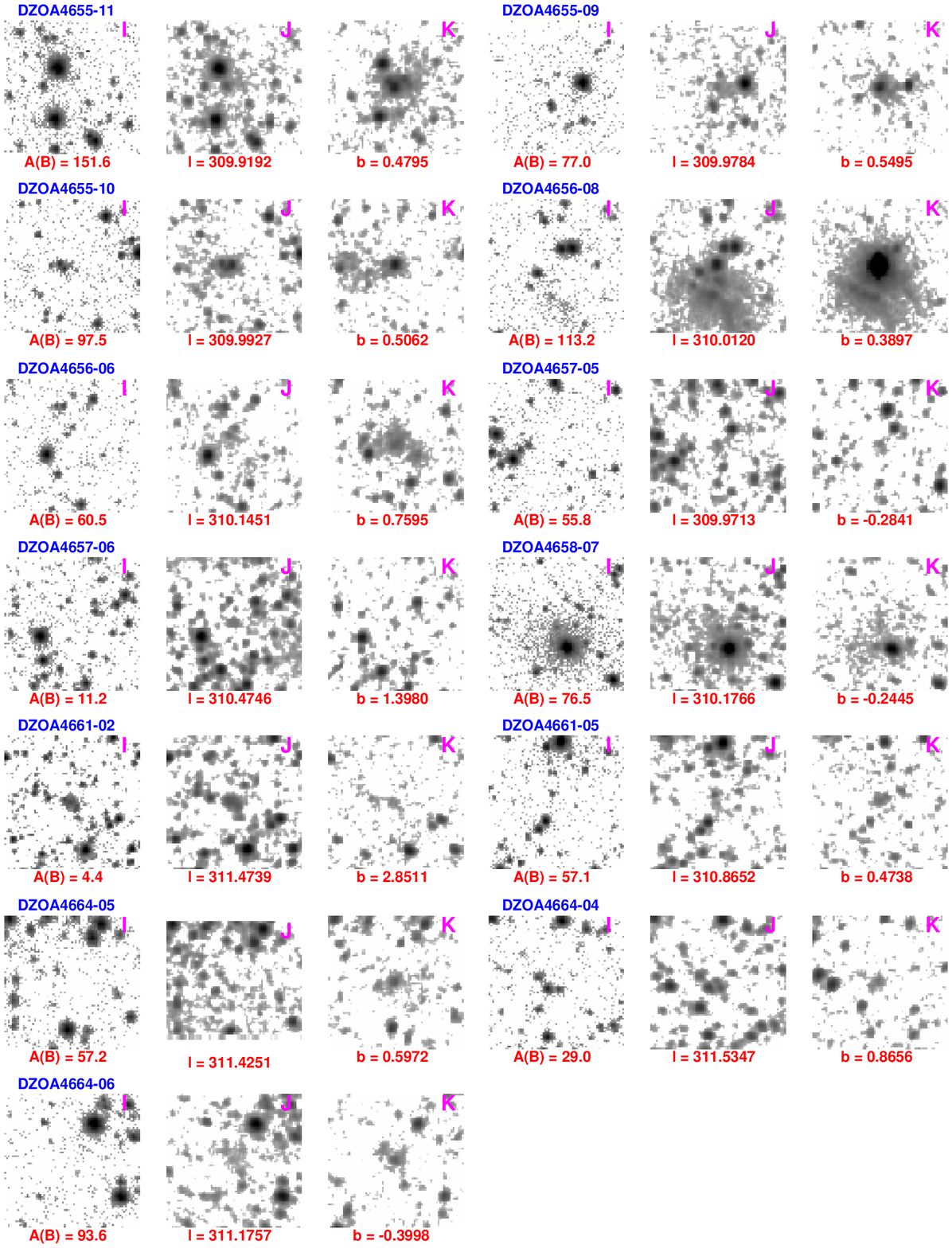}} 
\caption[]{continued.  } 
\end{figure*}

\end{document}

%% file: strips.tab
\begin{table*}[tb]
\caption[]{DENIS strips in the searched area. } \label{stripstab} 
\begin{center}
{\small 
\begin{tabular}{rrrcrccccc@{\hspace{1mm}}c@{\hspace{1mm}}c@{\hspace{1mm}}c@{\hspace{1mm}}}
\hline
\noalign{\medskip}
Slot & Strip & Date Obs. & R.A. (J2000) & \multicolumn{1}{c}{Image Nrs} & Bands & 
 \multicolumn{3}{c}{PSF (\IJK )} & Seeing & Weather & Astrom. & Prior. \\        
\multicolumn{1}{c}{(1)}&\multicolumn{1}{c}{(2)}&
\multicolumn{1}{c}{(3)}&\multicolumn{1}{c}{(4)}&
\multicolumn{1}{c}{(5)}&\multicolumn{1}{c}{(6)}&
\multicolumn{3}{c}{(7)}&\multicolumn{1}{c}{(8)}&
\multicolumn{1}{c}{(9)}&\multicolumn{1}{c}{(10)}&\multicolumn{1}{c}{(11)} \\
\noalign{\medskip}
\hline
\noalign{\medskip}
4638 &   7641 &   02/04/99 &  13 28 00.1  &$  1252\,420-456   $&	IJK &  3.48 & 3.47 & 3.47 & 2 & 1 & 1 & 1 \\  
4639 &   7659 &   05/04/99 &  13 29 20.3  &$  1256\,423-459   $&	IJK &  2.73 & 3.10 & 3.32 & 1 & 1 & 1 & 1 \\  
4640 &   7652 &   04/04/99 &  13 30 40.1  &$  1254\,833-869   $&	IJK &  3.08 & 3.39 & 3.57 & 2 & 1 & 1 & 1 \\  
4640 &   9426 &   27/04/00 &  13 30 40.0  &$  1631\,704-740   $&	IJK &  2.62 & 3.14 & 3.15 & 1 & 4 & 1 & 2 \\  
4641 &   3625 &   03/03/96 &  13 31 60.0  &$   435\,889-925   $&	IJK &  2.89 & 3.07 & 3.33 & 1 & 1 & 1 & 1 \\  
4642 &   2018 &   10/03/95 &  13 33 04.9  &$   151\,274-310   $&	-JK &  --   & --   & --   &-- &-- &-- & 3 \\  
4642 &   3823 &   02/04/96 &  13 33 20.0  &$   476\,115-151   $&	IJK &  2.91 & 3.85 & 3.43 & 2 & 1 & 1 & 2 \\  
4642 &   7747 &   18/04/99 &  13 33 20.1  &$  1275\,855-891   $&	IJK &  2.92 & 3.14 & 3.19 & 1 & 1 & 1 & 1 \\  
4643 &   3981 &   02/05/96 &  13 34 40.0  &$   508\,435-471   $&	IJK &  2.87 & 3.60 & 3.15 & 1 & 4 & 2 & 4 \\  
4643 &   9461 &   03/05/00 &  13 34 40.0  &$  1638\,871-907   $&	IJK &  3.68 & 3.93 & 3.96 & 3 & 1 & 1 & 2 \\  
4643 &   9523 &   18/05/00 &  13 34 40.0  &$  1652\,101-137   $&	IJK &  3.90 & 3.97 & 3.86 & 3 & 1 & 1 & 3 \\  
4643 &  10626 &   10/04/01 &  13 34 40.4  &$  1889\,552-588   $&	IJK &  2.88 & 3.20 & 3.26 & 1 & 1 & 1 & 1 \\ 
4644 &   3586 &   25/02/96 &  13 35 60.0  &$   428\,563-599   $&    IJ\,--\,&  2.89 & 3.02 &  --  &(1)& 2 & 2 & 2 \\ 
4644 &   6299 &   04/06/98 &  13 35 60.0  &$   965\,242-278   $&     I--\,K &  3.37 &  --  & 3.49 &(2)& 2 & 2 & 3 \\ 
4644 &   6371 &   01/07/98 &  13 35 59.3  &$   981\,749-785   $&	IJK &  2.47 & 3.03 & 3.52 & 1 & 1 & 1 & 1 \\ 
4645 &   7450 &   25/02/99 &  13 37 20.2  &$  1209\,560-596   $&	IJK &  3.50 & 3.59 & 3.36 & 3 & 1 & 1 & 2 \\ 
4645 &   7462 &   27/02/99 &  13 37 19.9  &$  1212\,136-172   $&	IJK &  2.52 & 3.22 & 3.21 & 1 & 1 & 1 & 1 \\ 
4646 &   3678 &   12/03/96 &  13 38 40.0  &$   446\,502-538   $&	IJK &  2.89 & 3.19 & 3.65 & 1 & 1 & 1 & 1 \\ 
4647 &   7514 &   08/03/99 &  13 40 00.0  &$  1223\,679-715   $&	IJK &  3.07 & 3.31 & 3.29 & 1 & 1 & 1 & 1 \\ 
4648 &   4145 &   03/06/96 &  13 41 20.0  &$   542\,355-391   $&	IJK &  2.73 & 3.11 & 3.24 & 1 & 1 & 1 & 1 \\ 
4649 &   1970 &   27/02/95 &  13 42 23.1  &$   142\,790-826   $&       --JK &  --   & 3.09 & 3.55 &-- &-- &-- & 3 \\ 
4649 &   3737 &   21/03/96 &  13 42 39.2  &$   458\,518-554   $&	IJK &  2.90 & 3.94 & 3.27 & 1 & 1 & 1 & 1 \\ 
4649 &   9418 &   26/04/00 &  13 42 40.0  &$  1630\,031-067   $&	IJK &  3.40 & 3.61 & 3.66 & 3 & 1 & 1 & 2 \\ 
4650 &   2056 &   23/03/95 &  13 43 44.9  &$   158\,110-146   $&       --JK &  --   & 3.20 & 3.56 &-- &-- &-- & 2 \\ 
4650 &   6306 &   07/06/98 &  13 43 59.4  &$   966\,517-553   $&        IJK &  2.80 & 3.34 & 3.36 & 2 & 1 & 1 & 1 \\ 
4651 &   4182 &   12/06/96 &  13 45 20.0  &$   549\,360-396   $&	IJK &  2.85 & 3.18 & 3.61 & 1 & 1 & 1 & 1 \\ 
4652 &   7527 &   10/03/99 &  13 46 40.1  &$  1226\,594-630   $&	IJK &  2.92 & 3.03 & 3.14 &(1)& 1 & 2 & 1 \\ 
4653 &   7532 &   11/03/99 &  13 48 00.0  &$  1227\,856-892   $&	IJK &  2.81 & 3.05 & 3.20 & 1 & 1 & 1 & 1 \\ 
4654 &   7503 &   06/03/99 &  13 49 21.0  &$  1221\,263-299   $&	IJK &  3.00 & 3.16 & 3.25 & 2 & 1 & 1 & 1 \\ 
4655 &   2019 &   10/03/95 &  13 50 24.2  &$   151\,478-514   $&       --JK &  --   & 2.91 & 3.10 &-- &-- &-- & 2 \\ 
4655 &   7497 &   05/03/99 &  13 50 40.6  &$  1219\,844-880   $&	IJK &  3.09 & 3.25 & 3.43 & 2 & 1 & 1 & 1 \\ 
4656 &   7491 &   04/03/99 &  13 52 00.7  &$  1218\,448-484   $&	IJK &  3.35 & 3.46 & 3.53 & 2 & 1 & 1 & 1 \\ 
4657 &   7484 &   03/03/99 &  13 53 20.1  &$  1216\,911-947   $&	IJK &  3.02 & 3.11 & 3.32 & 1 & 1 & 1 & 1 \\ 
4658 &   6052 &   05/03/98 &  13 54 41.1  &$   924\,158-196   $&	IJK &  2.56 & 3.06 & 3.44 &(1)& 4 & 2 & 1 \\ 
4659 &   5514 &   24/03/97 &  13 55 59.9  &$   812\,511-547   $&	IJK &  3.24 & 3.84 & 4.37 & 2 & 1 & 1 & 1 \\ 
4660 &   4136 &   01/06/96 &  13 57 20.0  &$   540\,905-941   $&	IJK &  2.62 & 3.30 & 3.32 & 1 & 1 & 1 & 1 \\ 
4661 &   3960 &   27/04/96 &  13 58 40.0  &$   504\,268-304   $&	IJK &  2.72 & 3.14 & 3.35 & 1 & 1 & 1 & 1 \\ 
4662 &   3969 &   28/04/96 &  14 00 00.0  &$   505\,951-987   $&	IJK &  2.66 & 3.38 & 3.31 &(1)& 2 & 2 & 1 \\ 
4663 &   5391 &   27/02/97 &  14 01 20.5  &$   789\,933-969   $&	IJK &  3.15 & 3.30 & 3.32 & 1 & 1 & 1 & 1 \\ 
4664 &   3676 &   11/03/96 &  14 02 40.0  &$   445\,934-970   $&	IJK &  3.71 & 3.75 & 4.01 & 2 & 1 & 1 & 1 \\ 
4665 &   3976 &   29/04/96 &  14 03 60.0  &$   507\,366-402   $&	IJK &  2.83 & 3.92 & 3.35 & 1 & 1 & 1 & 2 \\ 
4665 &   8071 &   24/06/99 &  14 04 00.2  &$  1345\,712-748   $&	IJK &  3.42 & 3.72 & 3.75 & 3 & 1 & 1 & 1 \\ 
4666 &   3952 &   26/04/96 &  14 05 20.0  &$   502\,766-802   $&	IJK &  2.48 & 3.28 & 3.29 & 1 & 1 & 1 & 1 \\ 
\noalign{\medskip}				 
\hline
\end{tabular}
}
\end{center}
\end{table*}

%% file: gal_denis.tex
\begin{landscape}  
\begin{table}[tb]
 \normalsize
\caption{Galaxies in the search area}
\label{galtab}
\scriptsize  
\begin{tabular*}{23.5cm}{
l  @{\extracolsep{2mm}}   l @{\extracolsep{2mm}} l@{\extracolsep{1.5mm}} l@{\extracolsep{2mm}}      
  c @{\extracolsep{1.5mm}} r @{\extracolsep{0.5mm}} r @{\extracolsep{2mm}} 
  l @{\extracolsep{1mm}} l @{\extracolsep{3mm}} l @{\extracolsep{3mm}} 
  r @{\extracolsep{0.5mm}} r @{\extracolsep{2mm}} r @{\extracolsep{0.5mm}} r @{\extracolsep{2mm}} r @{\extracolsep{0.5mm}} r @{\extracolsep{3mm}} 
  r @{\extracolsep{0.5mm}} c @{\extracolsep{2mm}} r @{\extracolsep{0.5mm}} c @{\extracolsep{2mm}} r @{\extracolsep{0.5mm}} c @{\extracolsep{1mm}} 
  r @{\extracolsep{1.5mm}}  r @{\extracolsep{1mm}} r @{\extracolsep{1mm}} r @{\extracolsep{2mm}}  
}
\noalign{\smallskip}
\hline
\noalign{\smallskip}
\multicolumn{1}{c}{Ident.} & N & \multicolumn{2}{c}{R.A.\, (J2000)\, Dec.}
 & Gal $\ell$ & Gal $b$ & $A_B$ & Class & Visibil. & Type & 
 \multicolumn{2}{c}{\II} & \multicolumn{2}{c}{\J} & \multicolumn{2}{c}{\K} &  
 \multicolumn{2}{c@{\extracolsep{1mm}}}{\ije} & \multicolumn{2}{c@{\extracolsep{1mm}}}{\ike} & \multicolumn{2}{c@{\extracolsep{1mm}}}{\jke} & 
 Phot. & $D_I$ & $D_J$ & $D_K$ \\
\noalign{\smallskip}
 & & \multicolumn{1}{c}{[$^{h}$ $^{m}$ $^{s}$]} & \multicolumn{1}{c}{[$\deg$ $\arcmin$ $\arcsec$]} 
 & [$\deg$] & [$\deg$] & [mag] & & & & 
 \multicolumn{2}{c}{[mag]} & \multicolumn{2}{c}{[mag]} & \multicolumn{2}{c}{[mag]} 
 & \multicolumn{2}{c}{[mag]} & \multicolumn{2}{c}{[mag]} & \multicolumn{2}{c}{[mag]} & 
 & [$\arcsec$] & [$\arcsec$] & [$\arcsec$] \\
\vspace{-1mm} \\
\multicolumn{1}{c}{(1)} & (2) & \multicolumn{1}{c}{(3a)} & \multicolumn{1}{c}{(3b)} & 
 (4a) & (4b) & \multicolumn{1}{c}{(5)} & (6) & \multicolumn{1}{c}{(7)} & (8) 
 & \multicolumn{2}{c}{(9)} & \multicolumn{2}{c}{(10)} & \multicolumn{2}{c}{(11)} & \multicolumn{2}{c}{(12)} 
 & \multicolumn{2}{c}{(13)} & \multicolumn{2}{c}{(14)} & (15) & (16)  & (17) & (18) \\
\noalign{\smallskip}
\hline
\noalign{\smallskip}
DZOA4638-04  &\phantom{0}1 & 13 27 20.3 & -57 52 08 & 307.71 & 4.67 &   2.8 & {\it DG} & 0 1 1 1 & SM   & 15.35 &$\pm\,0.10$& 13.87 &$\pm\,0.06$&   ... &       ... &  1.01 &$\pm\,0.10$&   ... &       ... &   ... &       ... & 000 & 11 & 14 &  5  \\
DZOA4638-09  &\phantom{0}1 & 13 27 25.1 & -58 20 28 & 307.65 & 4.20 &   2.6 & {\it DG} & 0 1 1 1 &E/SE  & 15.96 &$\pm\,0.07$& 14.02 &$\pm\,0.06$& 12.87 &$\pm\,0.28$&  1.06 &$\pm\,0.09$&  1.87 &$\pm\,0.20$&  0.80 &$\pm\,0.19$& 030 & 11 & 12 &  7  \\
DZOA4638-03  &\phantom{0}1 & 13 28 10.0 & -57 41 09 & 307.84 & 4.83 &   3.0 & {\it DG} & 0 1 1 1 &E/SE  & 16.10 &$\pm\,0.08$& 14.19 &$\pm\,0.06$& 12.98 &$\pm\,0.27$&  1.17 &$\pm\,0.09$&  2.08 &$\pm\,0.18$&  0.93 &$\pm\,0.17$& 000 & 10 & 13 &  5  \\
DZOA4638-11  &\phantom{0}1 & 13 28 10.3 & -60 22 59 & 307.46 & 2.17 &   5.0 & {\it DG} & 0 1 1 1 & SM   & 13.70 &$\pm\,0.02$& 11.92 &$\pm\,0.02$& 10.90 &$\pm\,0.06$&  0.27 &$\pm\,0.02$&  0.83 &$\pm\,0.04$&  0.56 &$\pm\,0.04$& 774 & 17 & 35 & 15  \\
DZOA4638-10  &\phantom{0}2 & 13 28 11.0 & -59 25 09 & 307.60 & 3.12 &   3.1 & {\it DG} & 0 1 1 0 & SM   & 16.40 &$\pm\,0.08$&   ... &        ...&   ... &       ... &   ... &       ... &   ... &       ... &   ... &       ... & 000 & 10 &  8 &...  \\
DZOA4638-06  &\phantom{0}2 & 13 28 15.4 & -57 55 21 & 307.82 & 4.60 &   2.7 & {\it DG} & 0 1 1 1 & SM   & 15.64 &$\pm\,0.05$& 14.19 &$\pm\,0.05$& 12.65 &$\pm\,0.25$&  0.88 &$\pm\,0.07$&  1.83 &$\pm\,0.23$&  0.95 &$\pm\,0.23$& 000 & 10 & 12 &  6  \\
DZOA4638-01  &\phantom{0}2 & 13 28 37.1 & -57 42 19 & 307.90 & 4.81 &   2.9 & {\it DG} & 1 1 1 1 & SE   & 13.96 &$\pm\,0.01$& 12.02 &$\pm\,0.02$& 10.78 &$\pm\,0.05$&  0.98 &$\pm\,0.02$&  1.90 &$\pm\,0.04$&  0.91 &$\pm\,0.04$& 000 & 22 & 25 & 17  \\
DZOA4639-07  &\phantom{0}3 & 13 28 44.4 & -58 03 32 & 307.87 & 4.46 &   2.8 & {\it UG} & 0 1 1 0 & SM   &   ... &      ...  &   ... &       ... &   ... &       ... &   ... &       ... &   ... &       ... &   ... &       ... & 000 &  7 &  8 &...  \\
DZOA4639-06  &\phantom{0}1 & 13 28 49.7 & -58 03 21 & 307.88 & 4.46 &   2.7 & {\it DG} & 1 1 1 1 & SM   & 15.48 &$\pm\,0.06$& 14.05 &$\pm\,0.06$&   ... &       ... &  1.02 &$\pm\,0.09$&   ... &       ... &   ... &       ... & 330 & 10 & 12 &  7  \\
DZOA4639-19  &\phantom{0}2 & 13 29 00.3 & -58 55 30 & 307.77 & 3.59 &   2.6 & {\it BG} & 1 1 1 0 & SL   &   ... &      ...  & 15.27 &$\pm\,0.18$&   ... &       ... &   ... &       ... &   ... &       ... &   ... &       ... & 000 &  6 &  9 &...  \\
DZOA4639-02  &\phantom{0}1 & 13 29 01.7 & -57 37 32 & 307.97 & 4.88 &   3.0 & {\it DG} & 0 1 1 0 & SM   & 16.30 &$\pm\,0.15$& 14.95 &$\pm\,0.13$&   ... &       ... &  0.99 &$\pm\,0.17$&   ... &       ... &   ... &       ... & 000 &  7 &  6 &...  \\
DZOA4639-14  &\phantom{0}1 & 13 29 10.4 & -59 42 27 & 307.68 & 2.82 &   4.1 & {\it UG} & 0 1 1 0 & S    &   ... &      ...  &   ... &       ... &   ... &       ... &   ... &       ... &   ... &       ... &   ... &       ... & 000 &  6 &  7 &...  \\
DZOA4639-01  &\phantom{0}1 & 13 29 14.7 & -57 37 06 & 308.00 & 4.88 &   3.0 & {\it UG} & 0 1 1 1 &E/SE  & 16.93 &$\pm\,0.12$& 14.71 &$\pm\,0.09$&   ... &       ... &  1.07 &$\pm\,0.16$&   ... &       ... &   ... &       ... & 000 &  9 & 13 &  7  \\
DZOA4639-03  &\phantom{0}1 & 13 29 16.1 & -57 39 56 & 307.99 & 4.83 &   2.9 & {\it DG} & 0 1 1 1 & SM   & 15.79 &$\pm\,0.06$& 13.60 &$\pm\,0.05$& 12.99 &$\pm\,0.25$&  0.95 &$\pm\,0.09$&  1.77 &$\pm\,0.22$&  0.81 &$\pm\,0.22$& 030 & 12 & 15 &  6  \\
DZOA4639-05  &\phantom{0}2 & 13 29 17.7 & -57 56 03 & 307.96 & 4.57 &   2.7 & {\it DG} & 1 1 1 1 &E/SE  & 14.30 &$\pm\,0.01$& 12.20 &$\pm\,0.02$& 11.20 &$\pm\,0.07$&  1.02 &$\pm\,0.02$&  2.04 &$\pm\,0.04$&  1.02 &$\pm\,0.04$& 040 & 20 & 22 & 14  \\
DZOA4639-10  &\phantom{0}1 & 13 29 25.8 & -59 07 11 & 307.80 & 3.39 &   2.6 & {\it DG} & 0 1 1 1 &E/SE  & 15.51 &$\pm\,0.05$& 13.53 &$\pm\,0.05$& 12.61 &$\pm\,0.22$&  0.98 &$\pm\,0.07$&  1.92 &$\pm\,0.16$&  0.94 &$\pm\,0.16$& 000 & 10 & 10 &  7  \\
DZOA4639-09  &\phantom{0}1 & 13 29 33.2 & -58 50 54 & 307.86 & 3.66 &   3.0 & {\it DG} & 1 1 1 1 & SM   & 14.25 &$\pm\,0.03$& 12.39 &$\pm\,0.03$& 11.39 &$\pm\,0.14$&  0.93 &$\pm\,0.05$&  1.57 &$\pm\,0.13$&  0.65 &$\pm\,0.13$& 330 & 17 & 25 & 11  \\
DZOA4639-08  &\phantom{0}1 & 13 29 40.9 & -58 49 30 & 307.87 & 3.68 &   3.1 & {\it DG} & 0 1 1 1 & SE   &   ... &       ... &   ... &       ... & 12.65 &$\pm\,0.15$&   ... &       ... &   ... &       ... &   ... &       ... & 000 & 15 & 13 &  7  \\
DZOA4639-16  &\phantom{0}6 & 13 29 50.5 & -60 44 05 & 307.61 & 1.79 &   5.8 & {\it UG} & 0 1 1 1 & SM   & 16.15 &$\pm\,0.06$& 13.83 &$\pm\,0.03$& 12.48 &$\pm\,0.10$&  0.81 &$\pm\,0.09$&  1.74 &$\pm\,0.10$&  0.93 &$\pm\,0.08$& 300 &  6 &  9 &  8  \\
DZOA4639-13  &\phantom{0}3 & 13 29 51.9 & -59 28 19 & 307.80 & 3.04 &   3.1 & {\it DG} & 1 1 1 1 & SM   & 15.57 &$\pm\,0.03$& 13.48 &$\pm\,0.03$& 12.50 &$\pm\,0.12$&  0.88 &$\pm\,0.04$&  1.79 &$\pm\,0.09$&  0.90 &$\pm\,0.09$& 440 & 11 & 12 &  8  \\
DZOA4640-05  &\phantom{0}2 & 13 30 13.4 & -59 08 34 & 307.90 & 3.35 &   3.1 & {\it DG} & 0 1 1 1 & SE   & 15.96 &$\pm\,0.07$& 14.45 &$\pm\,0.07$&   ... &       ... &  0.92 &$\pm\,0.10$&   ... &       ... &   ... &       ... & 000 &  9 & 12 &  6  \\
DZOA4640-03  &\phantom{0}2 & 13 30 34.6 & -58 29 24 & 308.04 & 3.99 &   3.4 & {\it DG} & 1 1 1 1 & SE   & 15.51 &$\pm\,0.06$& 13.64 &$\pm\,0.05$& 12.86 &$\pm\,0.21$&  1.07 &$\pm\,0.08$&  1.82 &$\pm\,0.16$&  0.75 &$\pm\,0.16$& 000 & 13 & 14 &  8  \\
DZOA4640-02  &\phantom{0}2 & 13 31 09.6 & -58 09 45 & 308.17 & 4.30 &   2.8 & {\it DG} & 0 1 1 1 & E    & 14.35 &$\pm\,0.02$& 12.69 &$\pm\,0.02$& 11.79 &$\pm\,0.09$&  0.95 &$\pm\,0.03$&  1.74 &$\pm\,0.07$&  0.79 &$\pm\,0.07$& 740 & 17 & 18 & 12  \\
DZOA4641-01  &\phantom{0}1 & 13 31 33.3 & -57 50 04 & 308.27 & 4.62 &   2.7 & {\it DG} & 1 1 1 1 & SM   & 13.53 &$\pm\,0.02$& 11.52 &$\pm\,0.02$& 10.70 &$\pm\,0.07$&  1.16 &$\pm\,0.03$&  2.19 &$\pm\,0.05$&  1.04 &$\pm\,0.05$& 000 & 40 & 42 & 21  \\
DZOA4641-04  &\phantom{0}1 & 13 31 36.6 & -60 22 37 & 307.88 & 2.11 &   5.8 & {\it UG} & 0 0 1 1 & SM   &   ... &       ... &   ... &       ... & 12.22 &$\pm\,0.22$&   ... &       ... &   ... &       ... &  0.79 &$\pm\,0.15$& 074 &... & 10 &  6  \\
DZOA4641-02  &\phantom{0}1 & 13 31 43.7 & -57 53 11 & 308.28 & 4.57 &   2.8 & {\it DG} & 1 1 1 1 & E    & 13.64 &$\pm\,0.02$& 11.83 &$\pm\,0.01$& 10.57 &$\pm\,0.06$&  1.03 &$\pm\,0.02$&  1.94 &$\pm\,0.04$&  0.91 &$\pm\,0.04$& 000 & 20 & 23 & 14  \\
DZOA4641-06  &\phantom{0}2 & 13 32 03.3 & -63 05 06 & 307.51 &-0.57 &  38.5 & {\it UG} & 0 0 0 1 & ?    &   ... &      ...  &   ... &       ... & 12.25 &$\pm\,0.08$&   ... &       ... &   ... &       ... &   ... &       ... & 000 &... &... &  7  \\
DZOA4642-04  &\phantom{0}2 & 13 33 11.8 & -58 49 22 & 308.32 & 3.61 &   4.0 & {\it DG} & 1 1 1 1 & SL   & 16.22 &$\pm\,0.08$& 14.26 &$\pm\,0.06$& 12.57 &$\pm\,0.24$&  0.84 &$\pm\,0.10$&  2.02 &$\pm\,0.18$&  1.18 &$\pm\,0.18$& 000 & 10 & 14 &  8  \\
DZOA4642-01  &\phantom{0}2 & 13 33 39.2 & -57 47 42 & 308.55 & 4.62 &   2.8 & {\it DG} & 1 1 1 1 & SE   & 13.50 &$\pm\,0.02$& 11.47 &$\pm\,0.01$& 10.13 &$\pm\,0.06$&  1.04 &$\pm\,0.02$&  1.95 &$\pm\,0.04$&  0.91 &$\pm\,0.04$& 000 & 28 & 30 & 19  \\
DZOA4642-06  &\phantom{0}2 & 13 33 47.6 & -59 03 07 & 308.36 & 3.37 &   4.0 & {\it DG} & 1 1 1 1 &SE/M  & 16.13 &$\pm\,0.07$& 14.24 &$\pm\,0.06$&   ... &       ... &  0.63 &$\pm\,0.10$&   ... &       ... &   ... &       ... & 030 &  9 &  9 &  6  \\
DZOA4642-02  &\phantom{0}6 & 13 33 58.2 & -58 00 29 & 308.56 & 4.40 &   3.2 & {\it DG} & 1 1 1 1 & SE   & 14.16 &$\pm\,0.01$& 12.97 &$\pm\,0.01$& 11.22 &$\pm\,0.02$&  0.85 &$\pm\,0.01$&  1.83 &$\pm\,0.03$&  0.98 &$\pm\,0.03$& 330 & 27 & 33 & 22  \\
DZOA4644-04  &          14 & 13 35 26.0 & -59 14 38 & 308.54 & 3.15 &   4.1 & {\it UG} & 0 1 1 1 & SM   & 15.60 &$\pm\,0.03$& 14.13 &$\pm\,0.03$&   ... &       ... &  0.36 &$\pm\,0.05$&   ... &       ... &   ... &       ... & 000 & 12 & 13 &  7  \\
DZOA4644-02  &\phantom{0}6 & 13 35 30.1 & -57 53 47 & 308.78 & 4.47 &   3.3 & {\it DG} & 0 1 1 0 & SL   & 15.97 &$\pm\,0.09$& 14.33 &$\pm\,0.15$&   ... &       ... &  0.94 &$\pm\,0.15$&   ... &       ... &   ... &       ... & 000 &  9 &  8 &...  \\
DZOA4644-01  &\phantom{0}3 & 13 36 08.1 & -57 37 48 & 308.91 & 4.72 &   2.7 & {\it DG} & 0 1 1 1 & SE   & 16.29 &$\pm\,0.15$& 15.02 &$\pm\,0.22$&   ... &       ... &  1.01 &$\pm\,0.18$&   ... &       ... &   ... &       ... & 300 &  7 &  9 &  6  \\
DZOA4645-01  &\phantom{0}2 & 13 37 05.0 & -58 02 40 & 308.96 & 4.29 &   3.1 & {\it DG} & 0 1 1 1 & SE   & 15.89 &$\pm\,0.05$& 14.08 &$\pm\,0.05$& 12.29 &$\pm\,0.20$&  1.05 &$\pm\,0.07$&  2.04 &$\pm\,0.17$&  0.98 &$\pm\,0.17$& 000 & 11 & 11 &  6  \\
DZOA4645-14  &\phantom{0}2 & 13 37 15.2 & -57 37 11 & 309.06 & 4.71 &   2.7 & {\it BG} & 1 1 1 0 &SM/L  &   ... &      ...  &   ... &       ... &   ... &       ... &   ... &       ... &   ... &       ... &   ... &       ... & 000 &  4 &  5 &...  \\
DZOA4645-13  &\phantom{0}2 & 13 37 20.7 & -63 28 12 & 308.03 &-1.05 &  24.3 & {\it UG} & 0 0 1 1 & ?    &   ... &      ...  &   ... &       ... & 10.44 &$\pm\,0.05$&   ... &       ... &   ... &       ... &   ... &       ... & 006 &... &  8 & 15  \\
DZOA4645-09  &\phantom{0}2 & 13 37 24.7 & -58 52 21 & 308.86 & 3.47 &   4.5 & {\it DG} & 1 1 1 1 & SE   & 13.10 &$\pm\,0.01$& 11.32 &$\pm\,0.01$& 10.28 &$\pm\,0.04$&  0.69 &$\pm\,0.01$&  1.38 &$\pm\,0.02$&  0.71 &$\pm\,0.02$& 000 & 47 & 43 & 22  \\
DZOA4645-04  &\phantom{0}2 & 13 37 31.9 & -58 08 01 & 309.00 & 4.19 &   3.1 & {\it DG} & 0 1 1 1 & E    & 16.59 &$\pm\,0.07$& 14.59 &$\pm\,0.07$&   ... &       ... &  1.15 &$\pm\,0.09$&   ... &       ... &   ... &       ... & 000 &  9 &  9 &  6  \\
DZOA4645-08  &\phantom{0}2 & 13 37 32.8 & -58 50 04 & 308.88 & 3.50 &   4.5 & {\it DG} & 1 1 1 1 & SL   & 15.00 &$\pm\,0.03$& 12.65 &$\pm\,0.03$& 12.38 &$\pm\,0.24$&  0.78 &$\pm\,0.05$&  1.35 &$\pm\,0.12$&  0.57 &$\pm\,0.12$& 003 & 17 & 22 & 10  \\
DZOA4645-10  &\phantom{0}4 & 13 37 32.9 & -58 54 14 & 308.87 & 3.44 &   4.5 & {\it DG} & 0 1 1 1 & SM   & 14.29 &$\pm\,0.02$& 11.87 &$\pm\,0.01$& 10.56 &$\pm\,0.03$&  1.41 &$\pm\,0.02$&  2.71 &$\pm\,0.02$&  1.30 &$\pm\,0.02$& 000 & 24 & 30 & 18  \\
DZOA4645-03  &\phantom{0}2 & 13 37 44.1 & -58 06 37 & 309.03 & 4.21 &   3.0 & {\it UG} & 0 1 1 1 &E/SE  & 16.75 &$\pm\,0.17$& 14.82 &$\pm\,0.12$&   ... &       ... &  1.19 &$\pm\,0.13$&   ... &       ... &   ... &       ... & 000 &  7 &  6 &  4  \\
DZOA4645-05  &\phantom{0}2 & 13 37 44.3 & -58 13 26 & 309.01 & 4.10 &   3.5 & {\it DG} & 1 1 1 0 & SL   &   ... &       ... &   ... &       ... &   ... &       ... &   ... &       ... &   ... &       ... &   ... &       ... & 000 &  9 & 15 &  5  \\
DZOA4645-02  &\phantom{0}2 & 13 37 51.3 & -58 00 59 & 309.06 & 4.30 &   2.9 & {\it DG} & 0 1 1 0 & S    & 17.13 &$\pm\,0.13$&   ... &       ... &   ... &       ... &   ... &       ... &   ... &       ... &   ... &       ... & 000 &  6 &  6 &...  \\
DZOA4645-07  &\phantom{0}3 & 13 37 59.2 & -58 30 55 & 308.99 & 3.81 &   4.7 & {\it DG} & 0 1 1 1 &SM/L  & 15.82 &$\pm\,0.06$& 12.52 &$\pm\,0.02$& 11.11 &$\pm\,0.06$&  1.04 &$\pm\,0.05$&  2.08 &$\pm\,0.07$&  1.04 &$\pm\,0.05$& 030 & 12 & 20 & 13  \\
DZOA4646-01  &\phantom{0}6 & 13 38 03.3 & -58 14 27 & 309.05 & 4.08 &   3.5 & {\it UG} & 0 1 1 1 & SM   &   ... &      ...  &   ... &       ... & 13.33 &$\pm\,0.18$&   ... &       ... &   ... &       ... &   ... &       ... & 000 &  7 &  9 &  5  \\
DZOA4646-03  &\phantom{0}6 & 13 38 08.5 & -58 45 19 & 308.97 & 3.57 &   4.1 & {\it DG} & 0 1 1 1 & SE   & 14.33 &$\pm\,0.01$& 12.21 &$\pm\,0.01$& 11.11 &$\pm\,0.04$&  0.89 &$\pm\,0.01$&  1.85 &$\pm\,0.03$&  0.96 &$\pm\,0.03$& 000 & 17 & 19 & 17  \\
DZOA4646-04  &\phantom{0}1 & 13 38 16.6 & -59 00 15 & 308.94 & 3.32 &   4.3 & {\it DG} & 0 1 1 1 &SE/M  & 15.74 &$\pm\,0.06$& 14.11 &$\pm\,0.05$& 12.72 &$\pm\,0.11$&  0.90 &$\pm\,0.07$&  1.62 &$\pm\,0.16$&  0.73 &$\pm\,0.15$& 000 & 12 & 15 &  6  \\
DZOA4646-06  &\phantom{0}1 & 13 38 21.7 & -60 17 02 & 308.72 & 2.06 &   9.0 & {\it DG} & 0 1 1 1 &E/SE  &   ... &       ... & 13.77 &$\pm\,0.07$& 12.41 &$\pm\,0.18$&   ... &       ... &   ... &       ... &  0.49 &$\pm\,0.14$& 040 &  7 & 10 &  9  \\
DZOA4647-03  &\phantom{0}4 & 13 39 17.6 & -59 04 46 & 309.06 & 3.22 &   4.2 & {\it UG} & 0 1 1 1 &SM/L  & 15.56 &$\pm\,0.03$& 13.89 &$\pm\,0.03$& 12.26 &$\pm\,0.14$&  0.29 &$\pm\,0.04$&  1.25 &$\pm\,0.11$&  0.96 &$\pm\,0.12$& 000 & 10 & 11 &  7  \\
DZOA4647-01  &\phantom{0}2 & 13 39 39.7 & -58 04 00 & 309.29 & 4.21 &   3.5 & {\it DG} & 0 1 1 1 & SM   & 16.03 &$\pm\,0.06$& 13.76 &$\pm\,0.06$& 12.43 &$\pm\,0.22$&  0.97 &$\pm\,0.09$&  2.03 &$\pm\,0.17$&  1.06 &$\pm\,0.17$& 000 & 10 & 12 &  5  \\
DZOA4647-02  &\phantom{0}1 & 13 39 52.7 & -57 42 17 & 309.39 & 4.56 &   2.6 & {\it DG} & 1 1 1 1 & SE   & 15.20 &$\pm\,0.06$& 13.80 &$\pm\,0.06$&   ... &       ... &  0.54 &$\pm\,0.08$&   ... &       ... &   ... &       ... & 000 & 12 & 14 &  5  \\
DZOA4649-02  &\phantom{0}3+& 13 41 54.8 & -58 48 28 & 309.44 & 3.42 &   3.9 & {\it DG} & 0 1 1 1 & SL   & 15.86 &$\pm\,0.08$& 13.32 &$\pm\,0.04$&   ... &       ... &  0.61 &$\pm\,0.12$&   ... &       ... &   ... &       ... & 000 & 25 & 25 &  6  \\
DZOA4649-07  &\phantom{0}3+& 13 42 09.8 & -61 08 18 & 309.02 & 1.13 &  12.0 & {\it DG} & 0 0 1 1 & ?    &   ... &       ... &   ... &       ... &   ... &       ... &   ... &       ... &   ... &       ... &   ... &       ... & 000 &... & 17 & 15  \\
DZOA4649-06  &\phantom{0}2+& 13 42 29.6 & -61 01 23 & 309.08 & 1.24 &  13.0 & {\it DG} & 0 0 1 1 & SE   &   ... &       ... & 13.69 &$\pm\,0.05$& 11.31 &$\pm\,0.04$&   ... &       ... &   ... &       ... &  1.11 &$\pm\,0.07$& 000 &... & 14 & 13  \\
\noalign{\smallskip}
\hline
\noalign{\smallskip}
\end{tabular*}
 \normalsize
\end{table}
\addtocounter{table}{-1}
\clearpage
\begin{table}[tb]
 \normalsize
\caption{ continued.}
\scriptsize  
\begin{tabular*}{23.5cm}{
l  @{\extracolsep{3mm}}   l @{\extracolsep{2mm}} l@{\extracolsep{1.5mm}} l@{\extracolsep{2mm}}      
  c @{\extracolsep{1.5mm}} r @{\extracolsep{0.5mm}} r @{\extracolsep{2mm}} 
  l @{\extracolsep{1mm}} l @{\extracolsep{3mm}} l @{\extracolsep{3mm}} 
  r @{\extracolsep{0.5mm}} r @{\extracolsep{2mm}} r @{\extracolsep{0.5mm}} r @{\extracolsep{2mm}} r @{\extracolsep{0.5mm}} r @{\extracolsep{3mm}} 
  r @{\extracolsep{0.5mm}} c @{\extracolsep{2mm}} r @{\extracolsep{0.5mm}} c @{\extracolsep{2mm}} r @{\extracolsep{0.5mm}} c @{\extracolsep{1mm}} 
  r @{\extracolsep{1.5mm}}  r @{\extracolsep{1mm}} r @{\extracolsep{1mm}} r @{\extracolsep{2mm}}  
}
\noalign{\smallskip}
\hline
\noalign{\smallskip}
\multicolumn{1}{c}{Ident.} & N & \multicolumn{2}{c}{R.A.\, (J2000)\, Dec.}
 & Gal $\ell$ & Gal $b$ & $A_B$ & Class & Visibil. & Type & 
 \multicolumn{2}{c}{\II} & \multicolumn{2}{c}{\J} & \multicolumn{2}{c}{\K} &  
 \multicolumn{2}{c@{\extracolsep{1mm}}}{\ije} & \multicolumn{2}{c@{\extracolsep{1mm}}}{\ike} & \multicolumn{2}{c@{\extracolsep{1mm}}}{\jke} & 
 Phot. & $D_I$ & $D_J$ & $D_K$ \\
\noalign{\smallskip}
 & & \multicolumn{1}{c}{[$^{h}$ $^{m}$ $^{s}$]} & \multicolumn{1}{c}{[$\deg$ $\arcmin$ $\arcsec$]} 
 & [$\deg$] & [$\deg$] & [mag] & & & & 
 \multicolumn{2}{c}{[mag]} & \multicolumn{2}{c}{[mag]} & \multicolumn{2}{c}{[mag]} 
 & \multicolumn{2}{c}{[mag]} & \multicolumn{2}{c}{[mag]} & \multicolumn{2}{c}{[mag]} & 
 & [$\arcsec$] & [$\arcsec$] & [$\arcsec$] \\
\vspace{-1mm} \\
\multicolumn{1}{c}{(1)} & (2) & \multicolumn{1}{c}{(3a)} & \multicolumn{1}{c}{(3b)} & 
 (4a) & (4b) & \multicolumn{1}{c}{(5)} & (6) & \multicolumn{1}{c}{(7)} & (8) 
 & \multicolumn{2}{c}{(9)} & \multicolumn{2}{c}{(10)} & \multicolumn{2}{c}{(11)} & \multicolumn{2}{c}{(12)} 
 & \multicolumn{2}{c}{(13)} & \multicolumn{2}{c}{(14)} & (15) & (16)  & (17) & (18) \\
\noalign{\smallskip}
\hline
\noalign{\smallskip}
DZOA4649-03  & 2+& 13 42 35.6 & -59 33 18 & 309.38 & 2.67 &   6.7 & {\it UG} & 0 1 1 0 &SM/L  & 16.98 &$\pm\,0.17$&   ... &       ... &   ... &       ... &  ... &       ... &  ... &       ... &  ... &       ... & 000 &  5 &  6 &...  \\
DZOA4649-01  & 2 & 13 42 59.9 & -57 39 06 & 309.81 & 4.53 &   2.5 & {\it DG} & 1 1 1 1 & SM   & 15.69 &$\pm\,0.06$&   ... &       ... &   ... &       ... &  ... &       ... &  ... &       ... &  ... &       ... & 000 & 13 & 13 &  7  \\
DZOA4650-09  & 1+& 13 44 03.7 & -60 19 35 & 309.40 & 1.88 &  10.4 & {\it DG} & 0 1 1 1 &E/SE  & 16.67 &$\pm\,0.13$& 12.64 &$\pm\,0.03$& 10.61 &$\pm\,0.04$& 0.87 &$\pm\,0.14$& 1.71 &$\pm\,0.14$& 0.85 &$\pm\,0.06$& 003 &  5 & 24 & 16  \\
DZOA4650-01  & 2+& 13 44 36.5 & -58 13 10 & 309.90 & 3.93 &   3.9 & {\it DG} & 0 1 1 1 & SM   & 14.98 &$\pm\,0.03$&   ... &       ... &   ... &       ... &  ... &       ... &  ... &       ... &  ... &       ... & 377 & 18 & 20 & 11  \\
DZOA4651-05  & 1 & 13 45 00.2 & -59 22 17 & 309.72 & 2.79 &   6.2 & {\it UG} & 0 1 1 0 &SM/L  &   ... &       ... &   ... &       ... &   ... &       ... &  ... &       ... &  ... &       ... &  ... &       ... & 000 &  2 &  3 &...  \\
DZOA4651-02  & 1 & 13 45 17.7 & -58 12 01 & 310.00 & 3.93 &   3.5 & {\it DG} & 0 1 1 0 & SL   &   ... &       ... & 14.88 &$\pm\,0.15$&   ... &       ... &  ... &       ... &  ... &       ... &  ... &       ... & 000 & 13 & 12 &...  \\
DZOA4651-08  & 1 & 13 45 25.3 & -60 29 14 & 309.54 & 1.69 &  11.9 & {\it UG} & 0 0 1 1 & ?    &   ... &       ... &   ... &       ... & 11.80 &$\pm\,0.09$&  ... &       ... &  ... &       ... &  ... &       ... & 003 &... & 12 & 12  \\
DZOA4651-06  & 2 & 13 45 50.7 & -60 09 05 & 309.66 & 2.01 &   8.2 & {\it DG} & 0 1 1 1 &E/SE  &   ... &       ... &   ... &       ... & 10.73 &$\pm\,0.06$&  ... &       ... &  ... &       ... &  ... &       ... & 004 & 12 & 21 & 13  \\
DZOA4652-01  & 1 & 13 46 40.3:& -57 39 50:& 310.29 & 4.42 &   2.3 & {\it UG} & 0 1 1 1 &SE/M  & 16.16 &$\pm\,0.09$&   ... &       ... &   ... &       ... &  ... &       ... &  ... &       ... &  ... &       ... & 000 &  9 &  8 &  5  \\
DZOA4652-04  & 2 & 13 46 48.9:& -60 24 29:& 309.72 & 1.73 &  12.3 & {\it DG} & 0 1 1 1 & E    & 15.74 &$\pm\,0.05$& 11.62 &$\pm\,0.01$&  9.11 &$\pm\,0.02$& 0.67 &$\pm\,0.05$& 1.40 &$\pm\,0.05$& 0.74 &$\pm\,0.02$& 000 & 12 & 38 & 32  \\
DZOA4652-02  & 1 & 13 46 57.3:& -58 10 19:& 310.22 & 3.91 &   3.3 & {\it DG} & 0 1 1 0 & SL   & 17.26 &$\pm\,0.17$&   ... &       ... &   ... &       ... &  ... &       ... &  ... &       ... &  ... &       ... & 000 &  5 &  7 &...  \\
DZOA4653-09  & 4 & 13 47 18.5 & -60 34 13 & 309.75 & 1.56 &  12.2 & {\it DG} & 0 0 1 1 &SE/M  &   ... &       ... & 12.65 &$\pm\,0.02$& 10.24 &$\pm\,0.02$&  ... &       ... &  ... &       ... & 0.82 &$\pm\,0.04$& 000 &... & 22 & 21  \\
DZOA4653-03  & 1 & 13 47 32.3 & -58 47 45 & 310.16 & 3.29 &   3.8 & {\it UG} & 0 1 1 0 & ?    &   ... &       ... &   ... &       ... &   ... &       ... &  ... &       ... &  ... &       ... &  ... &       ... & 000 & 10 &  8 &...  \\
DZOA4653-11  & 1 & 13 47 36.2 & -60 37 04 & 309.77 & 1.51 &  12.0 & {\it DG} & 0 1 1 1 &Sy\,2 & 16.77 &$\pm\,0.13$& 13.14 &$\pm\,0.04$& 10.56 &$\pm\,0.06$& 0.38 &$\pm\,0.14$& 1.63 &$\pm\,0.14$& 1.25 &$\pm\,0.05$& 000 &  7 & 14 & 16  \\
DZOA4653-04  & 1 & 13 47 38.2 & -58 52 15 & 310.15 & 3.21 &   4.3 & {\it UG} & 0 1 1 0 & S    &   ... &       ... &   ... &       ... &   ... &       ... &  ... &       ... &  ... &       ... &  ... &       ... & 000 &  4 &  2 &...  \\
DZOA4653-01  & 1 & 13 47 44.1 & -58 26 38 & 310.26 & 3.62 &   3.8 & {\it UG} & 0 1 1 1 & SM   &   ... &       ... &   ... &       ... &   ... &       ... &  ... &       ... &  ... &       ... &  ... &       ... & 000 & 15 & 17 & 10  \\
DZOA4653-07  & 1 & 13 48 27.5 & -60 11 47 & 309.97 & 1.89 &  10.7 & {\it DG} & 0 1 1 1 & S    & 16.23 &$\pm\,0.09$& 13.29 &$\pm\,0.04$& 11.46 &$\pm\,0.07$& 0.24 &$\pm\,0.11$& 0.75 &$\pm\,0.12$& 0.52 &$\pm\,0.08$& 030 &  8 & 18 & 11  \\
DZOA4654-03  & 1 & 13 49 04.0 & -58 27 34 & 310.43 & 3.57 &   3.2 & {\it UG} & 0 1 1 1 & SM   & 15.65 &$\pm\,0.07$& 13.92 &$\pm\,0.11$& 12.16 &$\pm\,0.19$& 0.91 &$\pm\,0.11$& 1.84 &$\pm\,0.19$& 0.93 &$\pm\,0.19$& 403 &  9 & 11 &  8  \\
DZOA4654-02  & 1 & 13 49 46.1 & -58 13 04 & 310.57 & 3.79 &   2.6 & {\it DG} & 1 1 1 1 & SM   & 14.61 &$\pm\,0.04$& 12.96 &$\pm\,0.04$& 11.81 &$\pm\,0.14$& 1.00 &$\pm\,0.06$& 1.91 &$\pm\,0.11$& 0.91 &$\pm\,0.11$& 000 & 16 & 16 & 10  \\
DZOA4654-01  & 1 & 13 49 49.9 & -57 37 25 & 310.71 & 4.36 &   2.5 & {\it DG} & 1 1 1 0 & SL   & 16.34 &$\pm\,0.12$& 14.59 &$\pm\,0.11$&   ... &       ... & 0.66 &$\pm\,0.21$&  ... &       ... &  ... &       ... & 000 & 12 &  7 &...  \\
DZOA4655-01  & 1+& 13 50 21.3 & -58 17 12 & 310.63 & 3.70 &   2.9 & {\it DG} & 0 1 1 1 & SE   & 14.91 &$\pm\,0.05$& 13.15 &$\pm\,0.05$& 12.51 &$\pm\,0.14$& 1.07 &$\pm\,0.05$& 1.82 &$\pm\,0.11$& 0.75 &$\pm\,0.10$& 330 & 14 & 14 &  8  \\
DZOA4655-04  & 1+& 13 50 47.0 & -59 23 08 & 310.43 & 2.62 &   6.1 & {\it DG} & 0 1 1 1 & SM   & 15.59 &$\pm\,0.13$& 12.86 &$\pm\,0.06$& 11.70 &$\pm\,0.12$& 1.12 &$\pm\,0.11$& 2.03 &$\pm\,0.13$& 0.91 &$\pm\,0.09$& 033 & 12 & 15 & 12  \\
DZOA4655-03  & 1+& 13 50 55.3 & -59 08 29 & 310.51 & 2.85 &   4.1 & {\it DG} & 0 1 1 1 &SM/L  & 15.70 &$\pm\,0.08$& 13.82 &$\pm\,0.06$&   ... &       ... & 0.80 &$\pm\,0.12$&  ... &       ... &  ... &       ... & 000 & 12 &  9 &  8  \\
DZOA4655-02  & 1+& 13 50 56.5 & -58 27 46 & 310.66 & 3.51 &   3.5 & {\it DG} & 0 1 1 1 & SM   & 15.88 &$\pm\,0.08$&   ... &       ... & 12.21 &$\pm\,0.11$&  ... &       ... & 2.26 &$\pm\,0.14$&  ... &       ... & 000 & 13 & 12 &  8  \\
DZOA4655-08  & 1+& 13 51 03.5 & -57 47 15 & 310.83 & 4.17 &   2.5 & {\it BG} & 1 0 0 0 & SL   &   ... &       ... &   ... &       ... &   ... &       ... &  ... &       ... &  ... &       ... &  ... &       ... & 000 &... &... &...  \\
DZOA4656-01  & 1 & 13 51 31.9 & -58 23 01 & 310.76 & 3.57 &   3.0 & {\it DG} & 0 1 1 1 & SE   & 15.75 &$\pm\,0.09$& 13.64 &$\pm\,0.07$& 12.85 &$\pm\,0.19$& 1.04 &$\pm\,0.12$& 2.12 &$\pm\,0.19$& 1.07 &$\pm\,0.18$& 000 &  8 & 12 &  7  \\
DZOA4656-04  & 1 & 13 51 33.9 & -60 07 17 & 310.36 & 1.88 &   9.2 & {\it UG} & 0 1 1 1 & S    &   ... &       ... & 13.45 &$\pm\,0.05$& 12.21 &$\pm\,0.20$&  ... &       ... &  ... &       ... & 0.31 &$\pm\,0.15$& 030 &  7 &  9 &  8  \\
DZOA4656-03  & 2 & 13 51 38.6 & -58 35 15 & 310.72 & 3.37 &   4.1 & {\it DG} & 1 1 1 1 & SE   & 12.77 &$\pm\,0.01$& 10.81 &$\pm\,0.01$&  9.14 &$\pm\,0.01$& 1.09 &$\pm\,0.01$& 1.99 &$\pm\,0.01$& 0.89 &$\pm\,0.01$& 000 & 48 & 54 & 37  \\
DZOA4656-02  & 1 & 13 51 39.8 & -58 26 48 & 310.76 & 3.51 &   3.2 & {\it UG} & 0 1 0 0 &SM/L  &   ... &       ... &   ... &       ... &   ... &       ... &  ... &       ... &  ... &       ... &  ... &       ... & 000 &  5 &... &...  \\
DZOA4657-03  & 1 & 13 52 55.4 & -58 09 59 & 310.99 & 3.74 &   2.3 & {\it UG} & 0 1 1 1 &E/SE  & 15.94 &$\pm\,0.09$& 14.61 &$\pm\,0.12$&   ... &       ... & 1.16 &$\pm\,0.14$&  ... &       ... &  ... &       ... & 000 &  6 &  9 &  5  \\
DZOA4657-04  & 1 & 13 53 08.7 & -58 12 59 & 311.00 & 3.69 &   2.4 & {\it UG} & 0 1 1 0 & SL   &   ... &       ... &   ... &       ... &   ... &       ... &  ... &       ... &  ... &       ... &  ... &       ... & 000 &  8 &  8 &  5  \\
DZOA4657-02  & 1 & 13 53 32.4 & -57 49 30 & 311.15 & 4.05 &   2.7 & {\it DG} & 0 1 1 1 &E/SE  &   ... &       ... &   ... &       ... &   ... &       ... &  ... &       ... &  ... &       ... &  ... &       ... & 000 & 12 & 11 &  7  \\
DZOA4657-01  & 1 & 13 53 34.1 & -57 49 10 & 311.15 & 4.06 &   2.7 & {\it DG} & 0 1 1 1 & SE   & 14.81 &$\pm\,0.04$& 13.26 &$\pm\,0.04$& 12.25 &$\pm\,0.14$& 0.91 &$\pm\,0.04$& 1.67 &$\pm\,0.10$& 0.77 &$\pm\,0.10$& 000 & 15 & 14 &  9  \\
DZOA4658-06  & 1 & 13 54 18.1:& -59 56 14:& 310.74 & 1.98 &  10.1 & {\it UG} & 0 1 1 0 & S    & 17.69 &$\pm\,0.18$& 14.28 &$\pm\,0.15$&   ... &       ... & 0.57 &$\pm\,0.20$&  ... &       ... &  ... &       ... & 999 &  4 &  8 &...  \\
DZOA4658-04  & 1 & 13 54 23.1:& -58 06 30:& 311.19 & 3.75 &   2.5 & {\it DG} & 0 1 1 1 & SM   &   ... &       ... &   ... &       ... &   ... &       ... &  ... &       ... &  ... &       ... &  ... &       ... & 999 &  9 & 12 &  8  \\
DZOA4658-03  & 1 & 13 54 27.3:& -58 07 35:& 311.19 & 3.73 &   2.6 & {\it DG} & 1 1 1 1 & SM   & 14.68 &$\pm\,0.03$& 12.69 &$\pm\,0.04$&   ... &       ... & 1.49 &$\pm\,0.06$&  ... &       ... &  ... &       ... & 999 & 14 & 15 & 12  \\
DZOA4658-01  & 1 & 13 54 38.5:& -57 41 11:& 311.32 & 4.15 &   2.3 & {\it DG} & 1 1 1 0 & SE   &   ... &       ... &   ... &       ... &   ... &       ... &  ... &       ... &  ... &       ... &  ... &       ... & 999 &  4 &  6 &...  \\
DZOA4658-05  & 1 & 13 54 39.5:& -58 31 07:& 311.12 & 3.34 &   3.4 & {\it UG} & 0 1 1 1 & SM   & 17.48 &$\pm\,0.12$& 14.38 &$\pm\,0.09$&   ... &       ... & 1.42 &$\pm\,0.20$&  ... &       ... &  ... &       ... & 999 &  6 &  9 &  3  \\
DZOA4659-13  & 2 & 13 55 18.4 & -58 05 32 & 311.31 & 3.74 &   2.8 & {\it UG} & 0 1 1 1 &E/SE  & 15.84 &$\pm\,0.13$& 14.48 &$\pm\,0.12$&   ... &       ... & 1.03 &$\pm\,0.13$&  ... &       ... &  ... &       ... & 000 &  9 &  9 &  5  \\
DZOA4659-11  & 1 & 13 55 29.0 & -57 38 59 & 311.44 & 4.16 &   2.4 & {\it BG} & 1 1 1 0 &SE/M  & 15.80 &$\pm\,0.09$& 14.48 &$\pm\,0.10$&   ... &       ... & 0.95 &$\pm\,0.17$&  ... &       ... &  ... &       ... & 300 &  6 & 13 &  3  \\
DZOA4659-10  & 1 & 13 55 29.5 & -57 35 26 & 311.46 & 4.22 &   2.4 & {\it BG} & 1 1 1 0 & S    &   ... &       ... &   ... &       ... &   ... &       ... &  ... &       ... &  ... &       ... &  ... &       ... & 000 &  5 &  3 &...  \\
DZOA4659-01  & 1 & 13 55 43.5 & -57 48 38 & 311.43 & 4.00 &   2.9 & {\it UG} & 0 1 1 1 & --   & 15.99 &$\pm\,0.07$& 13.04 &$\pm\,0.04$& 11.50 &$\pm\,0.15$& 1.24 &$\pm\,0.09$& 2.39 &$\pm\,0.15$& 1.15 &$\pm\,0.14$& 060 &  4 &  8 &  5  \\
DZOA4659-02  & 1 & 13 56 07.7 & -58 52 21 & 311.22 & 2.95 &   3.8 & {\it UG} & 0 1 1 1 & ?    & 14.75 &$\pm\,0.03$& 13.21 &$\pm\,0.04$& 13.09 &$\pm\,0.22$& 0.52 &$\pm\,0.05$& 0.88 &$\pm\,0.18$& 0.37 &$\pm\,0.18$& 664 & 11 & 17 &  6  \\
DZOA4659-05  & 1 & 13 56 09.7 & -59 46 14 & 311.00 & 2.08 &   8.1 & {\it UG} & 0 1 1 1 & S    & 16.85 &$\pm\,0.17$& 13.64 &$\pm\,0.05$& 12.65 &$\pm\,0.17$& 0.75 &$\pm\,0.14$& 1.15 &$\pm\,0.19$& 0.40 &$\pm\,0.15$& 000 & 10 & 10 &  5  \\
DZOA4660-02  & 1 & 13 56 52.3 & -57 51 53 & 311.57 & 3.91 &   2.5 & {\it DG} & 1 1 1 1 & SM   & 15.98 &$\pm\,0.08$& 14.50 &$\pm\,0.08$&   ... &       ... & 0.70 &$\pm\,0.13$&  ... &       ... &  ... &       ... & 000 & 10 & 12 &  6  \\
DZOA4660-05  & 1 & 13 57 01.5 & -59 12 35 & 311.25 & 2.60 &   5.4 & {\it DG} & 0 1 1 1 &E/SE  & 15.86 &$\pm\,0.06$& 13.05 &$\pm\,0.04$& 11.91 &$\pm\,0.13$& 0.91 &$\pm\,0.07$& 1.58 &$\pm\,0.11$& 0.67 &$\pm\,0.10$& 030 & 10 & 14 & 15  \\
DZOA4660-03  & 1 & 13 57 06.7 & -58 02 43 & 311.55 & 3.72 &   2.9 & {\it DG} & 0 1 1 1 &E/SE  & 16.22 &$\pm\,0.08$& 14.00 &$\pm\,0.06$& 12.98 &$\pm\,0.23$& 1.08 &$\pm\,0.10$& 2.26 &$\pm\,0.16$& 1.18 &$\pm\,0.15$& 000 &  9 & 12 &  8  \\
DZOA4660-01  & 2 & 13 57 08.2 & -57 44 09 & 311.63 & 4.02 &   2.6 & {\it UG} & 0 1 1 1 & SE   & 15.05 &$\pm\,0.03$& 13.94 &$\pm\,0.04$& 13.13 &$\pm\,0.20$& 0.56 &$\pm\,0.05$& 1.39 &$\pm\,0.16$& 0.84 &$\pm\,0.16$& 000 & 12 & 13 &  6  \\
DZOA4660-04  & 1 & 13 57 46.6 & -58 07 01 & 311.62 & 3.63 &   3.0 & {\it DG} & 0 1 1 1 & E    & 15.51 &$\pm\,0.06$& 13.98 &$\pm\,0.06$& 12.20 &$\pm\,0.15$& 0.99 &$\pm\,0.06$& 2.22 &$\pm\,0.10$& 1.22 &$\pm\,0.10$& 000 &  9 & 10 &  9  \\
DZOA4661-03  & 2 & 13 58 52.6 & -59 45 47 & 311.33 & 2.00 &   8.9 & {\it UG} & 0 1 1 1 & S    & 17.16 &$\pm\,0.09$& 13.91 &$\pm\,0.04$& 12.68 &$\pm\,0.12$& 0.66 &$\pm\,0.10$& 1.26 &$\pm\,0.13$& 0.59 &$\pm\,0.09$& 030 &  4 &  9 &  7  \\
DZOA4661-01  & 2 & 13 58 57.0 & -58 46 02 & 311.60 & 2.96 &   4.3 & {\it DG} & 1 1 1 1 & SM   & 13.99 &$\pm\,0.01$& 11.44 &$\pm\,0.01$& 10.05 &$\pm\,0.02$& 0.95 &$\pm\,0.01$& 1.81 &$\pm\,0.02$& 0.86 &$\pm\,0.02$& 030 & 26 & 27 & 22  \\
DZOA4662-02  & 2 & 13 59 35.7:& -58 25 49:& 311.77 & 3.27 &   4.5 & {\it UG} & 0 1 1 1 &E/SE  & 15.35 &$\pm\,0.03$& 13.35 &$\pm\,0.06$& 12.09 &$\pm\,0.09$& 0.72 &$\pm\,0.04$& 1.62 &$\pm\,0.07$& 0.90 &$\pm\,0.07$& 400 & 10 & 12 & 11  \\
DZOA4662-03  & 1 & 14 00 01.9:& -58 29 06:& 311.81 & 3.20 &   4.9 & {\it UG} & 0 1 1 0 & S    & 16.71 &$\pm\,0.11$&   ... &       ... &   ... &       ... &  ... &       ... &  ... &       ... &  ... &       ... & 000 & 10 &  7 &...  \\
DZOA4662-01  & 1 & 14 00 04.0:& -57 50 02:& 311.99 & 3.83 &   2.7 & {\it DG} & 1 1 1 1 & SM   & 15.51 &$\pm\,0.06$& 13.61 &$\pm\,0.05$& 12.03 &$\pm\,0.18$& 1.34 &$\pm\,0.09$& 2.24 &$\pm\,0.17$& 0.90 &$\pm\,0.16$& 000 & 17 & 21 & 18  \\
DZOA4662-04  & 1 & 14 00 10.9:& -59 12 35:& 311.64 & 2.50 &   5.2 & {\it UG} & 0 1 1 0 &SM/L  & 16.35 &$\pm\,0.11$&   ... &       ... &   ... &       ... &  ... &       ... &  ... &       ... &  ... &       ... & 000 &  5 &  8 &...  \\
\noalign{\smallskip}
\hline
\noalign{\smallskip}
\end{tabular*}
 \normalsize
\end{table}
\addtocounter{table}{-1}
\clearpage
\begin{table}[tb]
 \normalsize
\caption{ continued.}
\scriptsize  
\begin{tabular*}{23.5cm}{
l  @{\extracolsep{3mm}}   l @{\extracolsep{2mm}} l@{\extracolsep{1.5mm}} l@{\extracolsep{2mm}}      
  c @{\extracolsep{1.5mm}} r @{\extracolsep{0.5mm}} r @{\extracolsep{2mm}} 
  l @{\extracolsep{1mm}} l @{\extracolsep{3mm}} l @{\extracolsep{3mm}} 
  r @{\extracolsep{0.5mm}} r @{\extracolsep{2mm}} r @{\extracolsep{0.5mm}} r @{\extracolsep{2mm}} r @{\extracolsep{0.5mm}} r @{\extracolsep{3mm}} 
  r @{\extracolsep{0.5mm}} c @{\extracolsep{2mm}} r @{\extracolsep{0.5mm}} c @{\extracolsep{2mm}} r @{\extracolsep{0.5mm}} c @{\extracolsep{1mm}} 
  r @{\extracolsep{1.5mm}}  r @{\extracolsep{1mm}} r @{\extracolsep{1mm}} r @{\extracolsep{2mm}}  
}
\noalign{\smallskip}
\hline
\noalign{\smallskip}
\multicolumn{1}{c}{Ident.} & N & \multicolumn{2}{c}{R.A.\, (J2000)\, Dec.}
 & Gal $\ell$ & Gal $b$ & $A_B$ & Class & Visibil. & Type & 
 \multicolumn{2}{c}{\II} & \multicolumn{2}{c}{\J} & \multicolumn{2}{c}{\K} &  
 \multicolumn{2}{c@{\extracolsep{1mm}}}{\ije} & \multicolumn{2}{c@{\extracolsep{1mm}}}{\ike} & \multicolumn{2}{c@{\extracolsep{1mm}}}{\jke} & 
 Phot. & $D_I$ & $D_J$ & $D_K$ \\
\noalign{\smallskip}
 & & \multicolumn{1}{c}{[$^{h}$ $^{m}$ $^{s}$]} & \multicolumn{1}{c}{[$\deg$ $\arcmin$ $\arcsec$]} 
 & [$\deg$] & [$\deg$] & [mag] & & & & 
 \multicolumn{2}{c}{[mag]} & \multicolumn{2}{c}{[mag]} & \multicolumn{2}{c}{[mag]} 
 & \multicolumn{2}{c}{[mag]} & \multicolumn{2}{c}{[mag]} & \multicolumn{2}{c}{[mag]} & 
 & [$\arcsec$] & [$\arcsec$] & [$\arcsec$] \\
\vspace{-1mm} \\
\multicolumn{1}{c}{(1)} & (2) & \multicolumn{1}{c}{(3a)} & \multicolumn{1}{c}{(3b)} & 
 (4a) & (4b) & \multicolumn{1}{c}{(5)} & (6) & \multicolumn{1}{c}{(7)} & (8) 
 & \multicolumn{2}{c}{(9)} & \multicolumn{2}{c}{(10)} & \multicolumn{2}{c}{(11)} & \multicolumn{2}{c}{(12)} 
 & \multicolumn{2}{c}{(13)} & \multicolumn{2}{c}{(14)} & (15) & (16)  & (17) & (18) \\
\noalign{\smallskip}
\hline
\noalign{\smallskip}
DZOA4662-07  & 1 & 14 00 26.9:& -60 18 31:& 311.38 & 1.43 &  15.4 & {\it UG} & 0 0 1 1 & ?    &   ... &       ... & 14.37 &$\pm\,0.07$& 12.69 &$\pm\,0.15$&  ... &       ... &  ... &       ... & 0.16 &$\pm\,0.16$& 000 &... & 11 &  9  \\
DZOA4663-05  & 2 & 14 00 33.9 & -59 32 32 & 311.60 & 2.16 &   6.6 & {\it DG} & 0 1 1 1 &SE/M  & 16.53 &$\pm\,0.07$& 12.86 &$\pm\,0.02$& 11.45 &$\pm\,0.08$& 1.37 &$\pm\,0.08$& 2.27 &$\pm\,0.09$& 0.90 &$\pm\,0.06$& 000 &  6 & 16 & 11  \\
DZOA4663-02  & 1 & 14 00 55.3 & -57 37 23 & 312.15 & 4.00 &   2.9 & {\it DG} & 0 1 1 0 & SM   & 15.02 &$\pm\,0.03$& 14.22 &$\pm\,0.05$&   ... &       ... & 0.23 &$\pm\,0.06$&  ... &       ... &  ... &       ... & 000 & 13 & 15 &...  \\
DZOA4663-01  & 1 & 14 00 55.8 & -57 37 28 & 312.15 & 4.00 &   2.9 & {\it DG} & 0 1 1 0 & SM   & 15.27 &$\pm\,0.04$& 13.64 &$\pm\,0.04$&   ... &       ... & 0.74 &$\pm\,0.05$&  ... &       ... &  ... &       ... & 000 & 14 & 16 &...  \\
DZOA4663-04  & 1 & 14 01 25.6 & -58 37 37 & 311.95 & 3.01 &   4.4 & {\it DG} & 0 1 1 1 &SE/M  & 15.84 &$\pm\,0.07$& 13.43 &$\pm\,0.04$& 11.99 &$\pm\,0.10$& 1.11 &$\pm\,0.08$& 2.19 &$\pm\,0.14$& 1.08 &$\pm\,0.13$& 000 & 11 & 14 & 10  \\
DZOA4663-03  & 1 & 14 01 35.4 & -58 32 22 & 311.99 & 3.09 &   4.3 & {\it UG} & 0 1 1 0 & S    & 16.38 &$\pm\,0.14$& 15.39 &$\pm\,0.20$&   ... &       ... & 0.04 &$\pm\,0.20$&  ... &       ... &  ... &       ... & 000 &  7 &  8 &...  \\
DZOA4664-02  & 1 & 14 03 11.2 & -58 06 20 & 312.31 & 3.45 &   3.5 & {\it DG} & 0 1 1 1 & SL   & 15.03 &$\pm\,0.05$& 13.46 &$\pm\,0.05$&   ... &       ... & 0.83 &$\pm\,0.09$&  ... &       ... &  ... &       ... & 000 & 15 & 15 &  6  \\
DZOA4665-02  & 2 & 14 03 13.8 & -58 07 13 & 312.31 & 3.44 &   3.6 & {\it DG} & 0 1 1 1 & SM   & 14.50 &$\pm\,0.02$& 12.61 &$\pm\,0.02$& 11.67 &$\pm\,0.05$& 0.81 &$\pm\,0.03$& 1.66 &$\pm\,0.06$& 0.85 &$\pm\,0.06$& 030 & 18 & 21 & 12  \\
DZOA4665-03  & 1 & 14 03 57.5 & -58 09 49 & 312.40 & 3.37 &   4.0 & {\it UG} & 0 1 1 1 &SM/L  &   ... &       ... &   ... &       ... &   ... &       ... &  ... &       ... &  ... &       ... &  ... &       ... & 000 &  8 & 15 &  6  \\
DZOA4665-01  & 2 & 14 04 21.0 & -57 43 33 & 312.57 & 3.77 &   2.9 & {\it DG} & 0 1 1 1 & E    & 15.21 &$\pm\,0.03$& 13.29 &$\pm\,0.03$& 12.13 &$\pm\,0.10$& 1.10 &$\pm\,0.04$& 2.13 &$\pm\,0.06$& 1.03 &$\pm\,0.07$& 000 & 10 & 14 & 11  \\
DZOA4666-01  & 1 & 14 05 35.9 & -59 32 19 & 312.21 & 1.99 &   8.5 & {\it UG} & 0 1 1 1 &SM/L  & 17.00 &$\pm\,0.16$& 14.70 &$\pm\,0.13$&   ... &       ... & 0.09 &$\pm\,0.16$&  ... &       ... &  ... &       ... & 000 &  6 & 12 &  5  \\
\noalign{\smallskip}
\multicolumn{18}{l}{\it \B -band galaxies found to be non galaxian with DENIS:} \\
\noalign{\smallskip}
DZOA4654-04  & 1 & 13 49 41.3 & -57 48 40 & 310.65 & 4.18 &   2.6 & {\it NG} & 1 1 0 0 & --   &   ... &       ... &   ... &       ... &   ... &       ... &  ... &       ... &  ... &       ... &  ... &       ... & 000 &  4 &... &...  \\
DZOA4655-07  & 1 & 13 50 51.8 & -57 43 02 & 310.82 & 4.24 &   2.6 & {\it NG} & 1 1 1 0 & --   &   ... &       ... &   ... &       ... &   ... &       ... &  ... &       ... &  ... &       ... &  ... &       ... & 000 &  5 &  4 &...  \\
DZOA4660-06  & 2 & 13 56 42.9 & -59 11 16 & 311.22 & 2.63 &   5.1 & {\it NG} & 1 1 1 1 & --   & 15.95 &$\pm\,0.05$& 14.62 &$\pm\,0.06$&   ... &       ... & 0.08 &$\pm\,0.11$&  ... &       ... &  ... &       ... & 000 &... &... &...  \\
DZOA4661-04  & 1 & 13 58 40.7 & -58 18 13 & 311.69 & 3.42 &   3.5 & {\it NG} & 1 1 1 0 & --   &   ... &       ... &   ... &       ... &   ... &       ... &  ... &       ... &  ... &       ... &  ... &       ... & 000 &... &... &...  \\
DZOA4664-03  & 2 & 14 02 42.0 & -57 44 08 & 312.35 & 3.83 &   3.3 & {\it NG} & 1 1 0 0 & --   &   ... &       ... &   ... &       ... &   ... &       ... &  ... &       ... &  ... &       ... &  ... &       ... & 000 &... &... &...  \\
\noalign{\smallskip}
\hline
\noalign{\smallskip}
\end{tabular*}
\normalsize
\end{table}
\end{landscape}

%% file: neb_denis.tex
\begin{table*}[ht]
\normalsize
\caption{Galactic objects detected in the search area. }
\label{nebtab}
\scriptsize  
\begin{minipage}[t]{\textwidth}
\renewcommand{\footnoterule}{}  
\begin{tabular*}{18.1cm}{
  l  @{\extracolsep{3mm}} l @{\extracolsep{3mm}} l@{\extracolsep{3mm}} l@{\extracolsep{3mm}}      
  r @{\extracolsep{3mm}} r @{\extracolsep{3mm}}  r @{\extracolsep{3mm}} l @{\extracolsep{3mm}} l @{\extracolsep{3mm}} 
  l 
}
\noalign{\smallskip}
\hline
\noalign{\smallskip}
\multicolumn{1}{c}{Ident.} & N & \multicolumn{2}{c}{R.A.\, (J2000)\, Dec.}
 & Gal $\ell$ & Gal $b$ & $A_B^{tot}$ & Class & Visibil.  & Other name \\
\vspace{-1mm} \\
\multicolumn{1}{c}{(1)} & (2) & \multicolumn{1}{c}{(3a)} & \multicolumn{1}{c}{(3b)} & 
 (4a) & (4b) & (5) & (6) & \multicolumn{1}{c}{(7)} & (8) \\
\noalign{\smallskip}
\hline
\noalign{\smallskip}
DZOA4638-16  & 1  & 13 27 08.5 & -62 03 17  & 307.10 & 0.53 &   28.9 &  ...  & 0 0 0 1  & 2MASXJ13270813-6203201   \\
DZOA4638-13  & 2  & 13 27 10.4 & -62 05 05  & 307.10 & 0.50 &   28.1 &  ...  & 0 0 1 1  &   \\
DZOA4641-10\footnote{While this object shows two apparent point sources in
the \K -band, the \B - and \II -bands clearly show an elongated compact
extended emission with the brightest part positioned on one of the point sources.}
             & 2  & 13 32 12.4 & -60 25 33  & 307.94 & 2.05 &   12.5 &  ...  & 1 1 1 1  &   \\
DZOA4641-12  & 5  & 13 32 30.9 & -62 45 06  & 307.61 &-0.25 &  127.6 &  HII? & 0 1 1 1  &   \\
DZOA4642-09  & 6  & 13 32 31.4 & -63 05 22  & 307.56 &-0.59 &   50.0 &  HII  & 0 0 1 1  & PMNJ1332-6305   \\
DZOA4642-07  & 3  & 13 32 48.8 & -60 26 44  & 308.02 & 2.02 &   21.0 &  YSO  & 1 1 1 1  & 2MASXJ13324576-6026565,  AM1326-601  \\
DZOA4647-04  & 1  & 13 40 26.4 & -61 47 57  & 308.69 & 0.52 &  100.8 &  YSO  & 0 0 0 1  &   \\
DZOA4648-07  & 1  & 13 40 57.6 & -61 45 44  & 308.75 & 0.55 &  113.4 &  ...  & 0 0 1 1  & 2MASXJ13405761-6145447     \\
DZOA4649-10  & 3+ & 13 41 54.8 & -62 07 43  & 308.79 & 0.17 &   39.9 &  ...  & 0 1 1 1  &   \\
DZOA4650-12  & 4++& 13 44 39.3 & -62 05 31  & 309.12 & 0.14 &   94.2 &  HII  & 0 0 0 1  &   \\
DZOA4651-13  & 1  & 13 45 42.4 & -62 00 32  & 309.25 & 0.20 &   46.0 &  ...  & 0 0 0 1  &   \\
DZOA4652-10\footnote{This objects shows diffuse emission as well as clumps,
and possibly a dust lane. It could be a PN or a YSO seen on edge.}
             & 1  & 13 46 20.8:& -62 48 02: & 309.16 &-0.59 &   57.2 &  ...  & 0 1 1 1  & 2MASXJ13462058-6247597    \\
DZOA4652-09  & 1  & 13 46 37.4:& -62 39 27: & 309.22 &-0.46 &  111.9 &  HII? & 0 0 1 1  & 2MASXJ13463702-6239303    \\
DZOA4655-12  & 1  & 13 50 35.4 & -61 40 18  & 309.89 & 0.40 &  143.4 &  YSO  & 0 0 1 1  & 2MASXJ13503488-6140199,  (PMNJ1350-6141)  \\
DZOA4655-11  & 2  & 13 50 41.8 & -61 35 09  & 309.92 & 0.48 &  151.6 &  YSO  & 0 0 0 1  &   \\
DZOA4655-09\footnote{Very faint, it seems to show comparable features to
DZOA4652-10 and DZOA4656-08}
             & 1  & 13 51 02.6 & -61 30 15  & 309.98 & 0.55 &   77.0 &  ...  & 0 0 1 1  & 2MASXJ13510266-6130150    \\
DZOA4655-10  & 2  & 13 51 14.7 & -61 32 35  & 309.99 & 0.51 &   97.5 &  ...  & 0 0 1 1  &   \\
DZOA4656-08\footnote{An extremely bright point source in the \K -band, is also shows
diffuse and clumpy extended emission and possible a dust lane in all bands. }
             & 1  & 13 51 37.9 & -61 39 07  & 310.01 & 0.39 &  113.2 &  ...  & 0 1 1 1  &     \\
DZOA4656-06  & 1  & 13 51 59.8 & -61 15 41  & 310.15 & 0.76 &   60.5 &  HII? & 0 0 0 1  & 2MASXJ13515956-6115394    \\
DZOA4657-05  & 2  & 13 52 37.3 & -62 19 00  & 309.97 &-0.28 &   55.8 &  ...  & 0 0 1 1  &   \\
DZOA4657-06  & 2  & 13 53 23.3 & -60 33 48  & 310.48 & 1.40 &   11.2 &  PN   & 1 1 1 1  &   \\
DZOA4658-07  & 1  & 13 54 15.9:& -62 13 46: & 310.18 &-0.24 &   76.5 &  RN   & 1 1 1 1  & KK2000-60  \\
DZOA4661-02  & 2  & 13 58 13.8 & -58 54 31  & 311.48 & 2.85 &    4.4 &  PN   & 1 1 1 1  &   \\
DZOA4661-05  & 1  & 13 58 23.9 & -61 21 45  & 310.87 & 0.47 &   57.1 &  ...  & 0 0 0 1  &     \\
DZOA4664-05  & 2  & 14 02 36.2 & -61 05 44  & 311.43 & 0.60 &   57.2 &  HII? & 0 0 1 1  & 2MASXJ14023620-6105450  \\
DZOA4664-04  & 1  & 14 02 52.9 & -60 48 27  & 311.54 & 0.86 &   29.0 &  ...  & 0 0 1 1  &     \\
DZOA4664-06  & 1  & 14 02 52.9 & -62 07 22  & 311.18 &-0.40 &   93.6 &  HII? & 0 0 0 1  &     \\
\noalign{\smallskip}
\hline
\noalign{\smallskip}
\end{tabular*}
\end{minipage}
\normalsize
\end{table*}

%% file: gal_lit.tex
\begin{table*}[ht]
\normalsize
\caption{Cross-identifications with other catalogues }
\label{littab}
\scriptsize  
\begin{tabular*}{18.1cm}{
  l  @{\extracolsep{3mm}} l @{\extracolsep{2mm}}  l @{\extracolsep{3mm}}      
  r @{\extracolsep{3mm}} l @{\extracolsep{3mm}} l @{\extracolsep{3mm}} 
  l @{\extracolsep{2mm}} l @{\extracolsep{2mm}}  l  @{\extracolsep{2mm}}  l
}
\noalign{\smallskip}
\hline
\noalign{\smallskip}
\multicolumn{1}{c}{Ident.} & \multicolumn{2}{c}{R.A.\, (J2000)\, Dec.}
 & $A_B$ & Class & Visibil. 
& 2MASX & WKK & Other names & \\
\vspace{-1mm} \\
\multicolumn{1}{c}{(1)} & \multicolumn{1}{c}{(2a)} & \multicolumn{1}{c}{(2b)} & 
 (3) & (4) & \multicolumn{1}{c}{(5)} & (6) & (7) & (8) & \\
\noalign{\smallskip}
\hline
\noalign{\smallskip}
DZOA4638-04   &  13 27 20.3 & -57 52 08 &   2.8 & {\it DG} & 0 1 1 1 &  J13272018-5752081 &         & &   \\
DZOA4638-09   &  13 27 25.1 & -58 20 28 &   2.6 & {\it DG} & 0 1 1 1 &  J13272507-5820282 & 	     & &   \\
DZOA4638-03   &  13 28 10.0 & -57 41 09 &   3.0 & {\it DG} & 0 1 1 1 &  J13280986-5741102 & 	     & &   \\
DZOA4638-11   &  13 28 10.3 & -60 22 59 &   5.0 & {\it DG} & 0 1 1 1 &  J13281021-6022580 & 	     & &   \\
DZOA4638-10   &  13 28 11.0 & -59 25 09 &   3.1 & {\it DG} & 0 1 1 0 &  		   &	     & &   \\
DZOA4638-06   &  13 28 15.4 & -57 55 21 &   2.7 & {\it DG} & 0 1 1 1 &  		   &	     & &   \\
DZOA4638-01   &  13 28 37.1 & -57 42 19 &   2.9 & {\it DG} & 1 1 1 1 &  J13283697-5742188 & WKK\,2267 & &   \\
DZOA4639-07   &  13 28 44.4 & -58 03 32 &   2.8 & {\it UG} & 0 1 1 0 &  		   &	     & &   \\
DZOA4639-06   &  13 28 49.7 & -58 03 21 &   2.7 & {\it DG} & 1 1 1 1 &  J13284958-5803228 & WKK\,2271 & &   \\
DZOA4639-19   &  13 29 00.3 & -58 55 30 &   2.6 & {\it BG} & 1 1 1 0 &                    & WKK\,2274 & &   \\
DZOA4639-02   &  13 29 01.7 & -57 37 32 &   3.0 & {\it DG} & 0 1 1 0 &  		   &	     & &   \\
DZOA4639-14   &  13 29 10.4 & -59 42 27 &   4.1 & {\it UG} & 0 1 1 0 &  		   &	     & &   \\
DZOA4639-01   &  13 29 14.7 & -57 37 06 &   3.1 & {\it UG} & 0 1 1 1 &  		   &	     & &   \\
DZOA4639-03   &  13 29 16.1 & -57 39 56 &   2.9 & {\it DG} & 0 1 1 1 &  J13291609-5739562 & 	     & &   \\
DZOA4639-05   &  13 29 17.7 & -57 56 03 &   2.7 & {\it DG} & 1 1 1 1 &  J13291753-5756032 & WKK\,2278 & &   \\
DZOA4639-10   &  13 29 25.8 & -59 07 11 &   2.6 & {\it DG} & 0 1 1 1 &  		   &	     & &   \\
DZOA4639-09   &  13 29 33.2 & -58 50 54 &   3.0 & {\it DG} & 1 1 1 1 &  J13293316-5850552 & WKK\,2281 & &   \\
DZOA4639-08   &  13 29 40.9 & -58 49 30 &   3.1 & {\it DG} & 0 1 1 1 &  		   &	     & &   \\
DZOA4639-16   &  13 29 50.5 & -60 44 05 &   5.8 & {\it UG} & 0 1 1 1 &  		   &	     & &   \\
DZOA4639-13   &  13 29 51.9 & -59 28 19 &   3.1 & {\it DG} & 1 1 1 1 &                    & WKK\,2285 & &   \\
DZOA4640-05   &  13 30 13.4 & -59 08 34 &   3.1 & {\it DG} & 0 1 1 1 &  		   &	     & &   \\
DZOA4640-03   &  13 30 34.6 & -58 29 24 &   3.4 & {\it DG} & 1 1 1 1 &  J13303446-5829247 & WKK\,2292 & &   \\
DZOA4640-02   &  13 31 09.6 & -58 09 45 &   2.8 & {\it DG} & 0 1 1 1 &  J13310962-5809453 & 	     & &   \\
DZOA4641-01   &  13 31 33.3 & -57 50 04 &   2.7 & {\it DG} & 1 1 1 1 &  J13313318-5750054 & WKK\,2300 & &   \\
DZOA4641-04   &  13 31 36.6 & -60 22 37 &   5.8 & {\it UG} & 0 0 1 1 &  		   &	     & &   \\
DZOA4641-02   &  13 31 43.7 & -57 53 11 &   2.8 & {\it DG} & 1 1 1 1 &  J13314368-5753125 & WKK\,2303 & &   \\
DZOA4641-06   &  13 32 03.3 & -63 05 06 &  38.5 & {\it UG} & 0 0 0 1 &  		   &	     & &   \\
DZOA4642-04   &  13 33 11.8 & -58 49 22 &   4.0 & {\it DG} & 1 1 1 1 &                    & WKK\,2327 & &   \\
DZOA4642-01   &  13 33 39.2 & -57 47 42 &   2.8 & {\it DG} & 1 1 1 1 &  J13333914-5747422 & WKK\,2334 & &   \\
DZOA4642-06   &  13 33 47.6 & -59 03 07 &   3.9 & {\it DG} & 1 1 1 1 &                    & WKK\,2336 & &   \\
DZOA4642-02   &  13 33 58.2 & -58 00 29 &   3.2 & {\it DG} & 1 1 1 1 &                    & WKK\,2338 & &   \\
DZOA4644-04   &  13 35 26.0 & -59 14 38 &   4.1 & {\it UG} & 0 1 1 1 &  		   &	     & &   \\
DZOA4644-02   &  13 35 30.1 & -57 53 47 &   3.4 & {\it DG} & 0 1 1 0 &  		   &	     & &   \\
DZOA4644-01   &  13 36 08.1 & -57 37 48 &   2.7 & {\it DG} & 0 1 1 1 &  		   &	     & &   \\
DZOA4645-01   &  13 37 05.0 & -58 02 40 &   3.1 & {\it DG} & 0 1 1 1 &  		   &	     & &   \\
DZOA4645-14   &  13 37 15.2 & -57 37 11 &   2.7 & {\it BG} & 1 1 1 0 &                    & WKK\,2386 & &   \\
DZOA4645-13   &  13 37 20.7 & -63 28 12 &  24.3 & {\it UG} & 0 0 1 1 &                    &         & &   \\
DZOA4645-09   &  13 37 24.7 & -58 52 21 &   4.5 & {\it DG} & 1 1 1 1 &  J13372458-5852216 & WKK\,2390 & &   \\
DZOA4645-04   &  13 37 31.9 & -58 08 01 &   3.1 & {\it DG} & 0 1 1 1 &                    &         & &    \\
DZOA4645-08   &  13 37 32.8 & -58 50 04 &   4.5 & {\it DG} & 1 1 1 1 &  J13373272-5850056 & WKK\,2392 & &   \\
DZOA4645-10   &  13 37 32.9 & -58 54 14 &   4.5 & {\it DG} & 0 1 1 1 &  J13373282-5854136 &         & IRAS\,13342-5830 \ PMN\,J1337-5854  &   \\
DZOA4645-03   &  13 37 44.1 & -58 06 37 &   3.0 & {\it UG} & 0 1 1 1 &   		   &	     & &   \\
DZOA4645-05   &  13 37 44.3 & -58 13 26 &   3.5 & {\it DG} & 1 1 1 0 &                    & WKK\,2397 & &   \\
DZOA4645-02   &  13 37 51.3 & -58 00 59 &   2.9 & {\it DG} & 0 1 1 0 &  		   &	     & &   \\
DZOA4645-07   &  13 37 59.2 & -58 30 55 &   4.7 & {\it DG} & 0 1 1 1 &  J13375912-5830556 & 	     & &   \\
DZOA4646-01   &  13 38 03.3 & -58 14 27 &   3.5 & {\it UG} & 0 1 1 1 &  		   &	     & &   \\
DZOA4646-03   &  13 38 08.5 & -58 45 19 &   4.1 & {\it DG} & 0 1 1 1 &  J13380855-5845197 & 	     & &   \\
DZOA4646-04   &  13 38 16.6 & -59 00 15 &   4.3 & {\it DG} & 0 1 1 1 &  		   &	     & &   \\
DZOA4646-06   &  13 38 21.7 & -60 17 02 &   9.0 & {\it DG} & 0 1 1 1 &  		   &	     & &   \\
DZOA4647-03   &  13 39 17.6 & -59 04 46 &   4.3 & {\it UG} & 0 1 1 1 &  		   &	     & &   \\
DZOA4647-01   &  13 39 39.7 & -58 04 00 &   3.5 & {\it DG} & 0 1 1 1 &  J13393970-5804007 & 	     & &   \\
DZOA4647-02   &  13 39 52.7 & -57 42 17 &   2.6 & {\it DG} & 1 1 1 1 &                    & WKK\,2435 & &   \\
DZOA4649-02   &  13 41 54.8 & -58 48 28 &   3.9 & {\it DG} & 0 1 1 1 &  		   &	     & IRAS\,13386-5832  \ HIZOA\,J1341-58 &\\
DZOA4649-07   &  13 42 09.8 & -61 08 18 &  12.0 & {\it DG} & 0 0 1 1 &  		   &	     & &   \\
DZOA4649-06   &  13 42 29.6 & -61 01 23 &  13.0 & {\it DG} & 0 0 1 1 &  		   &	     & HIZSS\,082 \ HIZOA\,J1342-61 & \\
DZOA4649-03   &  13 42 35.6 & -59 33 18 &   6.7 & {\it UG} & 0 1 1 0 &  		   &	     & &   \\
DZOA4649-01   &  13 42 59.9 & -57 39 06 &   2.5 & {\it DG} & 1 1 1 1 &                    & WKK\,2483 & &   \\
DZOA4650-09   &  13 44 03.7 & -60 19 35 &  10.4 & {\it DG} & 0 1 1 1 &  J13440358-6019350 & 	     & &   \\
DZOA4650-01   &  13 44 36.5 & -58 13 10 &   3.9 & {\it DG} & 0 1 1 1 &  		   &	     & &   \\
DZOA4651-05   &  13 45 00.2 & -59 22 17 &   6.2 & {\it UG} & 0 1 1 0 &  		   &	     & &   \\
DZOA4651-02   &  13 45 17.7 & -58 12 01 &   3.5 & {\it DG} & 0 1 1 0 &  		   &	     & &   \\
DZOA4651-08   &  13 45 25.3 & -60 29 14 &  11.9 & {\it UG} & 0 0 1 1 &  		   &	     & NW04-03 &   \\
DZOA4651-06   &  13 45 50.7 & -60 09 05 &   8.2 & {\it DG} & 0 1 1 1 &  J13455154-6009067 & 	     & NW04-08 &   \\
DZOA4652-01   &  13 46 40.3:& -57 39 50:&   2.3 & {\it UG} & 0 1 1 1 &  		   &	     & &   \\
DZOA4652-04   &  13 46 48.9:& -60 24 29:&  12.3 & {\it DG} & 0 1 1 1 &  J13464910-6024299 &         & NW04-24 \ PKS\,1343-601 \ Centaurus B & \\
DZOA4652-02   &  13 46 57.3:& -58 10 19:&   3.3 & {\it DG} & 0 1 1 0 &  		   &	     & HIZOA\,J1347-58 &   \\
DZOA4653-09   &  13 47 18.5 & -60 34 13 &  12.2 & {\it DG} & 0 0 1 1 &  J13471848-6034133 & 	     & NW04-38 &   \\
DZOA4653-03   &  13 47 32.3 & -58 47 45 &   3.8 & {\it UG} & 0 1 1 0 &  		   &	     & &   \\
DZOA4653-11   &  13 47 36.2 & -60 37 04 &  12.0 & {\it DG} & 0 1 1 1 &                    &         & NW04-45 \ 4U\,1344-60 \ 1XMM\,J134736.1-603704 &  \\
DZOA4653-04   &  13 47 38.2 & -58 52 15 &   4.3 & {\it UG} & 0 1 1 0 &  		   &	     & &   \\
DZOA4653-01   &  13 47 44.1 & -58 26 38 &   3.8 & {\it UG} & 0 1 1 1 &  		   &	     & &   \\
DZOA4653-07   &  13 48 27.5 & -60 11 47 &  10.7 & {\it DG} & 0 1 1 1 &  		   &	     & NW04-51 &   \\
DZOA4654-03   &  13 49 04.0 & -58 27 34 &   3.2 & {\it UG} & 0 1 1 1 &  		   &	     & &   \\
DZOA4654-04   &  13 49 41.3 & -57 48 40 &   2.6 & {\it NG} & 1 1 0 0 &                    & WKK\,2559 & &   \\
DZOA4654-02   &  13 49 46.1 & -58 13 04 &   2.6 & {\it DG} & 1 1 1 1 &  J13494605-5813040 & WKK\,2562 & &   \\
DZOA4654-01   &  13 49 49.9 & -57 37 25 &   2.5 & {\it DG} & 1 1 1 0 &                    & WKK\,2564 & &   \\
DZOA4655-01   &  13 50 21.3 & -58 17 12 &   2.9 & {\it DG} & 0 1 1 1 &  J13502126-5817121 & 	     & &   \\
DZOA4655-04   &  13 50 47.0 & -59 23 08 &   6.0 & {\it DG} & 0 1 1 1 &  J13504691-5923083 & 	     & &   \\
DZOA4655-07   &  13 50 51.8 & -57 43 02 &   2.6 & {\it NG} & 1 1 1 0 &                    & WKK\,2586 & &   \\
\noalign{\smallskip}
\hline	      
\noalign{\smallskip}
\end{tabular*} 
 \normalsize
\end{table*}
\addtocounter{table}{-1}
\clearpage
\begin{table*}[ht]
\normalsize
\caption{continued.}
\scriptsize  
\begin{tabular*}{18.1cm}{
  l  @{\extracolsep{3mm}} l @{\extracolsep{2mm}}  l @{\extracolsep{3mm}}      
  r @{\extracolsep{3mm}} l @{\extracolsep{3mm}} l @{\extracolsep{3mm}} 
  l @{\extracolsep{2mm}} l @{\extracolsep{2mm}}  l  @{\extracolsep{2mm}}  l
}
\noalign{\smallskip}
\hline
\noalign{\smallskip}
\multicolumn{1}{c}{Ident.} & \multicolumn{2}{c}{R.A.\, (J2000)\, Dec.}
 & $A_B$ & Class & Visibil. 
& 2MASX & WKK & Other names & \\
\vspace{-1mm} \\
\multicolumn{1}{c}{(1)} & \multicolumn{1}{c}{(2a)} & \multicolumn{1}{c}{(2b)} & 
 (3) & (4) & \multicolumn{1}{c}{(5)} & (6) & (7) & (8) & \\
\noalign{\smallskip}
\hline
\noalign{\smallskip}
DZOA4655-03   &  13 50 55.3 & -59 08 29 &   4.1 & {\it DG} & 0 1 1 1 &  		   &	     & &   \\
DZOA4655-02   &  13 50 56.5 & -58 27 46 &   3.5 & {\it DG} & 0 1 1 1 &  		   &	     & &   \\
DZOA4655-08   &  13 51 03.5 & -57 47 15 &   2.6 & {\it BG} & 1 0 0 0 &                    & WKK\,2589 & &   \\
DZOA4656-01   &  13 51 31.9 & -58 23 01 &   3.0 & {\it DG} & 0 1 1 1 &  		   &	     & &   \\
DZOA4656-04   &  13 51 33.9 & -60 07 17 &   9.2 & {\it UG} & 0 1 1 1 &  		   &	     & &   \\
DZOA4656-03   &  13 51 38.6 & -58 35 15 &   4.1 & {\it DG} & 1 1 1 1 &  J13513848-5835153 & WKK\,2596 & IRAS\,13483-5820  \ HIZSS\,084 \ HIZOA\,J1351-58 &\\
DZOA4656-02   &  13 51 39.8 & -58 26 48 &   3.2 & {\it UG} & 0 1 0 0 &  		   &	     & &   \\
DZOA4657-03   &  13 52 55.4 & -58 09 59 &   2.3 & {\it UG} & 0 1 1 1 &  		   &	     & &   \\
DZOA4657-04   &  13 53 08.7 & -58 12 59 &   2.4 & {\it UG} & 0 1 1 0 &  		   &	     & &   \\
DZOA4657-02   &  13 53 32.4 & -57 49 30 &   2.7 & {\it DG} & 0 1 1 1 &  		   & 	     & &   \\
DZOA4657-01   &  13 53 34.1 & -57 49 10 &   2.7 & {\it DG} & 0 1 1 1 &  J13533399-5749098 &	     & &   \\
DZOA4658-06   &  13 54 18.1:& -59 56 14:&  10.1 & {\it UG} & 0 1 1 0 &  		   &   	     & &   \\
DZOA4658-04   &  13 54 23.1:& -58 06 30:&   2.5 & {\it DG} & 0 1 1 1 &  J13542319-5806299 & 	     & &   \\
DZOA4658-03   &  13 54 27.3:& -58 07 35:&   2.6 & {\it DG} & 1 1 1 1 &  J13542749-5807328 & WKK\,2660 & &   \\
DZOA4658-01   &  13 54 38.5:& -57 41 11:&   2.3 & {\it DG} & 1 1 1 0 &                    & WKK\,2669 & &   \\
DZOA4658-05   &  13 54 39.5:& -58 31 07:&   3.4 & {\it UG} & 0 1 1 1 &  		   &	     & &   \\
DZOA4659-13   &  13 55 18.4 & -58 05 32 &   2.8 & {\it UG} & 0 1 1 1 &  		   &	     & &   \\
DZOA4659-11   &  13 55 29.0 & -57 38 59 &   2.4 & {\it BG} & 1 1 1 0 &                    & WKK\,2684 & &   \\
DZOA4659-10   &  13 55 29.5 & -57 35 26 &   2.4 & {\it BG} & 1 1 1 0 &                    & WKK\,2686 & &   \\
DZOA4659-01   &  13 55 43.5 & -57 48 38 &   2.9 & {\it UG} & 0 1 1 1 &  J13554342-5748384 & 	     & &   \\
DZOA4659-02   &  13 56 07.7 & -58 52 21 &   3.8 & {\it UG} & 0 1 1 1 &  		   &	     & &   \\
DZOA4659-05   &  13 56 09.7 & -59 46 14 &   8.2 & {\it UG} & 0 1 1 1 &  		   &	     & &   \\
DZOA4660-06   &  13 56 42.9 & -59 11 16 &   5.1 & {\it NG} & 1 1 1 1 &                    & WKK\,2708 & &   \\
DZOA4660-02   &  13 56 52.3 & -57 51 53 &   2.5 & {\it DG} & 1 1 1 1 &                    & WKK\,2714 & &   \\
DZOA4660-05   &  13 57 01.5 & -59 12 35 &   5.4 & {\it DG} & 0 1 1 1 &  J13570135-5912362 & 	     & &   \\
DZOA4660-03   &  13 57 06.7 & -58 02 43 &   2.9 & {\it DG} & 0 1 1 1 &  		   &	     & &   \\
DZOA4660-01   &  13 57 08.2 & -57 44 09 &   2.6 & {\it UG} & 0 1 1 1 &  		   &	     & &   \\
DZOA4660-04   &  13 57 46.6 & -58 07 01 &   3.0 & {\it DG} & 0 1 1 1 &  J13574653-5807028 & 	     & &   \\
DZOA4661-04   &  13 58 40.7 & -58 18 13 &   3.6 & {\it NG} & 1 1 1 0 &                    & WKK\,2755 & &   \\
DZOA4661-03   &  13 58 52.6 & -59 45 47 &   8.9 & {\it UG} & 0 1 1 1 &  		   &	     & &   \\
DZOA4661-01   &  13 58 57.0 & -58 46 02 &   4.3 & {\it DG} & 1 1 1 1 &  J13585694-5846014 & WKK\,2764 & &    \\
DZOA4662-02   &  13 59 35.7:& -58 25 49:&   4.5 & {\it UG} & 0 1 1 1 &  J13593587-5825489 & 	     & &   \\
DZOA4662-03   &  14 00 01.9:& -58 29 06:&   4.8 & {\it UG} & 0 1 1 0 &  		   &	     & &   \\
DZOA4662-01   &  14 00 04.0:& -57 50 02:&   2.7 & {\it DG} & 1 1 1 1 &                    & WKK\,2785 & &   \\
DZOA4662-04   &  14 00 10.9:& -59 12 35:&   5.3 & {\it UG} & 0 1 1 0 &  		   &	     & &   \\
DZOA4662-07   &  14 00 26.9:& -60 18 31:&  15.4 & {\it UG} & 0 0 1 1 &  		   &	     & &   \\
DZOA4663-05   &  14 00 33.9 & -59 32 32 &   6.6 & {\it DG} & 0 1 1 1 &  		   &	     & &   \\
DZOA4663-02   &  14 00 55.3 & -57 37 23 &   2.9 & {\it DG} & 0 1 1 0 &  		   &	     & &   \\
DZOA4663-01   &  14 00 55.8 & -57 37 28 &   2.9 & {\it DG} & 0 1 1 0 &  J14005571-5737279 & 	     & &   \\
DZOA4663-04   &  14 01 25.6 & -58 37 37 &   4.4 & {\it DG} & 0 1 1 1 &  		   &	     & &   \\
DZOA4663-03   &  14 01 35.4 & -58 32 22 &   4.3 & {\it UG} & 0 1 1 0 &  		   &	     & &   \\
DZOA4664-03   &  14 02 42.0 & -57 44 08 &   3.3 & {\it NG} & 1 1 0 0 &                    & WKK\,2872 & &   \\
DZOA4664-02   &  14 03 11.2 & -58 06 20 &   3.6 & {\it DG} & 0 1 1 1 &  J14031094-5806201 & 	     & &   \\
DZOA4665-02   &  14 03 13.8 & -58 07 13 &   3.6 & {\it DG} & 0 1 1 1 &  J14031384-5807141 & 	     & &   \\
DZOA4665-03   &  14 03 57.5 & -58 09 49 &   4.0 & {\it UG} & 0 1 1 1 &  		   &	     & &   \\
DZOA4665-01   &  14 04 21.0 & -57 43 33 &   2.9 & {\it DG} & 0 1 1 1 &  J14042102-5743339 & 	     & &   \\
DZOA4666-01   &  14 05 35.9 & -59 32 19 &   8.5 & {\it UG} & 0 1 1 1 &  		   &	     & &   \\
\noalign{\smallskip}
\hline	      		
\noalign{\smallskip}	
\end{tabular*} 		
\normalsize		
\end{table*}		